\documentclass[prd,aps,showpacs,nofootinbib,eqsecnum]{revtex4}
\usepackage{amsmath}
\usepackage{amssymb}
\usepackage{amsfonts}
\usepackage{graphicx,bm}
\usepackage{dcolumn}
\usepackage{color,amsxtra}
\usepackage{epsf}
\usepackage{enumerate}
\usepackage{hhline}
\usepackage{array}
\usepackage{tabularx}
\usepackage{subfigure}
\usepackage{fancyhdr}
\usepackage{mathrsfs}

\newcommand{\be}{\begin{equation}}
\newcommand{\ee}{\end{equation}}
\newcommand{\bea}{\begin{eqnarray}}
\newcommand{\eea}{\end{eqnarray}}
\newcommand{\beaa}{\begin{eqnarray*}}
\newcommand{\eeaa}{\end{eqnarray*}}

\newcommand{\e}{\mathrm{e}}

\newcommand{\Eqn}[1]{&\hspace{-0.2em}#1\hspace{-0.2em}&}

\def\be{\begin{equation}}
\def\ee{\end{equation}}
\def\bea{\begin{eqnarray}}
\def\eea{\end{eqnarray}}

\def\e{\mathrm{e}}

\begin{document}


\title{Cosmic history of viable exponential gravity: Equation of state 
oscillations and growth index from inflation to dark energy era
}

\author{
Kazuharu Bamba$^{1, }$\footnote{
E-mail address: bamba@kmi.nagoya-u.ac.jp},
Antonio Lopez-Revelles$^{2, 3, }$\footnote{E-mail address: alopez@ieec.uab.es}, \\
R. Myrzakulov$^{4, }$\footnote{E-mail address: myrzakulov@gmail.com, 
rmyrzakulov@csufresno.edu},
S.~D. Odintsov$^{2, 4, 5, }$\footnote{E-mail address: odintsov@ieec.uab.es, also at TSPU, Tomsk} 
and 
L.~Sebastiani$^{3, 4, }$\footnote{E-mail
address: l.sebastiani@science.unitn.it}
}
\affiliation{
$^1$Kobayashi-Maskawa Institute for the Origin of Particles and the
Universe,
Nagoya University, Nagoya 464-8602, Japan\\ 
$^2$Consejo Superior de Investigaciones Cient\'{\i}ficas, ICE/CSIC-IEEC, 
Campus UAB, Facultat de Ci\`{e}ncies, Torre C5-Parell-2a pl, E-08193
Bellaterra (Barcelona), Spain\\
$^3$Dipartimento di Fisica, Universit\`a di Trento
and Istituto Nazionale di Fisica Nucleare, 
Gruppo Collegato di Trento, Italia\\
$^4$ Eurasian International Center for Theoretical Physics and Department of General \& Theoretical Physics, Eurasian National University, Astana 010008, Kazakhstan\\ 
$^5$Instituci\'{o} Catalana de Recerca i Estudis Avan\c{c}ats
(ICREA), Spain 
}



\begin{abstract}

A generic feature of viable $F(R)$ gravity is investigated:
It is demonstrated that during the matter dominated era the
large frequency oscillations of the effective dark energy may influence 
the behavior of higher derivatives of the Hubble parameter 
with the risk to produce some singular unphysical solutions at high redshift. 
This behavior is explicitly analyzed for realistic $F(R)$ models, 
in particular, exponential gravity and a power form model. 
To stabilize such oscillations, we consider the additional modification of 
the models via a correction term which does not destroy 
the viability properties. 
A detailed analysis on the future evolution of the universe and 
the evolution history of the growth index of the matter density perturbations 
are performed. 
Furthermore, 
we explore two applications of exponential gravity to the inflationary 
scenario. 
We show how it is possible to obtain different numbers of $e$-folds 
during the early-time acceleration by making different choices of the model 
parameters in the presence of ultrarelativistic matter, which destabilizes 
inflation and eventually leads to the exit from the inflationary stage. 
We execute the numerical analysis of inflation in two viable 
exponential gravity models. 
It is proved that at the end of the inflation, the effective 
energy density and curvature of the universe decrease and 
thus a unified description between inflation and the $\Lambda$CDM-like dark 
energy dominated era can be realized. 

\end{abstract}


\maketitle

\section{Introduction \label{SectI}}

The current cosmic acceleration is 
supported by various observations such as 
Supernovae Ia (SNe Ia)~\cite{SN1}, 
large scale structure (LSS)~\cite{LSS} with 
baryon acoustic oscillations (BAO)~\cite{Eisenstein:2005su}, 
cosmic microwave background (CMB) radiation~\cite{WMAP-Spergel, WMAP, 
Komatsu:2010fb} 
and weak lensing~\cite{Jain:2003tba}. 
There exist two representative procedures to solve this problem, namely, 
introducing ``dark energy'' in general relativity 
(for recent reviews in terms of dark energy, see~\cite{Li:2011sd, Kunz:2012aw, Bamba:2012cp}) and 
modifying the gravitational theory like $F(R)$ gravity 
(for recent reviews on modified gravity, 
see~\cite{Review-Nojiri-Odintsov, Viableconditions, Clifton:2011jh, 
Capozziello:2011et, Capozziello:2012hm}). 
In this paper, we adopt modified gravity approach to describe the inflation and dark energy eras.

There proposed several viable $F(R)$ gravity models have been constructed 
(for concrete viable models, see, e.g., the above reviews or~\cite{Bamba} and 
references therein). 
The conditions for the viability are summarized as follows: 
(i) Positive definiteness of the effective gravitational coupling. 
(ii) Matter stability condition~\cite{DEinflation, Dolgov:2003px, 
Faraoni:2006sy, Song:2006ej}. 
(iii) 
In the large curvature regime, the model is close to 
the $\Lambda$-Cold-Dark-Matter ($\Lambda$CDM) model asymptotically. 
(iv) Stability of the late-time de Sitter point~\cite{D-S-P, Amendola}. 
(v) The equivalence principle. 
(vi) Solar-system tests~\cite{DEinflation,Chiba:2003ir, SolarSystemconstraints}. 
It is considered to be one of the most significant issues on 
cosmology in the framework of $F(R)$ gravity 
to realize the unification of inflation with 
the late time cosmic acceleration \cite{Review-Nojiri-Odintsov,DEinflation}
(for the first proposal of $1/R$ gravity as gravitational alternative for dark energy, see~\cite{A,B}). 

In this paper, we study 
a generic feature of viable $F(R)$ gravity models, in particular, 
exponential gravity and a power form model. 
We find that 
the behavior of higher derivatives of the Hubble parameter may be 
influenced by large frequency oscillations of effective dark energy, 
which makes solutions singular and unphysical at a high redshift. 
Therefore, in order to stabilize such oscillations, 
we examine an additional correction term to the model 
and remove such an instability with keeping the viability properties. 
We also demonstrate the cosmological evolutions 
of the universe and growth index of the matter density perturbations 
in detail. 
Furthermore, 
by applying two viable models of exponential gravity to inflationary cosmology 
and executing the numerical analysis of the inflation process, 
we illustrate that the exit from inflation can be realized. 
Concretely, we demonstrate that 
different numbers of $e$-folds during inflation can be obtained 
by taking different model parameters in the presence of ultrarelativistic 
matter, the existence of which makes inflation end and leads to 
the exit from inflation. 
Indeed, we observe that at the end of the inflation, the effective energy density as well as the curvature of the universe decrease. 
Accordingly, a unified description between inflation and the late time cosmic acceleration is presented. 
We use units of $k_\mathrm{B} = c = \hbar = 1$ and denote the
gravitational constant $8 \pi G$ by 
${\kappa}^2 \equiv 8\pi/{M_{\mathrm{Pl}}}^2$ 
with the Planck mass of $M_{\mathrm{Pl}} = G^{-1/2} = 1.2 \times 10^{19}$GeV.

The paper is organized as follows. 
In Sec.~II, we briefly review the formulations of $F(R)$ gravity. 
We use the fluid representation of $F(R)$ gravity~\cite{cno06}. 
Here, in the Friedmann-Lema\^{i}tre-Robertson-Walker (FLRW) background, 
the equations of motion with the addition of an effective gravitational fluid 
are presented. 
In Sec.~III, 
we explain two well-known viable $F(R)$ gravity models and 
show those generic feature occurring in the matter dominated era, when large frequency oscillation of dark energy appears and influences on the behavior of higher derivatives of the Hubble parameter in terms of time 
with the risk to produce some divergence and to render the solution unphysical. Thus, we suggest a way to stabilize such oscillations by introducing an additive modification to the models. 
We also perform a numerical analysis of the matter dominated era. 
In Sec.~IV, 
we demonstrate that the term added to stabilize the dark energy oscillations 
in the matter dominated epoch does not cause any problem on the viability of 
the models, which satisfy the cosmological and local gravity constraints. 
We investigate their future evolution and show that the effective crossing of 
the phantom divide, which characterizes the de Sitter epoch, takes place 
in the very far future. 
We also analyze the growth index using three different ansatz choices.  
The second part of the paper is devoted to the study of $F(R)$ models for the 
unification of the early-time cosmic acceleration, i.e., inflation, and the late-time one. 
In Sec.~V, 
we explore two applications of exponential gravity for inflation. 
In particular, we show how it is possible to obtain different numbers of 
$e$-folds during inflation 
by making different choices of model parameters in the presence of 
ultrarelativistic matter in the early universe. 
In Sec.~VI, 
we execute the numerical analysis of inflation and illustrate that 
at the end of it 
the effective energy density and the curvature decrease and eventually 
the cosmology in the $\Lambda$CDM model can follow. 
Finally, the summary and outlook for this work are given in Sec.~VII. 
For reference, 
we also explain the procedure of conformal transformation 
in Appendix A and asymptotically phantom or quintessence modified gravity 
in Appendix B. 



\setcounter{equation}{0}

\section{$F(R)$ gravity and its dynamics in the FLRW universe: 
General overview}

In this section, we briefly review formulations in $F(R)$ gravity 
and derive the gravitational field equations in the FLRW space-time. 
The action describing $F(R)$ gravity is given by 
\begin{equation}
I=\int_{\mathcal{M}} d^4x\sqrt{-g}\left[
\frac{F(R)}{2\kappa^2}+\mathcal{L}^{\mathrm{(matter)}}\right]\,,
\label{action}
\end{equation}
where $F(R)$ is a generic function of the Ricci scalar $R$ only, 
$g$ is the determinant of the metric tensor $g_{\mu\nu}$, 
${\mathcal{L}}^{\mathrm{(matter)}}$ is the matter Lagrangian 
and $\mathcal{M}$ denotes the space-time manifold. 
In a large class of modified gravity models reproducing 
the standard cosmology in General Relativity (GR), 
i.e., 
$F(R)=R$, with a suitable correction to realize current acceleration 
and/or inflation, one represents 
\begin{equation}
F(R)=R+f(R)\,.
\label{actiontwo}
\end{equation}
Thus, the modification of gravity is encoded in the function $f(R)$, which is 
added to the classical term $R$ of the Einstein-Hilbert action 
in GR. 
In what follows, we discuss modified gravity in this form by explicitly 
separating the contribution of its modification from GR.  
The field equation simply reads
%
\begin{equation}
F'(R)\left(R_{\mu\nu}-\frac{1}{2}Rg_{\mu\nu}\right)=\kappa^2T^{{\mathrm{(matter)}}}_{\mu\nu}+\left[\frac{1}{2}g_{\mu\nu}\left(F(R)-RF'(R)\right)
+\left(\nabla_{\mu}\nabla_{\nu}-g_{\mu\nu}\Box\right)F'(R)\right]\,.
\label{Field equation}
\end{equation}
%
Here, ${\nabla}_{\mu}$ is the covariant derivative 
operator associated with $g_{\mu \nu}$, 
$\Box\phi\equiv g^{\mu\nu}\nabla_{\mu}\nabla_{\nu}\phi$ is the 
covariant d'Alembertian for a scalar field $\phi$, and 
$T^{\mu (\mathrm{matter})}_\nu = \mathrm{diag} \left(-\rho_{\mathrm{m}}, P_{\mathrm{m}}, P_{\mathrm{m}}, P_{\mathrm{m}} \right)$ 
is the contribution to the stress energy-momentum tensor from 
all ordinary matters, with 
$\rho_{\mathrm{m}}$ and $P_{\mathrm{m}}$ being the energy density and 
pressure of matter, respectively. 
Moreover, 
the prime denotes the derivative with respect to the curvature $R$. 

The flat FLRW space-time is described by the metric
%
$ds^{2}=-dt^{2}+a(t)^2 d \mathbf{x}^{2}
$,
%
where $a(t)$ is the scale factor of the universe. 
The Ricci scalar reads 
\begin{equation}
R=12H^2+6\dot H\,, 
\end{equation}
where $H=\dot{a}(t)/a(t)$ is the Hubble parameter and 
the dot denotes the time derivative of 
$\partial_t (\equiv \partial/\partial t)$. 
In the flat FLRW background, from the $(\mu,\nu)=(0,0)$ component
and the trace part of $(\mu,\nu)=(i,j)$ (with $i,j=1,\cdots,3$) components in
Eq.~(\ref{Field equation}), we obtain 
the gravitational field equations~\cite{Review-Nojiri-Odintsov} 
%
\begin{eqnarray}
\rho_{\mathrm{eff}} \Eqn{=} \frac{3}{\kappa^{2}}H^{2}\,, 
\label{EOM1bis}\\ 
P_{\mathrm{eff}} \Eqn{=} 
-\frac{1}{\kappa^{2}} \left( 2\dot H+3H^{2} \right)\,. 
\label{EOM2bis}
\end{eqnarray}
Here, $\rho_{\mathrm{eff}}$ and $P_{\mathrm{eff}}$ are 
the effective energy density and pressure of the universe, respectively, 
defined as
\begin{eqnarray}
\rho_{\mathrm{eff}} \Eqn{\equiv} 
\rho_{\mathrm{m}} + 
\frac{1}{2\kappa^{2}}
\left[ \left( F'R-F\right)-6H^2(F'-1)
-6H\dot F'
\right]\,,
\label{rhoeffRG} \\ 
P_{\mathrm{eff}} \Eqn{\equiv} 
P_{\mathrm{m}} +
\frac{1}{2\kappa^{2}} \Bigl[
-\left( F'R-F \right)+(4\dot{H}+6H^2)(F'-1)
+4H\dot F'+2\ddot F'
\Bigr]\,.
\label{peffRG}
\end{eqnarray}
In this way, we have a fluid representation of 
the so-called geometrical dark energy in $F(R)$ gravity 
with the energy density $\rho_{\mathrm{DE}}=\rho_{\mathrm{eff}}-\rho$ and 
pressure $P_{\mathrm{DE}}=P_{\mathrm{eff}}-P$. 
However, it is important for us to remember 
that gravitational terms enter in both left and right sides of 
Eqs.~(\ref{EOM1bis}) and (\ref{EOM2bis}). 
For general relativity in which $F(R)=R$, 
$\rho_{\mathrm{eff}} = \rho_{\mathrm{m}}$ and 
$P_{\mathrm{eff}} = P_{\mathrm{m}}$ and 
therefore Eqs.~(\ref{EOM1bis}) and (\ref{EOM2bis}) lead to 
Friedman equations. 

We also explain basic equations that we use to carry out our analysis. 
In order to study the dynamics of $F(R)$ gravity models in the flat FLRW 
universe, we may introduce the variable~\cite{Bamba, HuSaw}
\begin{equation}
y_H (z)\equiv\frac{\rho_{\mathrm{DE}}}{\rho_{\mathrm{m}(0)}}=\frac{H^2}{\tilde{m}^2}-(z+1)^3-\chi
(z+1)^{4}\,.
\label{y}
\end{equation}
Here, $\rho_{\mathrm{m}(0)}$ is the energy density of matter 
at the present time, 
$\tilde{m}^2$ is the mass scale, given by 
\begin{equation*}
\tilde{m}^2\equiv\frac{\kappa^2\rho_{\mathrm{m}(0)}}{3}\simeq 1.5 \times
10^{-67}\text{eV}^2\,,
\end{equation*}
and $\chi$ is defined as~\cite{WMAP} 
\begin{equation*}
\chi\equiv\frac{\rho_{\mathrm{r}(0)}}{\rho_{\mathrm{m}(0)}}\simeq 3.1 \times
10^{-4}\,,
\end{equation*}
where $\rho_{\mathrm{r}(0)}$ is the current energy density of radiation 
and $z=1/a(t)-1$ is the redshift. Here, we have taken the current value of 
the scale factor as unity. 
By using Eqs.~(\ref{EOM1bis}) and (\ref{y}), we find 
\begin{equation}
\frac{d^2 y_H(z)}{d z^2}+J_1\frac{d y_H(z)}{d z}+J_2
\left(y_H(z)\right)+J_3=0\,,
\label{superEq}
\end{equation}
where
\begin{eqnarray}
J_1 \Eqn{=} \frac{1}{(z+1)}\left[-3-\frac{1}{y_H+(z+1)^{3}+\chi (z+1)^{4}}\frac{1-F'(R)}{6\tilde{m}^2
F''(R)}\right]\,, 
\\ 
J_2 \Eqn{=} 
\frac{1}{(z+1)^2}\left[\frac{1}{y_H+(z+1)^{3}+\chi (z+1)^{4}}\frac{2-F'(R)}{3\tilde{m}^2 F''(R)}\right]\,, 
\\ 
J_3 \Eqn{=} 
-3 (z+1) 
\nonumber \\ 
&& 
-\frac{(1-F'(R))((z+1)^{3}+2\chi (z+1)^{4})
+(R-F(R))/(3\tilde{m}^2)}{(z+1)^2(y_H+(z+1)^{3}+\chi
(z+1)^{4})}\frac{1}{6\tilde{m}^2
F''(R)}\,.
\end{eqnarray}
%
Furthermore, 
the Ricci scalar is expressed as 
\begin{equation}
R=3\tilde{m}^2 \left[4y_H(z)-(z+1)\frac{d y_H(z)}{d z}+(z+1)^{3}\right]\,. 
\label{Ricciscalar}
\end{equation}
In deriving this equation, 
we have used 
the fact that $-(z+1)H(z) d/d z=H(t) d/d(\ln a(t))=d/d t$, 
where $H$ could be an explicit function of the red shift as $H=H(z)$, 
or an explicit function of the time as $H=H(t)$. 
In general, Eq.~(\ref{superEq}) can be solved in a numerical way, 
once we write the explicit form of an $F(R)$ gravity model.

\section{Generic feature of realistic $F(R)$ gravity models in the matter 
dominated era} 

In this section, 
we consider viable $F(R)$ gravity models representing 
a realistic scenario to account for dark energy, in particular, two well-known 
ones proposed in Refs.~\cite{HuSaw, Cognola:2007zu, Linder:2009jz, Battye, Starobinsky:2007hu, Tsujikawa:2007xu} 
(for more examples and detailed explanations on viable models, 
see, e.g.,~\cite{Bamba, Bamba:2010iy} and references therein). 
Here, we mention that in Ref.~\cite{Yang:2011cp}, 
the gravitational waves in viable $F(R)$ models have been studied, 
and that the observational constraints on exponential gravity 
have also been examined in Ref.~\cite{Yang:2010xq}. 
We show that for these models, large frequency oscillation of dark energy 
in the matter dominated era appears, and that it may influence on the behavior 
of higher derivatives of the Hubble parameter with respect to time. 
Such a oscillation has the risk to produce some divergence, 
and therefore we suggest a way to stabilize the frequency oscillation by 
performing the subsequent numerical analysis. 
In these models, 
a correction term 
to the Hilbert-Einstein action is added 
as $F(R)=R+f(R)$ in~(\ref{actiontwo}), 
so that 
the current acceleration of the universe can be reproduced in a simple way. 
Namely, 
a vanishing (or fast decreasing) cosmological constant in the flat limit 
of $R\rightarrow 0$ is incorporated, and a suitable, constant asymptotic 
behavior for large values of $R$ is exhibited.

\subsection{Realistic $F(R)$ gravity models} 

First, we explore the Hu-Sawicki model~\cite{HuSaw} 
(for the related study of such a model, see Ref.~\cite{HSbis}),
\begin{equation}
F(R)=R-\frac{\tilde{m}^{2}c_{1}(R/\tilde{m}^{2})^{n}}{c_{2}(R/\tilde{m}^{2})^{n}+1}=R-\frac{\tilde{m}^{2}c_{1}}{c_{2}}+\frac{\tilde{m}^{2}c_{1}/c_{2}}{c_{2}(R/\tilde{m}^{2})^{n}+1}\,,
\label{HuSawModel}
\end{equation}
where $\tilde{m}^{2}$ is the mass scale, $c_{1}$ and $c_{2}$ are positive parameters, and $n$ is a natural positive number. 
The model is very carefully constructed such that 
in the high curvature regime, 
$\tilde{m}^{2} c_{1}/c_{2}=2\Lambda$ can play a role of 
the cosmological constant $\Lambda$ and thus 
the $\Lambda$CDM model can be reproduced. 

Moreover, 
in Refs.~\cite{Cognola:2007zu, Linder:2009jz} 
another simple model which may easily be generalized to reproduce also 
inflation has been constructed 
\begin{equation}
F(R)=R-2\Lambda\left[1-\mathrm{e}^{-R/\left(b\,\Lambda\right)}\right]\,,
\label{model}
\end{equation}
where $b>0$ is a free parameter. 
Also in this model, in the flat space the solution of the Minkowski space-time 
is recovered, while at large curvatures 
the $\Lambda$CDM model is realized. 
This kind of models 
can satisfy the cosmological and local gravity constraints. 
Both of these models asymptotically approach the 
$\Lambda$CDM model in the high curvature regime. 
Indeed, however, the mechanisms work in two different manners, i.e., 
via a power function of $R$ (the first one) and 
via an exponential function of it (the second one). 
For our treatment, we reparameterize the model (\ref{HuSawModel}) by 
describing $c_{1}\tilde{m}^{2}/c_{2}=2\Lambda$ and 
$(c_2)^{1/n}\,\tilde{m}^2=b\,\Lambda$ with $b>0$, so that we can obtain 
\begin{equation}
F(R)=R-2\Lambda\left\{1-\frac{1}{\left[R/\left(b\,\Lambda\right)\right]^{n}+1}\right\}\,, 
\quad n=4\,.
\label{model2}
\end{equation}
Through this procedure, 
in both of these models the term $b\,\Lambda$ corresponds to 
the curvature for which the cosmological 
constant is ``switched on''. 
This means $b\ll 4$, so that $b\,\Lambda\ll 4\Lambda$ and hence $R=4\Lambda$ 
can be the curvature of de Sitter universe describing the current cosmic 
acceleration. 
In the mode in Eq.~(\ref{model2}), since 
$n$ has to be sufficiently large in order to reproduce the $\Lambda$CDM model, 
we have assumed $n=4$ and we keep only the parameter $b$ free.

\subsection{Dark energy oscillations in the matter dominated era}

Despite the fact that the models in Eqs.~(\ref{model}) and (\ref{model2}) 
precisely resemble the $\Lambda$CDM model, 
there is a problem that in the matter dominated era 
the higher derivatives of the Hubble parameter diverge and thus this can 
make the solutions unphysical. 
This problem originates from the stability conditions to be 
satisfied by these models~\cite{omegaDEnostrum} and from dark energy oscillations during the matter phase~\cite{Starobinsky:2007hu} 
in Ref.~\cite{Amendola}.
Since in matter dominated era $R=3\tilde m^2(z+1)^3$ and $y_H(z)\ll(1+z)^3$ and $\chi(1+z)^4\ll (z+1)^3$ in order 
for dark energy and radiation to vanish during this phase, one may locally solve Eq. (\ref{superEq}) around $z=z_0+(z-z_0)$, where $|z-z_0|\ll z$. The solution reads to the first order in terms of $(z-z_0)$, 
\begin{equation}
y_H''(z)+\frac{\alpha}{(z-z_0)}y'_H(z)+
\frac{\beta}{(z-z_0)^2}y_H(z)=
\zeta_0+
\zeta_1(z-z_0)\,,
\label{meq}
\end{equation}
where 
\begin{eqnarray}
\alpha \Eqn{=} -\frac{7}{2}-\frac{(1-F'(R_0))F'''(R_0)}{2F''(R_0)^2}\,,\nonumber\\
\beta \Eqn{=} 2+\frac{1}{R_0 F''(R_0)}+\frac{2(1-F'(R_0))F'''(R_0)}{F''(R_0)^2}\,,\label{alphabeta}
\end{eqnarray}
with $\zeta_0$ and $\zeta_1$ being 
constants and $R_0=3\tilde{m}^2(z_0+1)^3$. 
Thus, the solution of Eq.~(\ref{meq}) is derived as 
\begin{equation}
y_H(z)= a+b\cdot(z-z_0) + C_0 \cdot 
\exp{\frac{1}{2(z_0+1)}\left(-\alpha\pm\sqrt{\alpha^2-4\beta}\right)(z-z_0)}\,, \label{resultmatter}
\end{equation}
where $a$, $b$ and $C_0$ are constants. 
Now, for the two models in Eqs.~(\ref{model}) and (\ref{model2}), 
when $R\gg b\,\Lambda$, 
we find 
\begin{eqnarray}
F'(R) \Eqn{\simeq} 1\,, \nonumber\\
F''(R) \Eqn{\simeq} 0^+\,. 
\end{eqnarray}
These behaviors guarantee the occurrence of the realistic matter dominated era.
Furthermore, 
since in the expanding universe $(z-z_0)<0$, it turns out that the dark energy perturbations in Eq.~(\ref{resultmatter}) remain small around $R_0$, 
and that we acquire 
\begin{equation}
\frac{\left(1-F'(R_0)\right)F'''(R_0)}{2F''(R_0)^2}>-\frac{7}{2}\,,
\quad
\frac{1}{R_0F''(R_0)}>12\,,\label{q}
\end{equation}
for both these models. 
Owing to the fact that $F''(R)$ is very close to $0^+$, the discriminant in the square root of Eq.~(\ref{resultmatter}) is negative and dark energy oscillates 
as
%
\begin{equation}
y_H(z)= \frac{\Lambda}{3\tilde{m}^2}+\mathrm{e}^{-\frac{\alpha_{1,2}(z-z_0)}{2(z_0+1)}}\left[A\sin\left(\frac{\sqrt{\beta_{1,2}}}{(z_0+1)}(z-z_0)\right)+
B\cos\left(\frac{\sqrt{\beta_{1,2}}}{(z_0+1)}(z-z_0)\right)\right]\,.
\label{matterDE}
\end{equation}
%
Here, $A$ and $B$ are constants and $\alpha_{1,2}$ and $\beta_{1,2}$ are given by Eq. (\ref{alphabeta}), so they correspond to two models under investigation. In particular, $\alpha_1=-3$ for the model in Eq. (\ref{model}) and $\alpha_2\simeq -29/10$ for the model in Eq. (\ref{model2}), while $\beta_{1,2}\simeq 1/(R_0F''(R_0))$, i.e., 
%
\begin{equation}
\beta_1\simeq\left(\frac{b^2\Lambda\mathrm{e}^{\frac{R_0}{\tilde{R}}}}{2 R_0}\right)\,,  
\end{equation}
in case of exponential model in Eq.~(\ref{model}) and 
\begin{equation}
\beta_2\simeq\frac{R_0\left[1+\left(\frac{R_0}{b\Lambda}\right)^n\right]^3\left(\frac{b\Lambda}{R_0}\right)^n}{2\Lambda\,n\left\{1+n\left[\left(\frac{R_0}{b\Lambda}\right)^n-1\right]+\left(\frac{R_0}{b\Lambda}\right)^n\right\}}\simeq\frac{R_0}{2\Lambda\,n(n+1)}\left(\frac{R_0}{b\Lambda}\right)^n\,, 
\end{equation}
in case of model in Eq.~(\ref{model2}). 
This means that 
the frequency of dark energy oscillations increases as the curvature (and redshift) becomes large. Moreover, the effects of such oscillations are amplified in 
the derivatives of the dark energy density, namely, 
\begin{equation}
\left| \frac{d^n}{d t^n}y_H(t_0)\right|\propto 
\left(\mathcal{F}(z_0)\right)^n\,,
\label{frequency}
\end{equation}
where $\mathcal{F}(z)\simeq \left(R*F''(R)\right)^{-1/2}/(z+1)$ is the oscillation frequency and 
$t_0$ is the cosmic time corresponding to the redshift $z_0$. 
This is for example the case of the EoS parameter 
for dark energy defined as\footnote{Throughout this paper, we describe 
the EoS parameter by ``$\omega$'' and not ``$w$''.} 
\begin{equation}
\omega_{\mathrm{DE}}(z)\equiv\frac{P_{\mathrm{DE}}}{\rho_{\mathrm{DE}}}=-1+\frac{1}{3}(z+1)\frac{1}{y_H(z)}\frac{d y_H(z)}{d (z)}\,.
\label{oo}
\end{equation}
%
%
%
%
For large values of the redshift, the dark energy density oscillates 
with a high frequency and also its derivatives become large, showing a different feature of the dark energy EoS parameter in the models in Eqs.~(\ref{model}) 
and (\ref{model2}) compared with the case of the cosmological constant in GR. 
During the matter dominated era, the Hubble parameter behaves as 
\begin{equation}
H(z)\simeq \sqrt{\tilde{m}^2}\left[(z+1)^{3/2}+\frac{y_H(z)}{2(z+1)^{3/2}}\right]\,.
\label{eq:FR5-15-3.14}
\end{equation}
If the frequency $\mathcal{F}(z_0)$ in Eq.~(\ref{frequency}) is extremely large, the derivatives of dark energy density could become dominant in some higher derivatives of the Hubble parameter 
%
%
which may approach an effective singularity and therefore make 
the solution unphysical. 
We see it for specific cases. 
In Refs.~\cite{Bamba, twostep, altri}, the cosmological evolutions in exponential gravity and the Hu-Sawicki model have carefully been explored. 
It has explicitly been demonstrated that the late-time cosmic acceleration 
which follows the matter dominated era can occur, according with astrophysical 
data. 
A reasonable choice is to take $b=1$ for both these models. 
We also put $\Lambda = 7.93 \tilde{m}^2$~\cite{WMAP}. 
We can solve Eq.~(\ref{superEq}) numerically\footnote{We have used 
Mathematica 7 \textcopyright.} by taking the initial conditions at 
$z=z_i$, where $z_i\gg 0$ is the redshift at the initial time to 
execute the numerical calculation, as follows: 
\begin{eqnarray*}
\frac{d y_H(z)}{d (z)}\Big\vert_{z_i} \Eqn{=} 0\,, \\ 
y_H(z)\Big\vert_{z_i} \Eqn{=} \frac{\Lambda}{3\tilde{m}^2}\,. 
\end{eqnarray*}
Here, we have used the fact that at a high redshift the universe should be 
very close to the $\Lambda$CDM model. 
We have set $z_i=2.80$ for the model in Eq.~(\ref{model}) and $z_i=4.5$ for the model in Eq.~(\ref{model2}), such that $R\,F''(R)\sim 10^{-8}$ at $R=3\tilde m^2(z_i+1)^3$. 
We note that it is hard to extrapolate 
the numerical results to the higher redshifts because of the large frequency of dark energy oscillations. 

\begin{figure}[!h]
\subfigure[]{\includegraphics[width=0.3\textwidth]{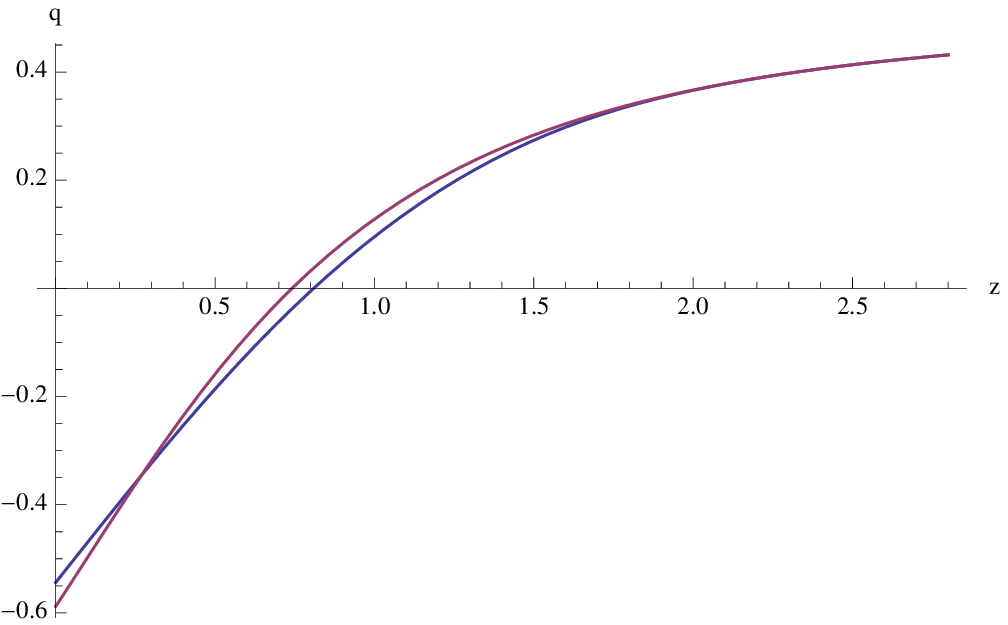}}
\qquad
\subfigure[]{\includegraphics[width=0.3\textwidth]{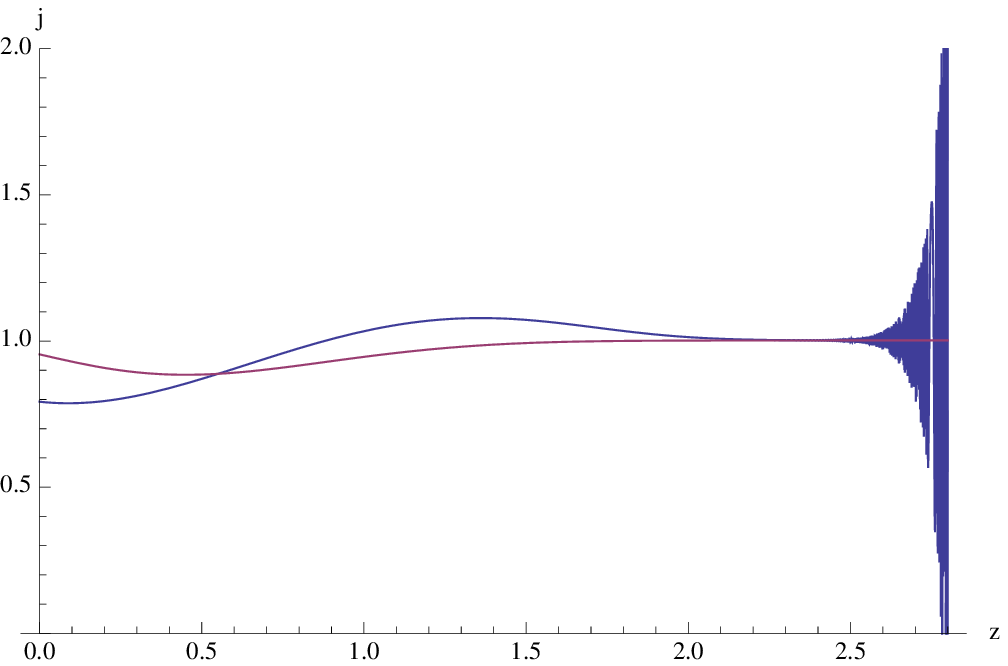}}
\qquad
\subfigure[]{\includegraphics[width=0.3\textwidth]{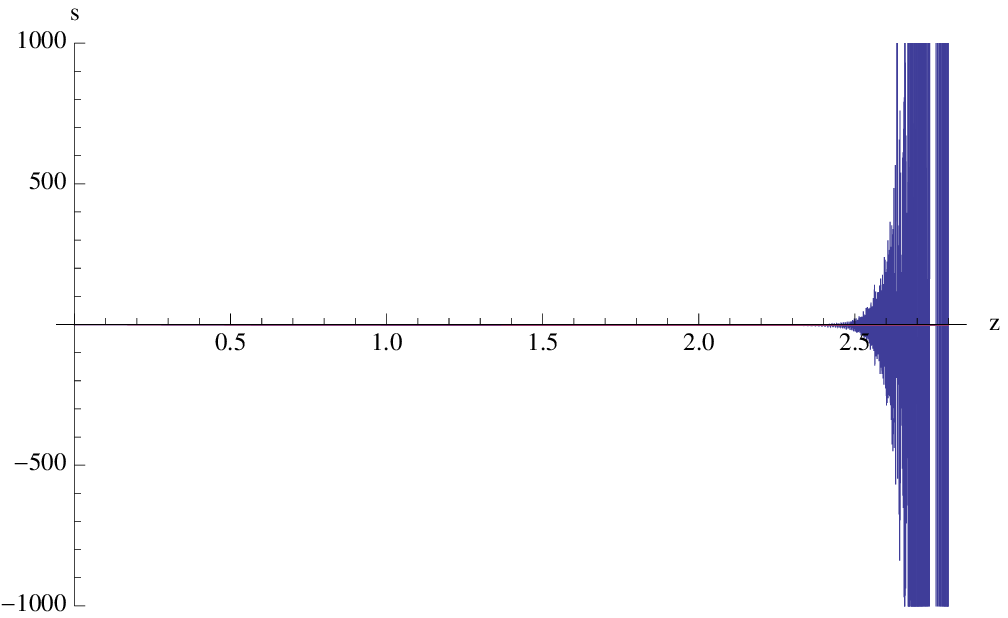}}
\qquad
\subfigure[]{\includegraphics[width=0.3\textwidth]{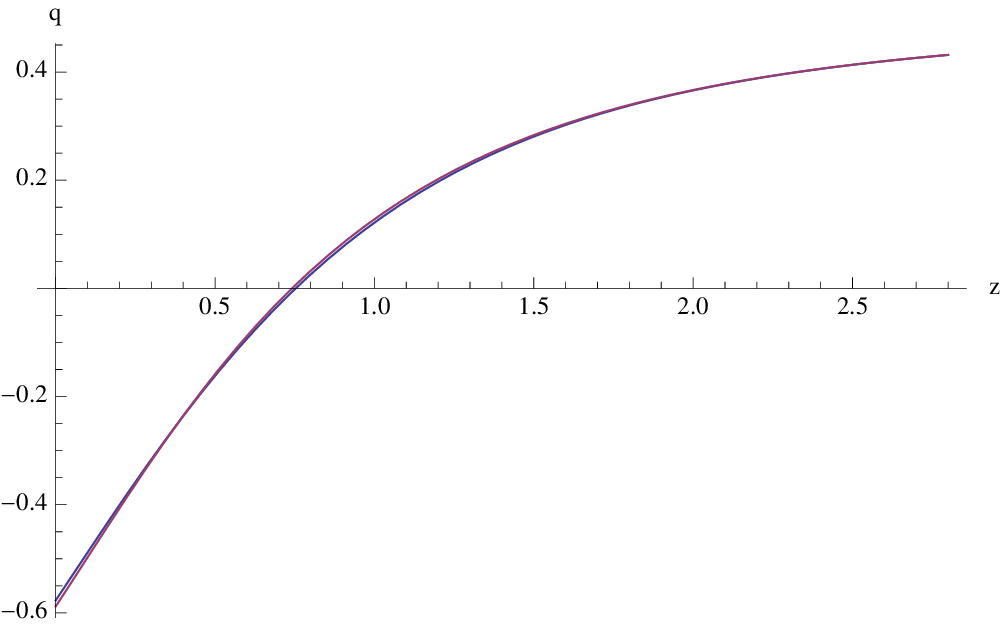}}
\qquad
\subfigure[]{\includegraphics[width=0.3\textwidth]{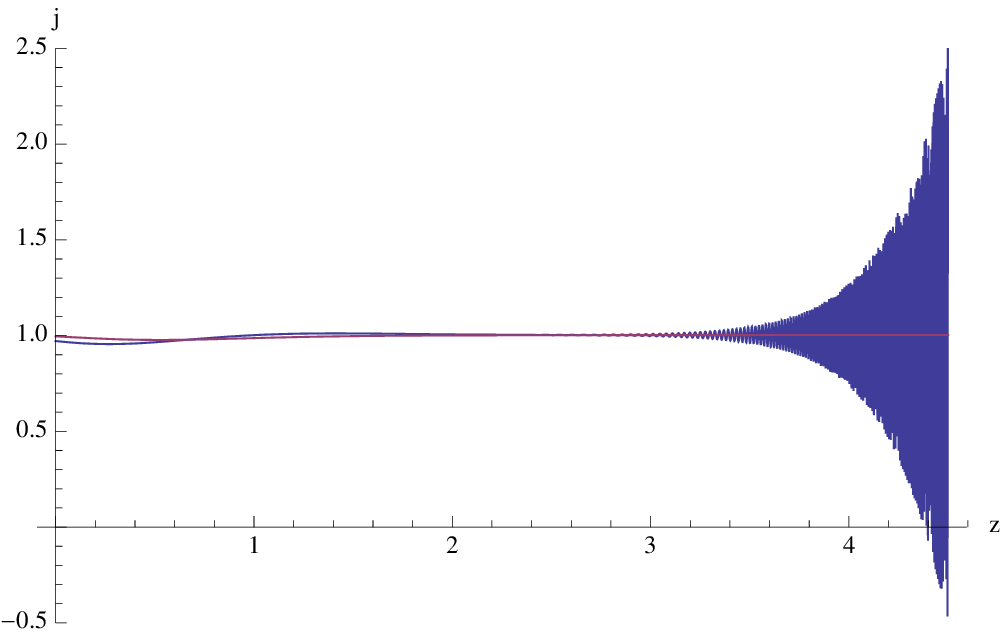}}
\qquad
\subfigure[]{\includegraphics[width=0.3\textwidth]{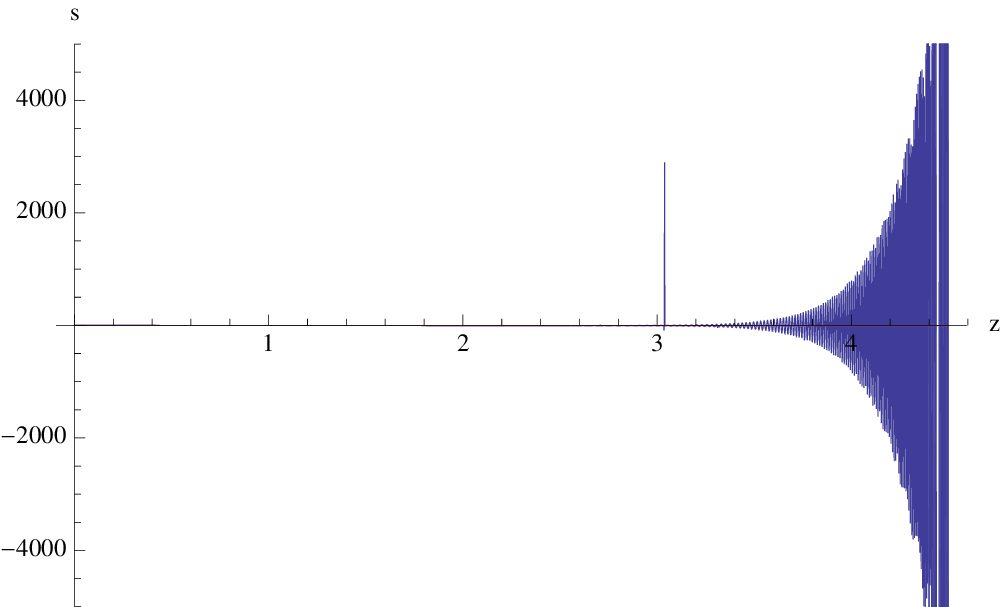}}
\caption{
Cosmological evolutions of $q(z)$ [(a) and (d)], $j(z)$ [(b) and (e)] and $s(z)$ [(c) and (f)] parameters as functions of the redshift $z$ for Model I 
[(a)--(c)] and Model II [(d)--(f)] in the region of $z>0$. 
\label{parameters}}
\end{figure}

Using Eq.~(\ref{oo}) with $y_{H}$, we derive $\omega_\mathrm{DE}$. 
In addition, by using Eq.~(\ref{Ricciscalar}) we obtain $R$ as a function of the redshift. 
We can also execute the extrapolation in terms of 
the behavior of $\Omega_{\mathrm{DE}}$, given by 
\begin{equation}
\Omega_\mathrm{DE}(z)\equiv\frac{\rho_\mathrm{DE}}{\rho_\mathrm{eff}}
=\frac{y_H}{y_H+\left(z+1\right)^3+\chi\left(z+1\right)^4}\,.
\end{equation} 
The numerical extrapolation to the present universe leads to the 
following results: 
For the model~(\ref{model}), 
$y_H(0)=2.736$, $\omega_{\mathrm{DE}}(0)=-0.950$, $\Omega_{\mathrm{DE}}(0)=0.732$ and $R(z=0)=4.365$, 
whereas for the model~(\ref{model2}), 
$y_H(0)=2.652$, $\omega_{\mathrm{DE}}(0)=-0.989$, $\Omega_{\mathrm{DE}}(0)=0.726$ and $R(z=0)=4.358$. 
These resultant data are in accordance with the last and very accurate 
observations of our current universe~\cite{WMAP}, which are 
\begin{eqnarray}
\omega_\mathrm{DE} \Eqn{=} -0.972^{+0.061}_{-0.060}\,, \nonumber \\ 
\Omega_\mathrm{DE} \Eqn{=} 0.721\pm 0.015\,.
\label{data}
\end{eqnarray}
Next, 
we introduce the deceleration $q$, jerk $j$ and snap $s$ 
parameters~\cite{Chiba:1998tc, Sahni:2002fz}
\begin{eqnarray}
q(t) \Eqn{\equiv} -\frac{1}{a(t)}\frac{d^2 a(t)}{d t^2}\frac{1}{H(t)^2}=-\frac{\dot H}{H^2}-H^2\nonumber\\
j(t) \Eqn{\equiv} \frac{1}{a(t)}\frac{d^3 a(t)}{d t^3}\frac{1}{H(t)^3}=\frac{\ddot H}{H^3}-3q-2\nonumber\\
s(t) \Eqn{\equiv} \frac{1}{a(t)}\frac{d^4 a(t)}{d t^4}\frac{1}{H(t)^4}=\frac{\dddot H}{H^4}+4j+3q(q+4)+6\,.
\label{cosmpar}
\end{eqnarray}
In what follows, 
we show the values of these cosmological parameters at the present time 
($z=0$) as the result of numerical extrapolation in our two models, 
which we called Model I in Eq.~(\ref{model}) and Model II 
in Eq.~(\ref{model2}), and the calculation in the $\Lambda$CDM model: 
\begin{eqnarray*}
q(z=0) \Eqn{=} -0.650\, (\Lambda \mathrm{CDM})\,, 
\,-0.544\, (\mathrm{Model\, I})\,, 
\,-0.577\, (\mathrm{Model\, II})\\
j(z=0) \Eqn{=} 1.000\, (\Lambda \mathrm{CDM})\,, 
\,0.792 (\mathrm{Model\, I})\,, 
\,0.972\, (\mathrm{Model\,II})\\
s(z=0) \Eqn{=} -0.050\, (\Lambda \mathrm{CDM})\,, 
\,-0.171 (\mathrm{Model\, I})\,, 
-0.152\,\, (\mathrm{Model\,II})\,.
\end{eqnarray*}
The deviations of the parameters in Models I and II from those in the 
$\Lambda$CDM model are small at the present. 
However, since these parameters depend on the time derivatives of 
the Hubble parameter, it is interesting to analyze those behaviors at high 
curvature. 
Therefore, 
in Fig.~\ref{parameters} we plot the cosmological evolutions of $q$, $j$ and $s$ as functions of the redshift $z$. {}From this figure, we see that there 
exist overlapped regions for Models I and II with those in 
the $\Lambda$CDM model.

The deceleration parameter in Models I and II remains very close to the value in the $\Lambda$CDM model, because in the first time derivative of the Hubble 
parameter the contribution of dark energy is still negligible. 
Hence, it guarantees the correct cosmological evolution of these models. 
However, it is clearly seen that in the jerk and snap parameters 
the derivatives of the dark energy density become relevant and the parameters 
grow up with an oscillatory behavior. 
Since the frequency of such oscillations strongly increases in the redshift, 
it is reasonable to expect that some divergence occurs in the past. 
We also remark that 
if from one side at high redshifts 
the exponential Model I is more similar to the $\Lambda$CDM model 
because of the faster decreasing of exponential function in comparison with 
the power function of Model II, from the other side it involves stronger 
oscillations in the matter dominated era. 

It may be stated that the closer the model is to 
the $\Lambda$CDM model (i.e., as much $F''(R)$ is close to zero), 
the bigger the oscillation frequency of dark energy becomes. 
As a consequence, despite the fact that the dynamics of the universe 
depends on the matter and the dark energy density remains very small, 
some divergences in the derivatives of the Hubble parameter can occur. 
In the models in Eqs.~(\ref{model}) and (\ref{model2}), 
although the approaching manners to a model with the cosmological constant 
are different from each other, 
it may be interpreted that these models 
show a generic feature of realistic $F(R)$ gravity models, in which 
the cosmological evolutions are similar to those in a model with 
the cosmological constant. 
The corrections to the Einstein's equations in the small curvature regime 
lead to undesired effects in the high curvature regime. 
Thus, we need to investigate additional modifications.

\subsection{Proposal of a correction term}

In order to remove the divergences in the derivatives of the Hubble parameter, 
we introduce a function $g(R)$ for which the oscillation frequency of 
the dark energy density in Eq.~(\ref{matterDE}) acquires a constant value 
$1/\sqrt{\delta}$, where $\delta>0$, for a generic curvature $R\gg b\Lambda$, 
and we stabilize the oscillations of dark energy during the matter dominated 
era with the use of a correction term. 
Since in the matter dominated era, i.e., 
$z+1=\left[R/(3\tilde{m}^2)\right]^{1/3}$, 
we have to require
\begin{eqnarray}
\frac{(3\tilde{m}^2)^{2/3}}{R^{5/3}\,g''(R)} \Eqn{=} 
\frac{1}{\delta} \nonumber\\
g(R) \Eqn{=} -\tilde{\gamma}\,\Lambda\left(\frac{R}{3\tilde m^2}\right)^{1/3}\,, 
\quad 
\tilde{\gamma}>0\,, 
\label{eq:FR5-15-3.18}
\end{eqnarray}
where $\tilde{\gamma}\equiv (9/2) \delta (3\tilde m^2/\Lambda)=1.702\,\delta$. 
We explore 
the models in Eqs.~(\ref{model}) and (\ref{model2}) 
with adding these correction as 
\begin{eqnarray}
F_{1}(R) \Eqn{=} 
R-2\Lambda(1-\mathrm{e}^{-\frac{R}{b\,\Lambda}})-\tilde{\gamma}\,\Lambda\left(\frac{R}{3\tilde{m}^2}\right)^{1/3}\,, 
\label{F3exp} \\ 
F_{2}(R) \Eqn{=} R-2\Lambda\left[1-\frac{1}{(R/b\, \Lambda)^4+1}\right]-\tilde{\gamma}\,\Lambda\left(\frac{R}{3\tilde{m}^2}\right)^{1/3}\,. 
\label{F3HS}
\end{eqnarray}
We note that in both cases $F_{1,2}(0)=0$ and therefore we still have 
the solution of the flat space in the Minkowski space-time. 
The effects of the last term vanish in the de Sitter epoch, when 
$R=4L$ and these models resemble to a model with an effective 
cosmological constant, provided that $\tilde{\gamma}\ll (\tilde{m}^2/\Lambda)^{1/3}$. 
We may also evaluate the dark energy density at high redshifts 
by deriving $\rho_{\mathrm{DE}}=\rho_{\mathrm{eff}}-\rho_\mathrm{m}$ 
from Eq.~(\ref{rhoeffRG}) 
and by putting $R=3\tilde{m}^2(z+1)^3$ such that 
\begin{equation}
y_H(z)\simeq\frac{\Lambda}{3\tilde{m}^2}\left[1+\tilde{\gamma}(1+z)\right]\,. 
\label{densmatter}
\end{equation}
According to the observational data of our universe, 
the current value of dark energy amount is estimated 
as $y_H\equiv\Lambda/(3\tilde{m})=2.643$. 
With the reasonable choice $\tilde{\gamma}\sim1/1000$, 
the effects of modification of gravity on the dark energy density 
begin to appear at a very high redshift (for example, at $z=9$, 
$y_H(9)= 1.01\times y_H(0)$), and hence the universe seems to be very close to the 
$\Lambda$CDM model. 
However, while the pure models in Eqs.~(\ref{model}) and (\ref{model2}) mimic 
an effective cosmological constant, the models in Eqs.~(\ref{F3exp}) and 
(\ref{F3HS}) mimic (for the matter solution) a quintessence fluid. 
Equation (\ref{oo}) leads to 
\begin{equation}
\omega_{\mathrm{DE}}(z)\simeq -1+\frac{(1+z)\tilde{\gamma}}{3(1+(1+z)\tilde{\gamma})}\,, 
\end{equation}
so that when $z\gg\tilde{\gamma}^{-1}$, $\omega_{\mathrm{DE}}(z)\simeq -2/3$. 

Thus, it is simple to verify that all the cosmological 
constraints~\cite{Viableconditions} are still satisfied. 
Since $|F_{1,2}'(R \gg b\Lambda)-1| \ll 1$, 
the effective gravitational coupling $G_{\mathrm{eff}}=G_/F_{1,2}'(R)$ is 
positive, and hence the models are protected against the 
anti-gravity during the cosmological evolution until the de Sitter solution 
($R_{\mathrm{dS}} = 4\Lambda$) of the current universe is realized. 
Thus, thanks to the fact that $|F_{1,2}''(R\gg b\Lambda)>0|$, 
we do not have any problem in terms of the existence of a stable matter. 
In Sec.~{\ref{tests}}, 
we also analyze the local constraints in detail, 
and we see that our modifications do not destroy the feasibility of the 
models in the solar system. 
It should be stressed that the energy density preserves 
its oscillation behavior in the matter dominated era, 
but that owing to the correction term reconstructed here, 
such oscillations keep a constant frequency 
$\mathcal{F}=\sqrt{1.702/\tilde{\gamma}}$ and do not diverge. 
Despite the small value of $\tilde{\gamma}$, in this way 
the high redshift divergences and possible effective singularities 
are removed. 

{}From the point of view of the end of inflation, 
there is another resolution of this problem. 
It is well known that 
the scalar begins to oscillate once the mass $m$ becomes larger than 
the Hubble parameter, $H < m$. 
Indeed, for a canonical scalar, the energy density sloshes between the potential energy ($w=-1$, where $w$ is the equation of state of the canonical scalar) 
and the kinetic energy ($w=+1$). 
What is done usually is that the oscillations enough rapidly 
(i.e., those with $m \gg H$) can be averaged over giving an effective 
energy-momentum tensor with $w=0$, i.e., dust. 
The same procedure should be performed here, 
once the oscillations are rapid enough. 
In this interpretation, there would be no problem with any 
strange rapidly oscillating contributions to the energy momentum tensor. 
A solution is to choose the potential effectively so that 
the mass can not increase as the matter energy density increases. 

Furthermore, 
it is significant to remark that 
in a number of models of $F(R)$ gravity for dark energy, 
there exists a well-known problem that 
positions in the field space are a finite distance away from the minimum of 
the effective potential, so that a curvature singularity 
in the Jordan frame could appear. 
This means that large excursions of the scalar could result in a singularity 
forming in a solution. 
It is known that the solution for this problem is also adding the higher 
powers of $R$ so that the behavior at large curvatures can be soften. 
The oscillations are extremely large at small curvatures too, 
and the higher power of $R$ or $R$ itself do not change 
in this range of detection. 
We also note that this argument is applicable to 
the so-called type I, II and III finite-time 
future singularities (where $R$ diverges), which has been classified 
in Ref.~\cite{Nojiri:2005sx}, while for a kind of singularities in our work, 
$R$ does not become singular, and hence the argument would become 
different from the above. 

\subsection{Analysis of exponential and power-form models with correction terms 
in the matter dominated era 
\label{matterstudy}}

\begin{figure}[!h]
\subfigure[]{\includegraphics[width=0.3\textwidth]{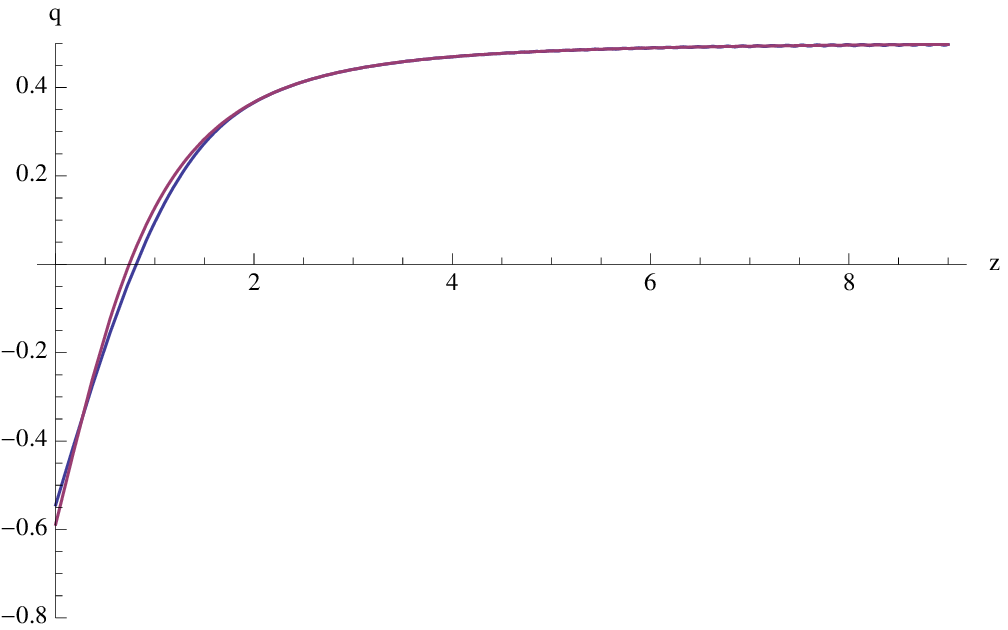}}
\qquad
\subfigure[]{\includegraphics[width=0.3\textwidth]{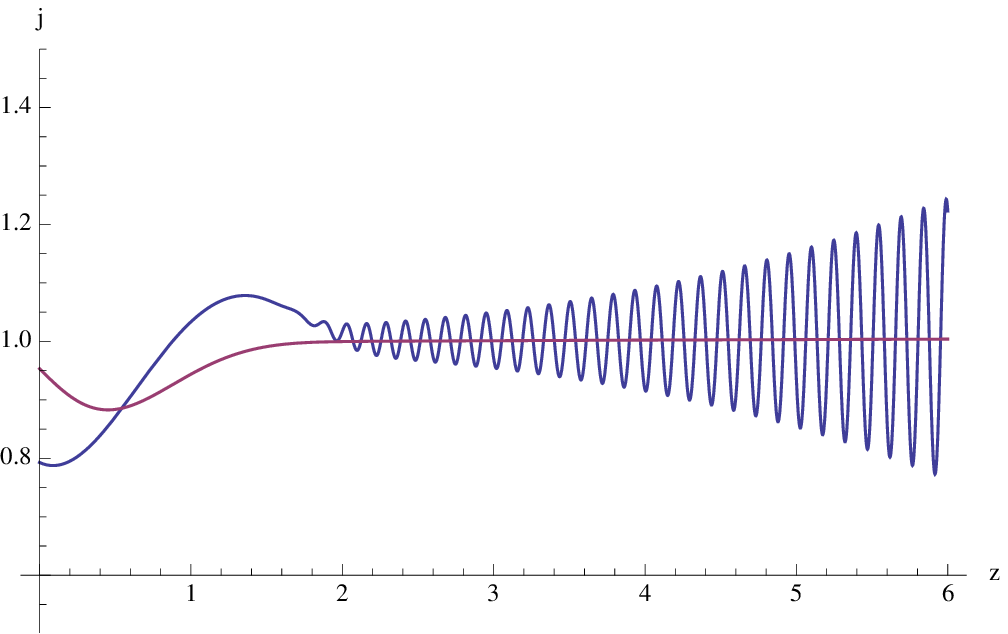}}
\qquad
\subfigure[]{\includegraphics[width=0.3\textwidth]{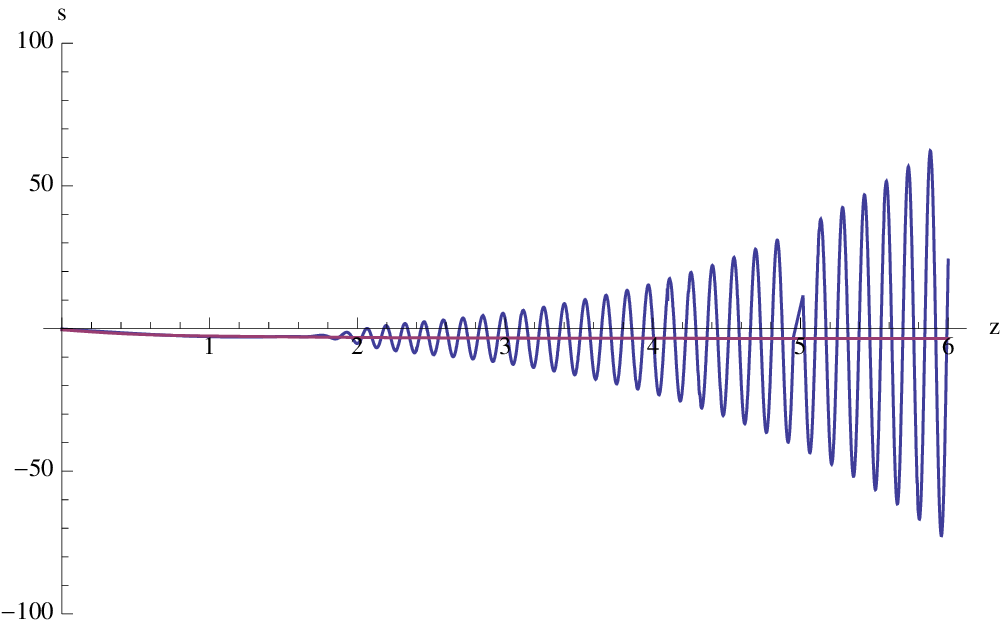}}
\qquad
\subfigure[]{\includegraphics[width=0.3\textwidth]{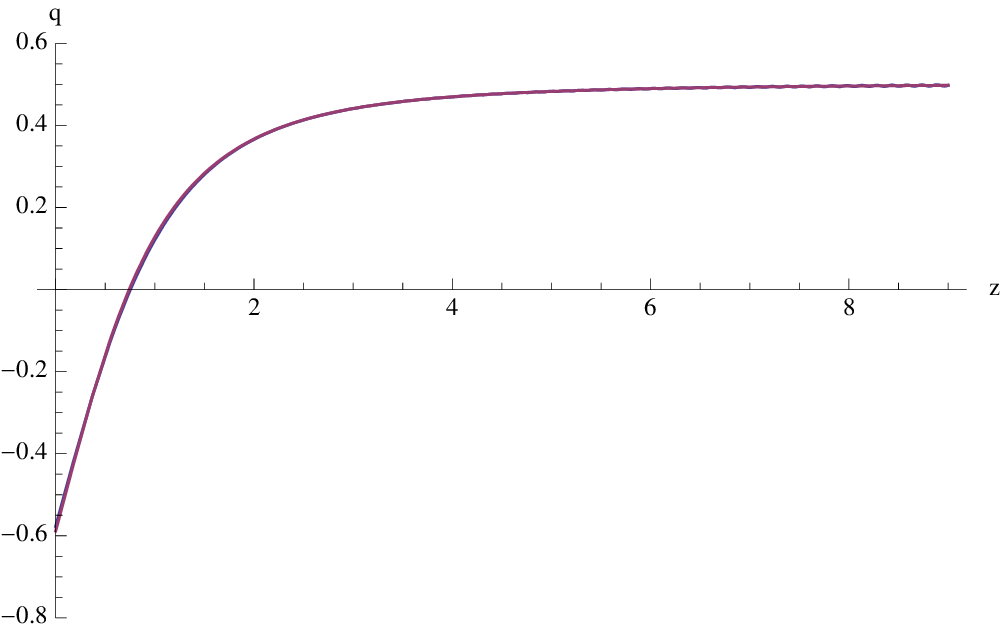}}
\qquad
\subfigure[]{\includegraphics[width=0.3\textwidth]{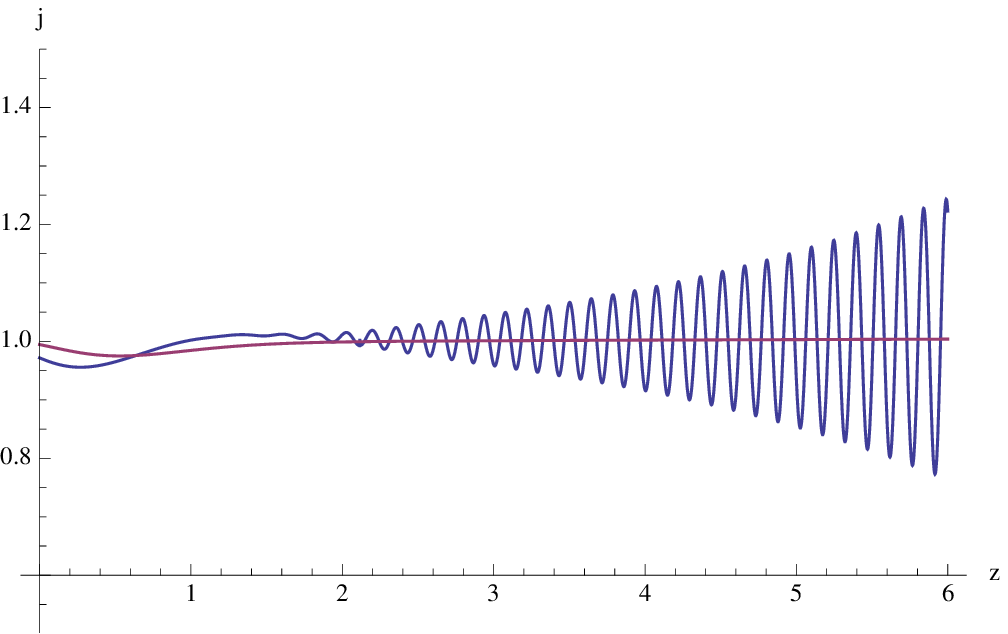}}
\qquad
\subfigure[]{\includegraphics[width=0.3\textwidth]{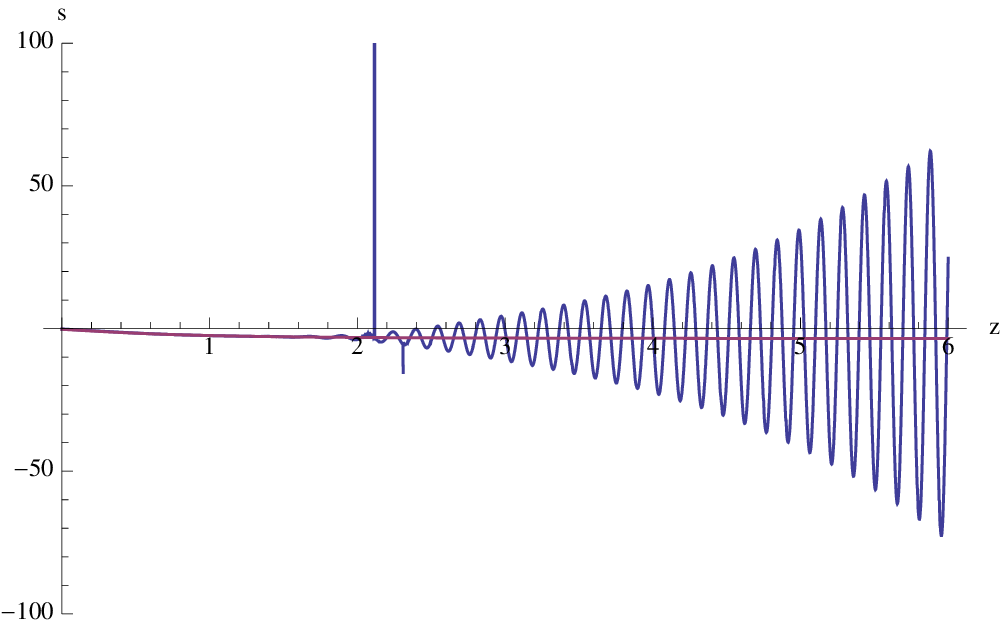}}
\caption{
Cosmological evolutions of $q(z)$ [(a) and (d)], $j(z)$ [(b) and (e)] and $s(z)$ [(c) and (f)] parameters as functions of the redshift $z$ for 
the model $F_1(R)$ [(a)--(c)] and the model $F_2(R)$ [(d)--(f)] 
in the region of $z>0$. 
\label{3}}
\end{figure}

In this subsection, we carry out the numerical analysis of the models 
in Eqs.~(\ref{F3exp}) and (\ref{F3HS}). 
In both cases, 
we assume $b=1$ and $\tilde{\gamma}=1/1000$ and solve 
Eq.~(\ref{superEq}) in a numerical way, 
by taking accurate initial conditions at $z=z_i$ 
so that $z_i\gg2$. 
By using Eq.~(\ref{densmatter}), we acquire 
\begin{eqnarray*}
\frac{d y_H(z)}{d (z)}\Big\vert_{z_i} \Eqn{=} 
\frac{\Lambda}{3\tilde{m}^2}\gamma\,,\\ 
y_H(z)\Big\vert_{z_i} \Eqn{=} \frac{\Lambda}{3\tilde{m}^2}\left(1+\gamma\,(z_i+1)\right)\,, 
\end{eqnarray*}
where we have set $z_i=9$. 
The feature of the models in Eqs.~(\ref{F3exp}) and (\ref{F3HS}) at the present time is very similar to those of the models in Eqs.~(\ref{model}) and 
(\ref{model2}). 
With the numerical extrapolation to the current universe, 
for the model in Eq.~(\ref{F3exp}) we have 
$y_H(0)=2.739$, $\omega_{\mathrm{DE}}(0)=-0.950$, $\Omega_{\mathrm{DE}}(0)=0.732$ and $R(z=0)=4.369$, 
while for the model in Eq.~(\ref{F3HS}), we find 
$y_H(0)=2.654$, $\omega_{\mathrm{DE}}(0)=-0.989$, $\Omega_{\mathrm{DE}}(0)=0.726$ and $R(z=0)=4.361$. 
We analyze those behaviors in the matter dominated era. 
It follows from the initial conditions 
$y_H(9)=2.670$ and $\omega_{\mathrm{DE}}(9)=-0.997$ 
that the universe is extremely close to the $\Lambda$CDM model 
also at high redshifts. 
We see how the dynamical correction of 
the Einstein's equation, which corresponds to, roughly speaking, 
the fact of having ``a dynamical cosmological constant'', 
introduces an oscillatory behavior of dark energy density. 
Thanks to the contribution of the correction term, 
we obtain a constant frequency of such oscillations without changing 
the cosmological evolution described by the theory. 
In Fig.~\ref{3}, we show the cosmological evolutions of 
the deceleration, jerk and snap parameters as functions of the redshift $z$ 
in these models. 
There is overlapped region of the evolutions with those in the $\Lambda$CDM 
model. 
We may compare the graphics in Fig.~\ref{3} 
with the corresponding ones in Fig.~\ref{parameters} 
of the models in Eqs.~(\ref{model}) and (\ref{model2}) without the correction 
term analyzed in Sec.~III B. 
At high redshifts, 
the deceleration parameter is not influenced by dark energy and hence 
the behavior in 
both these models in Eqs.~(\ref{F3exp}) and (\ref{F3HS}) are 
the same as that in the $\Lambda$CDM model. 
On the other hand, 
in terms of the jerk and snap parameters, 
the derivatives of dark energy density become relevant 
and accordingly these parameters oscillate with the same 
frequency as that of dark energy, 
showing a different behavior in comparison with the case of GR 
with the cosmological constant. 
However, here such oscillations have a constant frequency and do not diverge. 
The predicted value of the oscillation frequency is 
$\mathcal{F}\equiv\sqrt{1.702/\tilde{\gamma}}=41.255$. 
The oscillation period is $T=2\pi/\mathcal{F}\simeq 0.152$. 
Thus, the numerical data are in good accordance with the predicted ones. 
(We can also appreciate the result by taking into account the fact that 
the number of crests per units of the redshift has to be $1/T\simeq7$). 

Consequently, 
we have shown in both analytical and numerical ways that 
increasing oscillations of dark energy in the past approach to 
effective singularities. 
It is not ``a rapid oscillating system'' but a system which becomes singular. 
The effects of such oscillations are evident especially in the higher 
derivative of the Hubble parameter. 
It is not a case that if all the numerical simulations presented in the 
literature start from small redshifts, at higher redshifts this singular 
problem appears.  
Eventually, the oscillations may influence also on the behavior of 
the Ricci scalar (which depends on the first derivative of 
the Hubble parameter, see Eq.~(\ref{eq:FR5-15-3.14}) 
and $\left| d^n H(t)|_{t_0}/d t^n \right|\propto 
\left(\mathcal{F}(z_0)\right)^n$ with $n\gg 1$, following 
from Eq.~(\ref{frequency})). 
Of course, the average value of the dark energy density remains negligible, 
but the oscillations around this value become huge. 
Thus, the Ricci scalar may have an oscillatory behavior. 
We have also evaluated the frequency of the oscillations, so that 
the result can match with the numerical simulations, and therefore 
all the analyses in this work are consistent. 
This behavior of realistic $F(R)$ gravity models has recently been studied 
also in Ref.~\cite{Lee:2012dk}. 

We remark that 
if the mass of the additional scalar degree of freedom, 
the so-called scalaron, is too large, 
the predictability could be lost~\cite{Upadhye:2012vh}. 
Clearly, the mass of the scalaron in the two models in Eqs.~(\ref{F3exp}) and 
(\ref{F3HS}) is not bounded, and thus it would diverge 
in very dense environment. 
We have confirmed that 
in the large curvature regime compared with the current curvature 
the correction term $g(R)$ in~(\ref{eq:FR5-15-3.18}) 
in these two models do not strongly affect the scalaron potential 
in the Einstein frame, namely, the correction term would not be 
the leading term in the form of the scalaron potential, 
and thus the scalaron mass is not changed very much. 
The model parameter of the correction term $g(R)$ 
mainly related to the scalaron mass as well as its potential 
is $\tilde{\gamma}$. 
In the limit that the energy density of the environment 
becomes infinity, since the contribution of the correction term 
to the scalaron mass, it would be impossible to constrain 
the values of $\tilde{\gamma}$, for which the divergence 
of the scalaron mass can be avoided. 

\section{Cosmological constraints and future evolution\label{tests}}

In this section, first we show that the models in Eqs.~(\ref{F3exp}) and 
(\ref{F3HS}) satisfy the cosmological and local gravity 
constraints~\cite{SolarSystemconstraints}, and that the term added to 
stabilize the dark energy oscillations in the matter dominated epoch 
does not cause any problem to these proprieties. 
The confrontation of $F(R)$ models with SNIa, BAO, CMB radiation 
and gravitational lensing has been executed in the past several 
works~\cite{listone}. 
We have just seen that the models with the choice of $b=1$ can be consistent 
with the observational data of the universe. 
Here, we examine the range of $b$ in which the models are compatible with 
the observations and analyze the behavior of the models near to local (matter) 
sources in order to check possible Newton law corrections or matter 
instabilities. 
Then, we concentrate on the future evolution of the universe in the models 
and demonstrate that the effective crossing of the phantom divide which characterizes the de Sitter epoch takes place in the very far future. 

In the way of trying to explain the several aspects that characterize our 
universe, there exists the problem of distinguishing different theories. 
It has been revealed that sometimes the study of the expansion history of the universe is not enough because different theories can achieve the same expansion history. 
Fortunately, theories with the same expansion history can have a different cosmic growth history. This fact makes the growth of the large scale structure in the universe an important tool in order to discriminate among the different theories proposed. Thus, the characterization of growth of the matter density perturbations become very significant. In order to execute it, the so-called growth index $\gamma$~\cite{Linder:2005in} is useful. Therefore, in the second part of this section we study the evolution of the matter density perturbation for our $F(R)$ gravity model. 

Again, we clearly state the main purpose of this section. 
Since the original models, i.e., 
the Hu-Sawicki model~\cite{HuSaw} in Eq.~(\ref{model2}) and 
exponential gravity~\cite{Cognola:2007zu, Linder:2009jz} in 
Eq.~(\ref{model}), have been studied well, 
we concentrate on the question 
whether the corrected models in Eqs.~(\ref{F3HS}) and (\ref{F3exp}) 
lead to any difference in the observables. 
These modified models have been constructed 
in order not to alter the background evolution 
significantly except the oscillatory effect.  
In Refs.~\cite{HuSaw, Linder:2009jz} and many follow-up studies of these 
pioneering works, 
the cosmological background evolutions and the growth of structures in 
the two unmodified models in Eqs.~(\ref{model2}) 
and (\ref{model}) have been investigated. 
In order make this work self consistent study of modified gravity, 
we explicitly demonstrate the cosmological background evolutions and 
the growth of the matter density perturbations in 
the modified models in Eqs.~(\ref{F3HS}) and (\ref{F3exp}). 
It is meaningful to investigate these behaviors in the modified models 
even though the modifications on the observable quantities are small. 

\subsection{Cosmological and local constraints}

\begin{figure}[!h]
\subfigure[]{\includegraphics[width=0.4\textwidth]{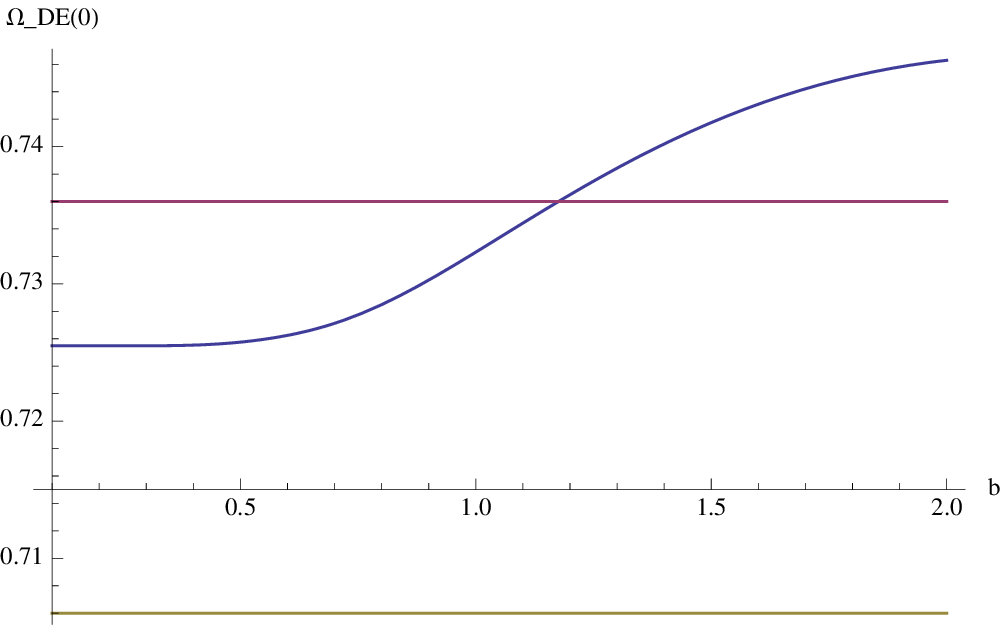}}
\qquad
\subfigure[]{\includegraphics[width=0.4\textwidth]{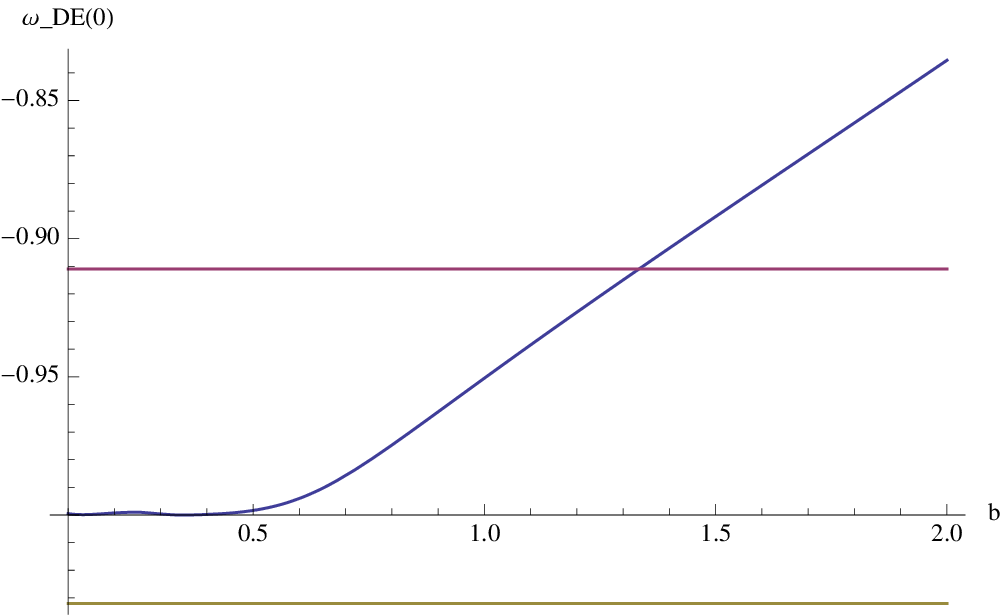}}
\caption{Behaviors 
of $\omega_\mathrm{DE}(z=0,b)$ and of $\Omega_\mathrm{DE}(z=0,b)$ 
as functions of $b$ for exponential model. 
The observational data bounds (horizontal lines) are also shown.
\label{b-1}}
\end{figure}

\begin{figure}[!h]
\subfigure[]{\includegraphics[width=0.4\textwidth]{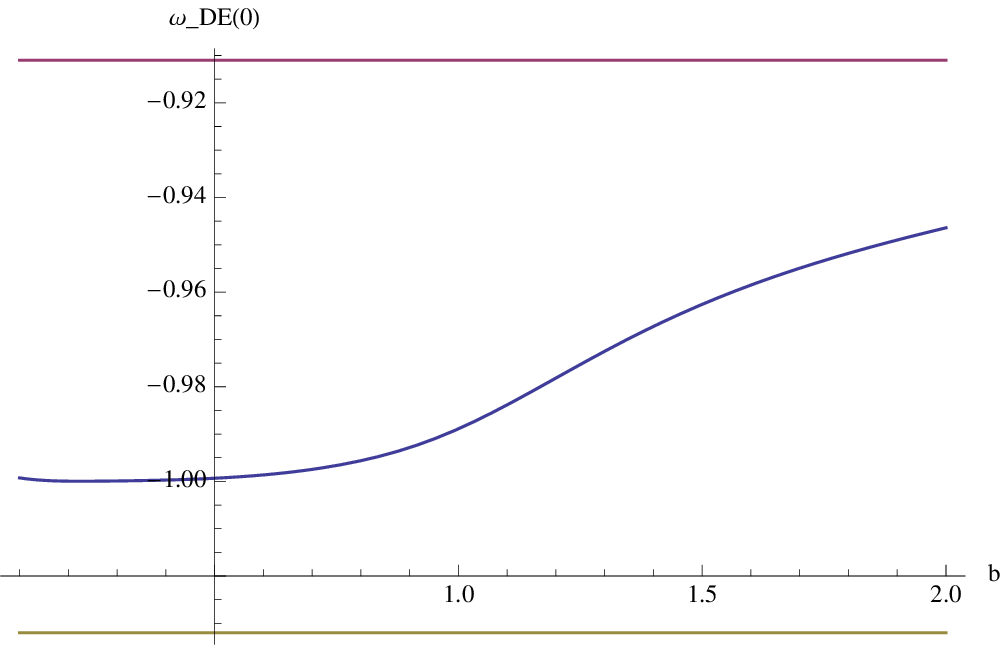}}
\qquad
\subfigure[]{\includegraphics[width=0.4\textwidth]{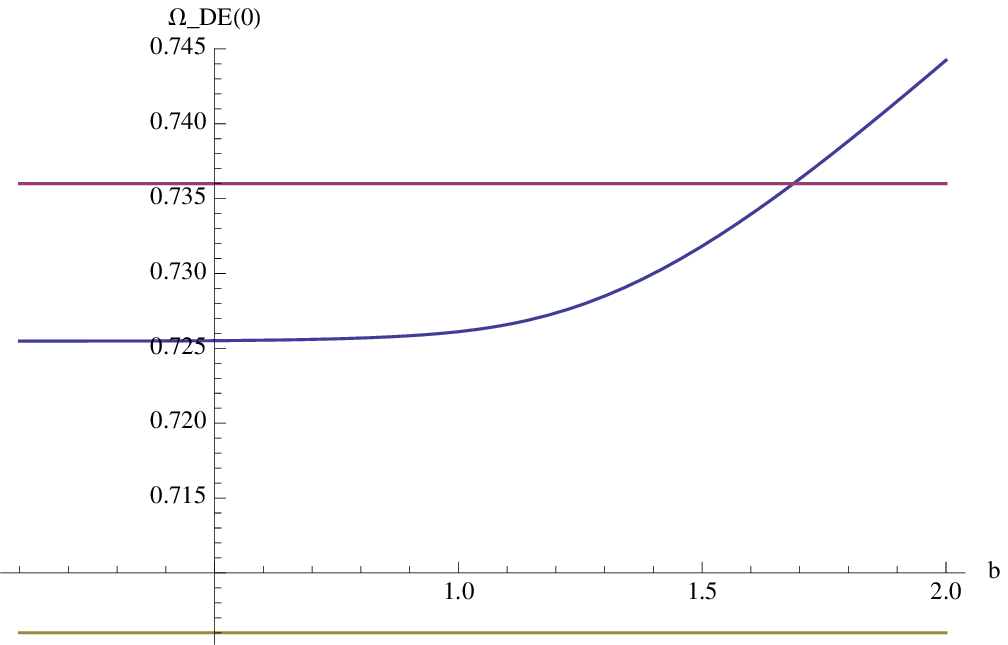}}
\caption{Behaviors 
of $\omega_\mathrm{DE}(z=0,b)$ and of $\Omega_\mathrm{DE}(z=0,b)$ 
as functions of $b$ for power-law model. 
Legend is the same as Fig.~\ref{b-1}. 
\label{b-2}}
\end{figure}

We take $\tilde{\gamma}=1/1000$ in the models in Eqs.~(\ref{F3exp}) and (\ref{F3HS}), keeping the parameter $b$ free. Now, the dark energy density is a function of $z$ and $b$, i.e., $ y_H(z,b)$. 
We can again solve Eq.~(\ref{superEq}) numerically, taking the initial conditions at $z_i=9$ as 
\begin{eqnarray*}
\frac{d y_H(z,b)}{d (z)}\Big\vert_{z_i} \Eqn{=} 
\frac{\Lambda}{3\tilde{m}^2}\tilde{\gamma}\,,\\ 
y_H(z,b)\Big\vert_{z_i} \Eqn{=} \frac{\Lambda}{3\tilde{m}^2}\left(1+\tilde{\gamma}\,(z_i+1)\right)\,,
\end{eqnarray*}
as we did in the previous section. 
We take $0.1<b<2$. In Figs.~\ref{b-1} and \ref{b-2}, 
we display the resultant values of dark energy EoS parameter $\omega_\mathrm{DE}(z=0,b)$ and 
$\Omega_\mathrm{DE}(z=0,b)$ at the present time as functions of $b$ for the two models. 
We also show the bounds of cosmological data in Eq.~(\ref{data}), namely, 
the lines in rose denote the upper bounds, while the lines in yellow do 
the lower ones. 
By matching the comparison between the two graphics of every model, 
we find that in order to correctly reproduce the universe where we live with exponential gravity in Eq.~(\ref{F3exp}), $0.1<b<1.174$, with power-law model in Eq.~(\ref{F3HS}), $0.1<b<1.699$. The results are consistent with the choices 
in Sec.~\ref{matterstudy}.

\subsubsection*{Newton law corrections and stability on a planet surface}

In Ref.~\cite{matterinstabilityplanet}, 
it has been shown that some realistic models of $F(R)$ gravity may lead to 
significant Newton law corrections at large cosmological scales. 
We briefly review this result. 
{}From the trace of the field equation (\ref{Field equation}), 
if we consider the constant background of $R=R_0$, such that $2F(R_0)-R_0 F'(R_0)=0$, by performing a 
variation with respect to $R=R^{(0)}+\delta R$ 
and supposing the presence of a matter point source (like a planet), that is, 
$T^{\mathrm{(matter)}}=T_0\,\delta(x)$, where $\delta(x)$ is the Dirac's 
distribution, 
we find, to first order in $\delta R$, 
\begin{equation}
\left(\Box-m^2\right)\delta R=\frac{\kappa^2}{3F''(R_0)}T_0\delta(x)\,,
\end{equation}
with 
\begin{equation}
m^2=\frac{1}{3}\left(\frac{F'(R_0)}{F''(R_0)}-R_0\right)\,.
\end{equation}
The solution is given by 
\begin{equation}
\delta R=\frac{\kappa^2}{3F''(R_0)}T_0\,G(m^2,|x|)\,,
\end{equation}
where $G(m^2,|x|)$ is the correlation function which satisfies 
\begin{equation}
\left(\Box-m^2\right)G(m^2,|x|)=\delta(x)\,.
\end{equation}
Hence, if $m^2 < 0$, there appears a tachyon and thus there could be some 
instability. Even if $m^2 > 0$, when $m^2$ is small compared with $R_0$, 
$\delta R\neq 0$ at long ranges, which generates the large correction to 
the Newton law. For the pure exponential model in Eq.~(\ref{model}) without 
correction terms, when $R_0\gg b\,\Lambda$, $m^2$ reads
\begin{equation}
m^2\simeq\frac{(b^2\Lambda)}{6}\mathrm{e}^{\frac{R_0}{b\Lambda}}\,. \label{ttt}
\end{equation}
Therefore, in general $m^2/R_0$ is very large effectively. 
The same thing happens in the model in Eq.~(\ref{model2}). 
Next, for 
the models in Eqs.~(\ref{F3exp}) and (\ref{F3HS}) with correction terms, 
we have
\begin{equation}
m^2\simeq\frac{3^{4/3}\tilde m^2 R}{2\Lambda\tilde{\gamma}}\left(\frac{R}{\tilde m^2}\right)^{2/3}\,. 
\label{**}
\end{equation}
Despite the fact that in this case $m^2$ is smaller than in Eq.~(\ref{ttt}), 
it still remains sufficiently large and the correction to the Newton law is 
very small. 
For example, the typical value of the curvature in the solar system is 
$R_0\simeq 10^{-61} \text{eV}^2$ 
(it corresponds to one hydrogen atom per cubic centimeter). In this case, 
{}from Eq.~(\ref{**}) we obtain $m^2/R_0\simeq 2\times 10^{6}$. 

Concerning the matter instability~\cite{Dolgov:2003px, Faraoni:2006sy}, this might also occur when the curvature is 
rather large, as on a planet ($R\simeq 10^{-38} \text{eV}^2$), 
as compared with the average curvature of the 
universe today ($ R\simeq 10^{-66} \text{eV}^2$). 
In order to arrive at a stability condition, 
we can perturb again Eq.~(\ref{scalaroneeqbis}) around $R=R_{b}$, where $R_{b}$ is the curvature of the planet surface and the perturbation $\delta R$ is given by the curvature difference between the internal and the external solution. 
The curvature $R_b=-\kappa^2 T^{\mathrm{(matter)}}$ depends on the radial 
coordinate $r$. By assuming $\delta R$ depending on time only, we acquire 
\begin{equation}
-\partial_{t}^2(\delta R)\sim U(R_{b})\delta R\,,
\end{equation}
where
\begin{eqnarray}
U(R_{b})\Eqn{=}\left[\left(\frac{F'''(R_{b})}{F''(R_{b})}\right)^2
-\frac{F'''(R_{b})}{F''(R_{b})}\right]g^{rr}\nabla_{r}R_{b}\nabla_{r}R_{b}
-\frac{R_{b}}{3}+\frac{F'(R_{b})}{3F''(R_{b})} 
\nonumber \\
&&
\frac{F'''(R_{b})}{3(F''(R_{b}))^2}(2F(R_{b})-R_{b}F'(R_{b})-R_{b})\,.
\label{U}
\end{eqnarray}
Here, $g_{\mu\nu}$ is the diagonal metric describing the planet. 
If $U(R_{b})$ is negative, then the perturbation $\delta R$ becomes 
exponentially large and the whole system becomes unstable. Thus, the 
planet stability condition is 
\begin{equation}
U(R_{b})>0\,.
\end{equation}
For our models in Eqs.~(\ref{F3exp}) and (\ref{F3HS}), $U(R_{b})\simeq m^2$, 
where $m^2$ is given by Eq.~(\ref{**}) again. 
Also in this case, we do not have any particular problem. For example, 
by putting $R_{b}\simeq 10^{-38}\text{eV}^2$, we find 
$U(R_{b})/R_b\simeq 4\times 10^{21}$. 
Thus, the models under consideration easily pass these local tests.

We mention that 
in the past, the non-linear effects on the scalar 
are much more important, owing to the mechanism of the chameleon 
effect~\cite{Mota:2003tc, Chameleon mechanism}, 
and that only at late times the linear evolution is a good approximation. 
For example, 
if a high-curvature solution is achieved, 
the Solar-System test is the examination whether 
the solution is stable against the Dolgov-Kawasaki 
instability~\cite{Dolgov:2003px}. 
This is not the same as whether the high-curvature solution can 
at all be achieved, which is a much more subtle issue 
and discussed at length by Hu and Sawicki in Ref.~\cite{HuSaw}.

\subsection{Future universe evolution\label{dS}}

\begin{figure}[!h]
\subfigure[]{\includegraphics[width=0.4\textwidth]{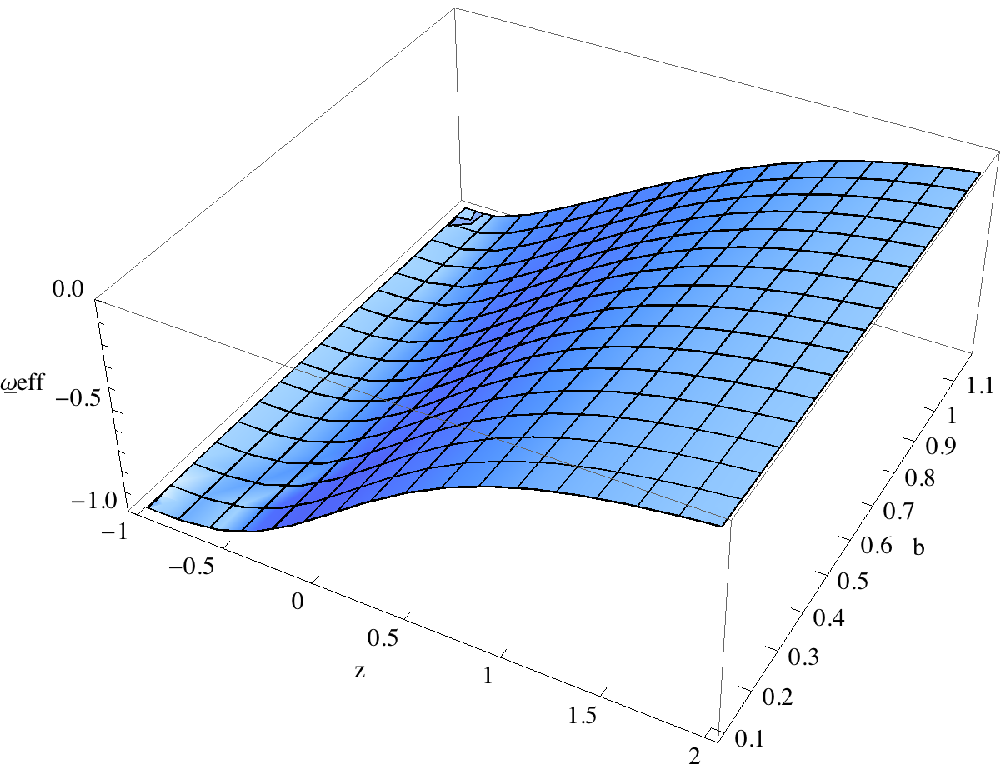}}
\qquad
\subfigure[]{\includegraphics[width=0.4\textwidth]{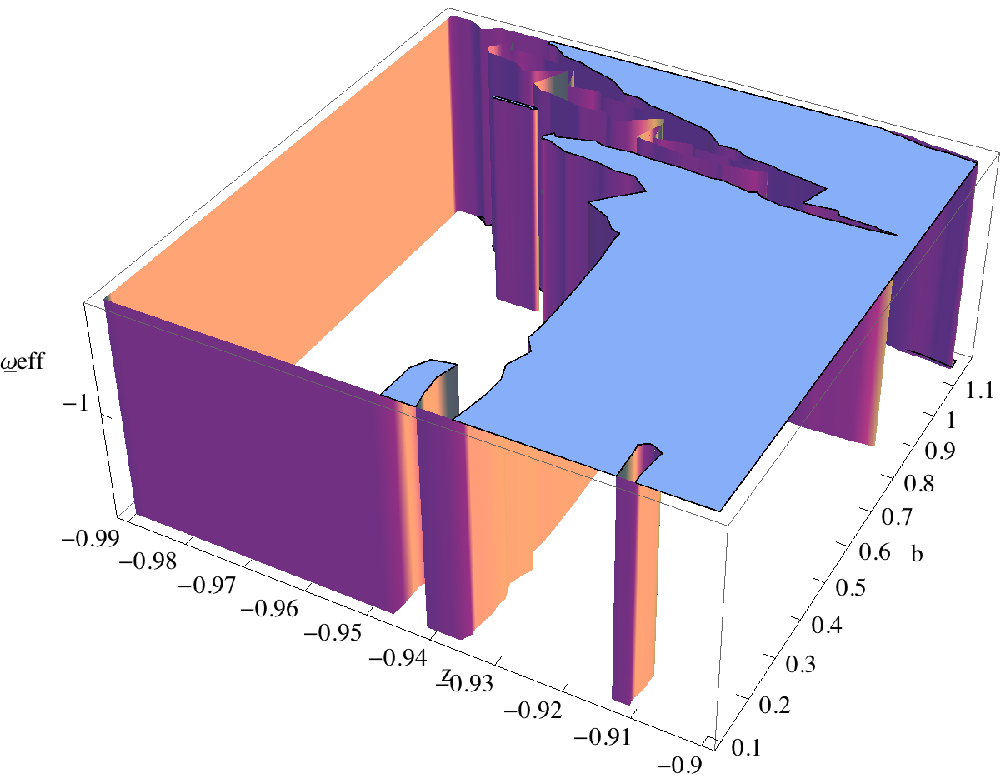}}
\caption{Cosmological evolution of $\omega_{\mathrm{eff}}$ as a function of the red shift $z$ and the $b$ parameter for the model in Eq.~(\ref{F3exp}). 
The left panel plots it for $-1<z<2$ and the right one displays 
around $z=-1$.
\label{33}}
\end{figure}

\begin{figure}[!h]
\qquad
\subfigure[]{\includegraphics[width=0.4\textwidth]{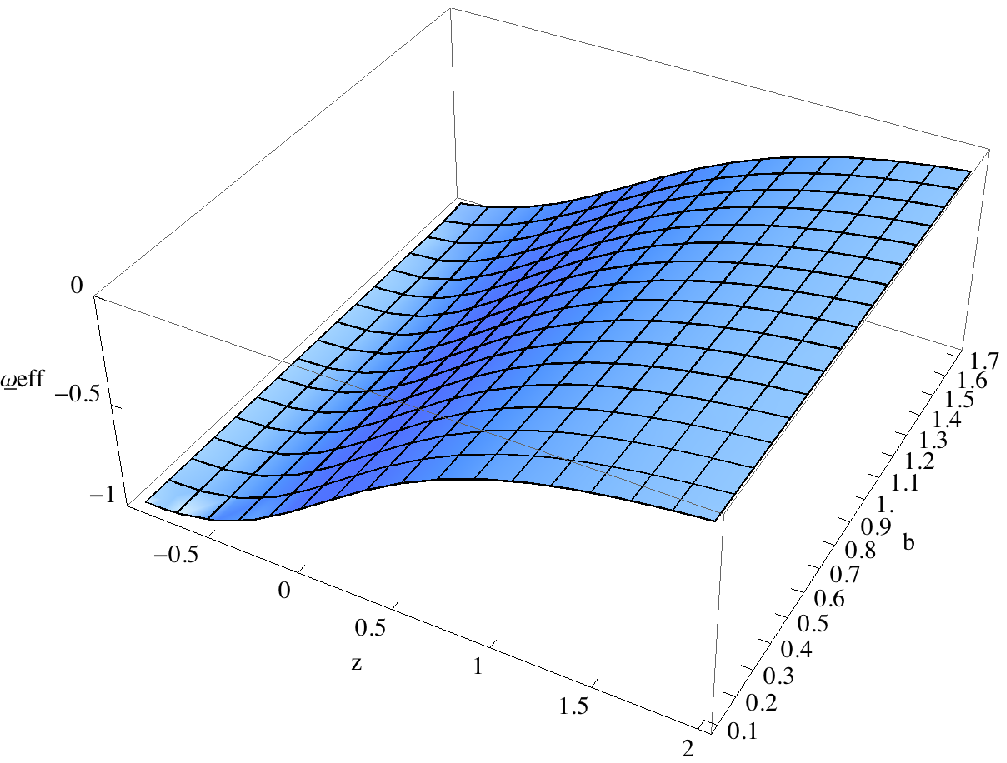}}
\qquad
\subfigure[]{\includegraphics[width=0.4\textwidth]{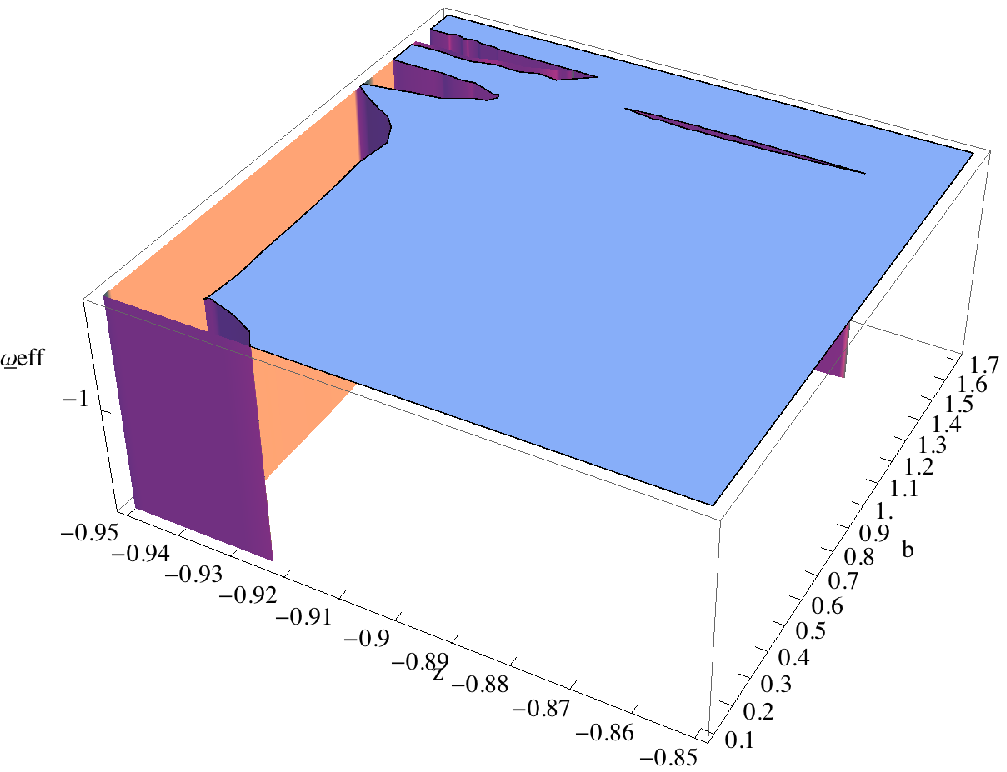}}
\caption{Cosmological evolution of $\omega_{\mathrm{eff}}$ as a function of the red shift $z$ and the $b$ parameter for the model in Eq.~(\ref{F3HS}). 
Legend is the same as Fig.~\ref{33}.
\label{33bis}}
\end{figure}

In de Sitter universe, we have $R=R_{\mathrm{dS}}$, where $R_{\mathrm{dS}}$ is 
the constant curvature given by the constant dark energy density $y_H=y_{0}$, 
such that $y_0=R_{\mathrm{dS}}/12\tilde{m}^2$. 
Starting from Eq.~(\ref{superEq}), we are able to study perturbations around 
the de Sitter solution in the models~(\ref{F3exp}) and (\ref{F3HS}) 
which provide this solution for $R_{\mathrm{dS}}=4\Lambda$ and well satisfied 
the de Sitter condition 
$2F(R_{\mathrm {dS}})=R_{{\mathrm dS}}F'(R_{\mathrm{dS}})$ 
as a consequence of 
the trace of the field equation 
in vacuum. 
Performing the variation with respect to $y_H(z)=y_0+y_1(z)$ with $|y_1(z)|\ll 1$ and assuming the contributions of radiation and matter to be much smaller 
than $y_0$, at the first order in $y_1(z)$ Eq.~(\ref{superEq}) reads 
\begin{equation}
\frac{d^2 y_1(z)}{d z^2}+\frac{\alpha}{(z+1)}\frac{d y_1(z)}{d z}+\frac{\beta}{(z+1)^2}
y_1(z)=4\zeta(z+1)\,,
\label{superEqbis}
\end{equation}
where
%
\begin{equation}
\alpha=-2\,,\quad
\beta= -4+\frac{4F'(R_{\mathrm{dS}})}{RF''(R_{\mathrm{dS}})}\,,\quad
\zeta=1+\frac{1-F'(R_{\mathrm{dS}})}{R_{\mathrm{dS}}F''(R_{\mathrm{dS}})}\,.\
\end{equation}
%
The solution of Eq.~(\ref{superEqbis}) is given by 
\begin{eqnarray}
y_H(z) \Eqn{=} y_0+y_1(z)\,, \\
y_1(z) \Eqn{=} C_0(z+1)^{\frac{1}{2}\left(1-\alpha\pm\sqrt{(1-\alpha)^2-4\beta}\right)} + \frac{4\zeta}{\beta}(z+1)^3\,,
\label{result}
\end{eqnarray}
where $C_0$ is a constant.
The well-known stability condition for the de Sitter space-time, $F'(R_\mathrm{dS})/((R_\mathrm{dS})F''(R_\mathrm{dS}))>1$, is also valid. 
It has also been demonstrated that since in realistic $F(R)$ gravity models for the de Sitter universe $F''(R)\rightarrow 0^+$, 
$F'(R_{\mathrm{dS}})/(R_{\mathrm{dS}}F''(R_{\mathrm{dS}}))>25/16$ \cite{Staro} giving negative the discriminant of Eq.~(\ref{result}) and an oscillatory behavior to the dark energy density during this phase. 
Thus, in this case 
the dark energy EoS parameter $\omega_\mathrm{DE}$~(\ref{oo}) becomes 
%
\begin{eqnarray}
\label{omegaoscillating}
&&\omega_{\mathrm{DE}}(R=R_{\mathrm{dS}})\simeq-1+4\tilde{m}^2\frac{(z+1)^{\frac{3}{2}}}{R_{\mathrm{dS}}}\times 
\\\nonumber
&&\hspace{-10mm}\left[A_0\cos\left(\sqrt{\left(\frac{4}{R_{\mathrm{dS}}F''(R_{\mathrm{dS}})}\right)}
\log(z+1)\right)+B_0\sin\left(\sqrt{\left(\frac{4}{R_{\mathrm{dS}}F''(R_{\mathrm{dS}})}\right)}\log(z+1)\right)\right]\,,
\end{eqnarray}
and oscillates infinitely often around the line of the phantom divide 
$\omega_\mathrm{DE}=-1$~\cite{Staro}. 
According to various recent observational data, 
the crossing of the phantom divide occurred 
in the near past~\cite{Observational-status}.
These models possess one crossing in the recent past~\cite{Bamba}, after the end of the matter dominated era, and infinite crossings in the future 
(for detailed investigations on the future crossing of the phantom divide, 
see~\cite{Bamba:2010iy}), but the amplitude of such crossings decreases as $(z+1)^{3/2}$ and it does not cause any serious problem to the accuracy of the cosmological evolution during the de Sitter epoch which is in general the final 
attractor of the system~\cite{Bamba, twostep}. 
However, the existence of a phantom phase can give some undesirable effects 
such as 
the possibility to have the Big Rip~\cite{Caldwell} as an alternative scenario of the universe (in such a case, the model may suddenly exit from $\Lambda$CDM description) or 
the disintegration of bound structures which does not necessarily require to having the final (Big Rip) singularity~\cite{lRip1, Rip2}. 
In this subsection, we show that in the models in Eqs.~(\ref{F3exp}) and (\ref{F3HS}) the effective EoS parameter of the universe 
(for an alternative study, see~\cite{ultimo}) 
defined as 
\begin{equation}
\omega_{\mathrm{eff}}\equiv\frac{\rho_{\mathrm{eff}}}{P_{\mathrm{eff}}}=-1+\frac{2(z+1)}{3H(z)}\frac{dH(z)}{dz} 
\end{equation}
never crosses the phantom divide line in the past, and that only when $z$ is very close to $-1$ (this means in the very far future), 
it coincides with $\omega_{\mathrm{DE}}$ and the crossings occur. 
We remark that $\rho_{\mathrm{eff}}$ and $P_{\mathrm{eff}}$ correspond 
to the total energy density and pressure of the universe, 
and hence that if dark energy strongly dominates over ordinary matter, 
we can consider $\omega_{\mathrm{eff}} \approx \omega_{\mathrm{DE}}$. 
In both of the models under investigation, 
we take again $\tilde{\gamma}=1/1000$ and keep the parameter $b$ free, 
such that $0.1<b<1.174$ (model in Eq.~(\ref{F3exp})) and $0.1<b<1.699$ 
(model in Eq.~(\ref{F3HS})), according to the realistic representation 
of current universe. 
The numerical evaluation of Eq.~(\ref{superEq}) leads to 
$H(z)$, given by 
\begin{equation}
H(z)=\sqrt{\tilde m^2\left[y_H(z)+(z+1)^3+\chi(z+1)^4\right]}\,,
\end{equation}
and therefore $\omega_{\mathrm{eff}}(z)$. 
We depict the cosmological evolution 
of $\omega_{\mathrm{eff}}$ as a function of the red shift $z$ and 
the $b$ parameter in Fig.~\ref{33} for the model in Eq.~(\ref{F3exp}) 
and in Fig.~\ref{33bis} for the model in Eq.~(\ref{F3HS}). 
On the left panels, we plot the effective EoS parameter for $-1<z<2$. 
We can see that for both of the models, independently on the choice of $b$, 
$\omega_{\mathrm{DE}}$ starts from zero in the matter dominated 
era and asymptotically approaches -1 without any appreciable deviation. 
Only when $z$ is very close to $-1$ and the matter contribution to 
$\omega_{\mathrm{eff}}$ is effectively zero, 
we have the crossing of the phantom divide due to the oscillation behavior of 
dark energy. 
On the right panels, we display the behavior of the effective EoS parameter 
around $z=-1$. 
Here, we focused on the phantom divide line and we excluded the graphic area out of the range $-1.0001<\omega_{\mathrm{eff}}<-0.9999$. 
The blue region indicates that $\omega_{\mathrm{eff}}$ is still in the quintessence phase. We note that especially in the model in Eq.~(\ref{F3HS}), the first crossing of phantom divide is very far in the future. 
For example, with the scale factor $a(t)=\exp \left(H_0 t\right)$, 
where $H_0\simeq 6.3\times 10^{-34}\text{eV}^{-1}$ is the Hubble parameter of 
the de Sitter universe, $z=-0.90$ (when 
the crossing of the phantom divide may begin to appear 
in the exponential models) corresponds to $10^{26}$ years. 

\subsubsection*{Avoidance of the phantom crossing with (inhomogeneous) fluid}

It may be of some interest to check if it is possible to avoid the crossing of 
the phantom divide by adding a suitable (compensating) fluid in the future cosmological 
scenario described by the models~(\ref{F3exp}) and (\ref{F3HS}). Here, 
we indicate a possible realization of it. 
We examine an inhomogeneous fluid with its energy density $\rho$, pressure $P$ and the Eos parameter $\omega$ as a function of $\rho$, i.e., 
$\omega=\omega(\rho)$. 
The EoS is expressed as 
\begin{equation}
\frac{d}{dz}\log\rho=\frac{3}{(z+1)}(\omega(\rho)+1)\,.
\end{equation}
We explore the simple case 
\begin{equation}
\omega(\rho)=A_0\sigma(z)\rho^{\alpha-1}-1\,,
\label{11}
\end{equation}
where $\alpha$ is a constant and $A_0$ is a positive parameter. 
Moreover, 
$\sigma(z)=-1$ when $z\geq 0$ and $\sigma(z)=1$ when $z<0$, such that 
the fluid is in the phantom region for $z\geq 0$ and in the quintessence 
region for $z<0$. 
The fluid energy density reads 
\begin{equation}
\rho=\rho_0\left(B_0-\sigma(z)\log(z+1)\right)^\frac{1}{(1-\alpha)}\,, 
\label{fluid}
\end{equation}
where $\rho_0=\left[3(\alpha-1)A_0\right]^{1/(1-\alpha)}$ and $B_0$ are positive parameters depending on the initial conditions. 

We note that one 
can choose $B_0=1$ without the loss of generality 
and in this way the energy density is defined as a positive quantity. 
If we take $\alpha>1$, when $z\rightarrow +\infty$ or $z\rightarrow-1^+$, 
the energy density asymptotically tends to zero. For $z=0$, we have a maximum, 
$\rho(z=0)=\rho_0$, so that we should require $\rho_0\ll \Lambda/\kappa^2$, 
namely, 
the fluid energy density is always small with respect to the dark energy density given by our models for the cosmological constant. 
A fluid in the form of Eq.~(\ref{fluid}) may asymptotically produce a (Big Rip) singularity $H(t)\sim (t_0-t)^{\beta}$, where $t<t_0$ and $\beta>1$ (for general study of singularities in modified gravity, see~\cite{lavorisingol}), 
only for $\beta=1/(2\alpha-1)$~\cite{Alessia}, but in our case $\alpha>1$, 
so that this kind of divergence can never appear. 
If we add this fluid in the scenario described by $F(R)$ gravity models 
in Eqs.~(\ref{F3exp}) and (\ref{F3HS}), when $z\rightarrow-1$ we find 
%
\begin{equation}
\omega_{\mathrm{eff}}=\frac{P_{\mathrm{DE}}+P}{\rho_{\mathrm{DE}}+\rho}\simeq -1+\frac{A_0\rho^\alpha}{\Lambda/\kappa^2}\,.
\end{equation}
This means that owing to the presence of fluid, the oscillations of the 
effective EoS parameter realize 
not around the phantom divide 
but around $\omega_{\mathrm{eff}}$ given by the last equation, 
namely in the quintessence region. 
With an accurate fitting of the parameters, in this way 
we may avoid the crossing of the phantom divide. 

We can also add a fluid to the cosmological scenario in order to have an 
asymptotical phantom phase without the Big Rip singularity. 
To this purpose, we investigate 
the EoS parameter of the fluid as in Eq.~(\ref{11}) with $A_0>0$ and 
$\sigma(z)=-1$, which describes a phantom fluid. 
The fluid energy density is given by Eq.~(\ref{fluid}). 
We put $B_0=0$  and $\alpha<1$ such that $1/(1-\alpha)$ can be an even 
number and one can have the energy density defined as a positive quantity. 
In this way, the fluid energy density decreases until $z=0$ and then it starts 
to grow up. We can take $\rho_0$ sufficiently small so that the fluid 
contribution can become dominant only in the asymptotical limit, 
when $z$ is close to $-1$, avoiding the quintessence region in the final cosmological evolution of our $F(R)$ gravity models. {}From the equation of motion 
$3H^2/\kappa^2=\rho$, we obtain 
\begin{equation}
t=-\int^{z(t)}_{0} \sqrt{\frac{3}{\kappa^2\rho(z')}}\frac{d z'}{(z'+1)}\,.
\end{equation}
In our case, it is easy to verify that 
$t\sim |\log(z+1)^{(2\alpha-1)/(2\alpha-2)}|$ and if $\alpha\leq 1/2$, 
when $z(t)\rightarrow -1$ the integral diverges and $t\rightarrow +\infty$, 
avoiding the Big Rip at a finite time. 
In this kind of models, the fluid energy density increases with time, but $\omega\rightarrow -1$ asymptotically, so that there can be no future singularity. 
However, in Ref.~\cite{lRip1} 
a careful investigation on 
the conditions necessary to produce this evolution has
been done, and 
it has been demonstrated that this fluid can rapidly expand in the future, 
leading to the disintegration of all bound structures 
(this is the so-called ``Little Rip''). 
For example, a planet in an orbit of radius $\bar{R}$ around a star of mass $M$ will become unbound when $-(4\pi/3)(\rho + 3P)\bar{R}^3 \simeq M$. 
In our case, $-(\rho + 3P)=A_0 \rho^\alpha$ and in the future every 
gravitationally bound system will be disintegrated~\cite{Caldwell}.

\subsection{Growth of the matter density perturbations: growth index}

\begin{figure}[!h]
\subfigure[]{\includegraphics[width=0.45\textwidth]{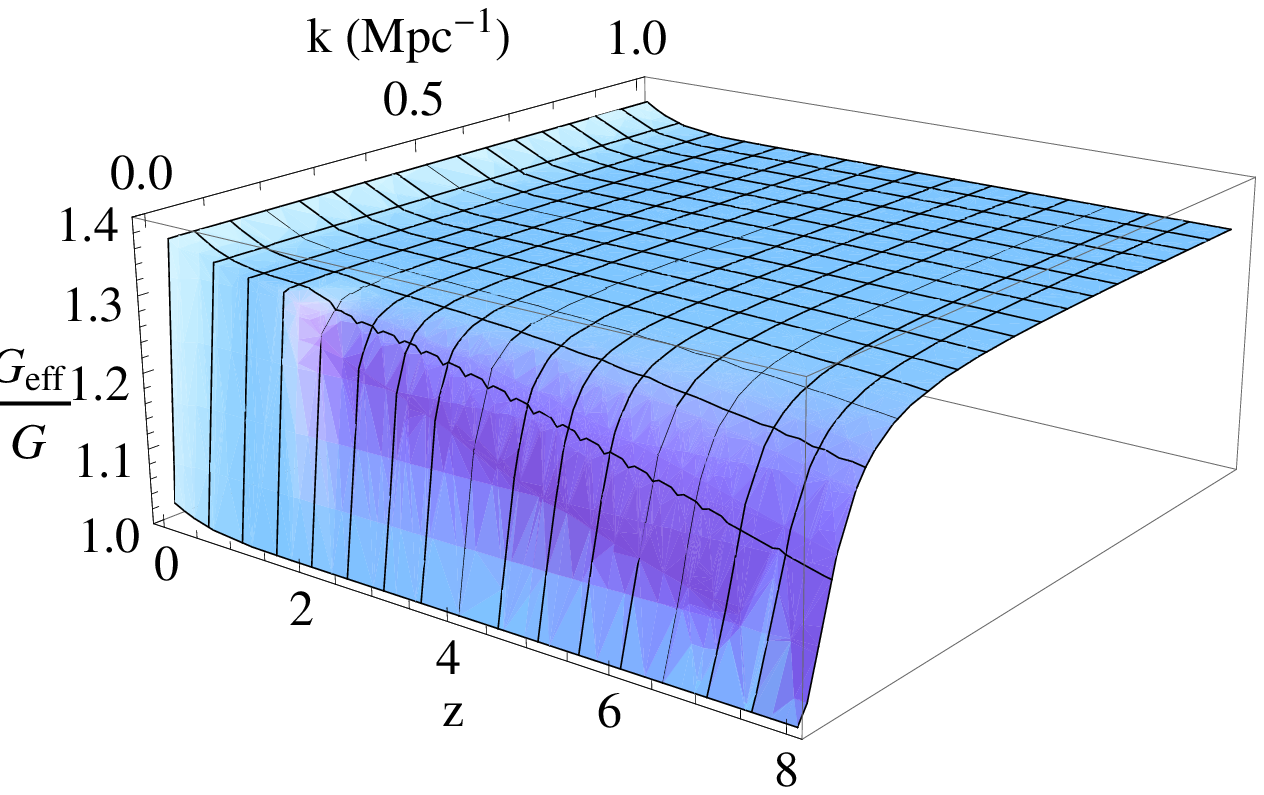}}
\qquad
\subfigure[]{\includegraphics[width=0.45\textwidth]{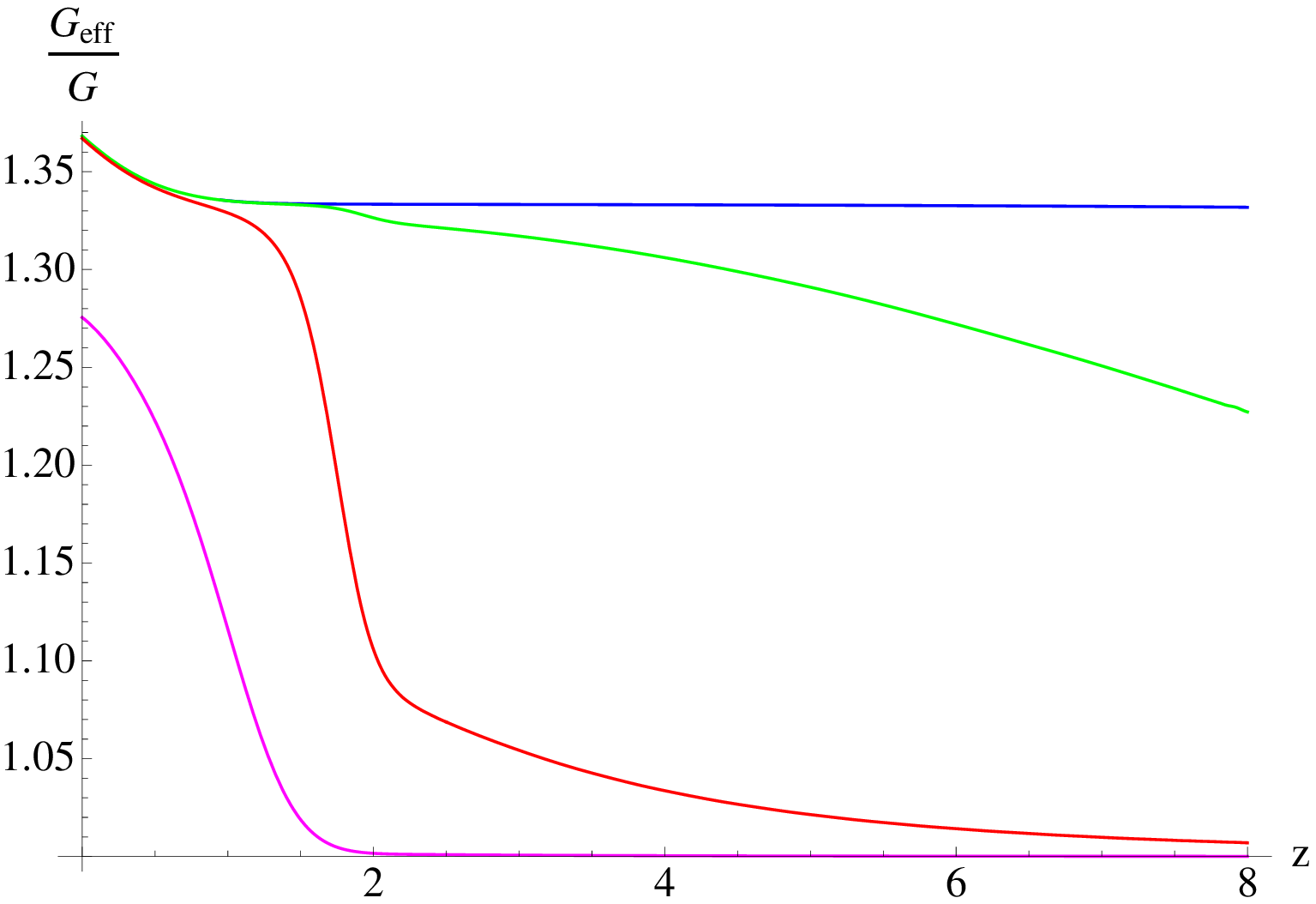}}
\caption{(a) Cosmological evolution as a function of $z$ and the scale dependence on $k$ of the effective gravitational constant $G_\mathrm{eff}$ 
for the model $F_1(R)$ with $b=1$ and $\tilde{\gamma}=1/1000$. (b) Cosmological evolution of $G_\mathrm{eff}$ as a function of $z$ in the model $F_1(R)$ with $b=1$ and $\tilde{\gamma}=1/1000$ 
for $k = 1 \mathrm{Mpc}^{-1}$ (blue), $k = 0.1 \mathrm{Mpc}^{-1}$ (green), $k = 0.01 \mathrm{Mpc}^{-1}$ (red) and $k = 0.001 \mathrm{Mpc}^{-1}$ (fuchsia).}
\label{EM_eff_grav_const}
\end{figure}

\begin{figure}[!h]
\subfigure[]{\includegraphics[width=0.45\textwidth]{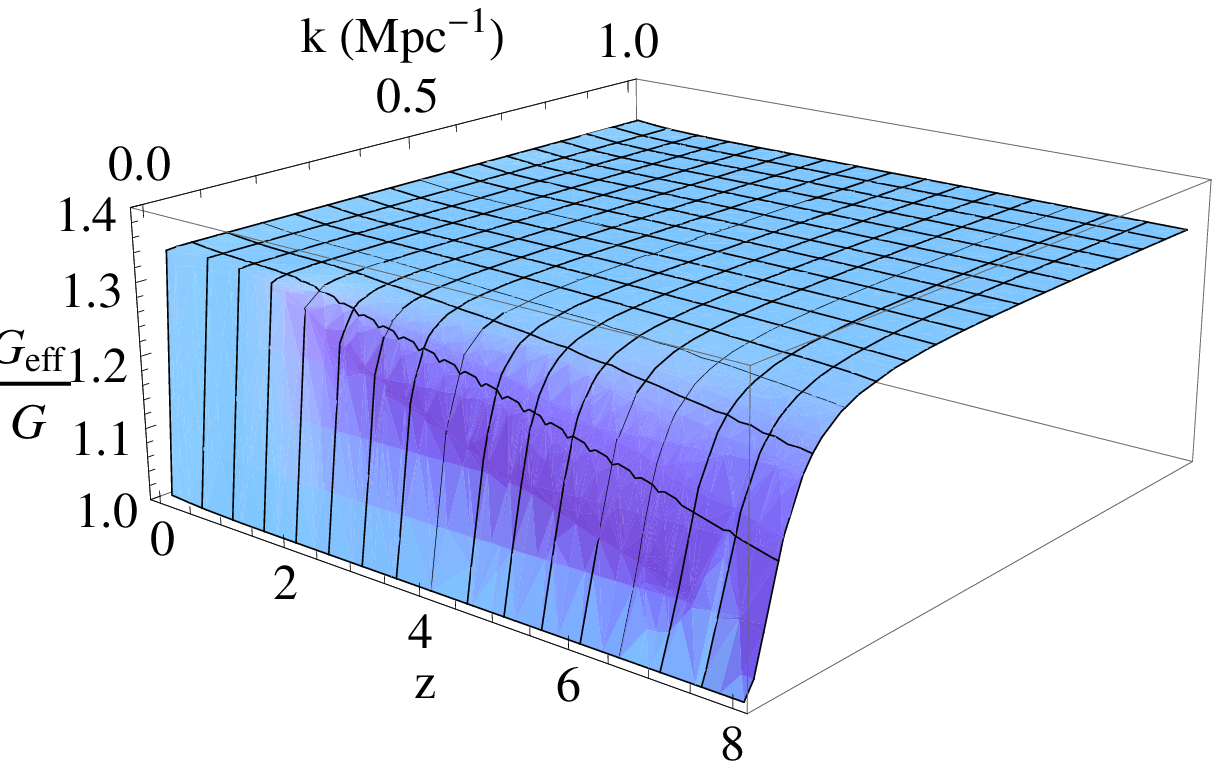}}
\qquad
\subfigure[]{\includegraphics[width=0.45\textwidth]{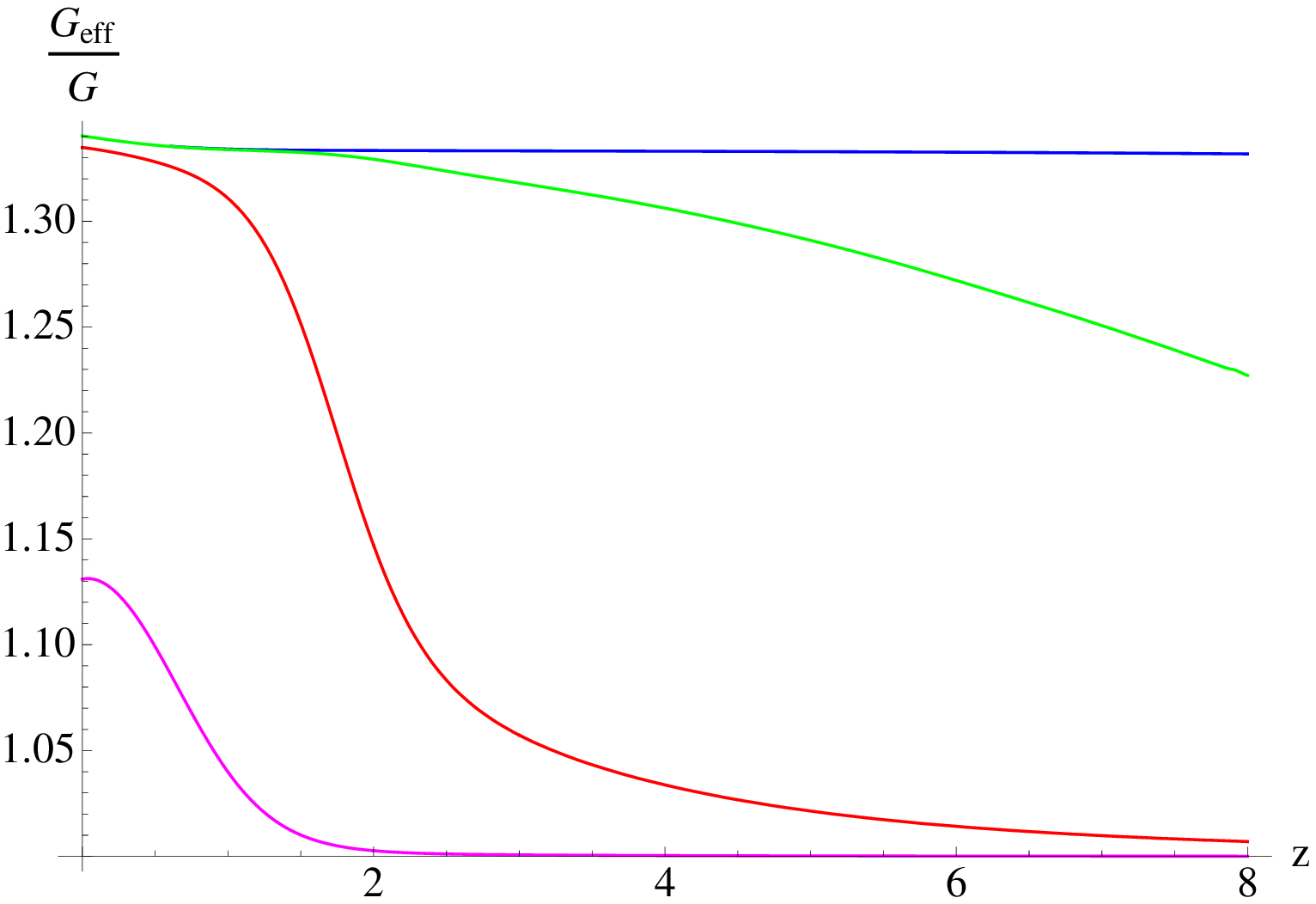}}
\caption{(a) Cosmological evolution as a function of $z$ and the scale dependence on $k$ of the effective gravitational constant $G_\mathrm{eff}$ 
for the model $F_2(R)$ with $b=1$ and $\tilde{\gamma}=1/1000$. (b) Cosmological evolution of $G_\mathrm{eff}$ as a function of $z$ for the model $F_2(R)$ with $b=1$ and $\tilde{\gamma}=1/1000$. 
Legend is the same as Fig.~\ref{EM_eff_grav_const}.}
\label{HS_eff_grav_const}
\end{figure}

In this subsection, we study the matter density perturbations. 
The equation that governs the evolution of the matter density perturbations for $F(R)$ gravity has been derived in the literature 
(see, for example,~\cite{Tsujikawa:2007gd} and references therein). 
Under the subhorizon approximation (for the case without 
such an approximation, see~\cite{BMN-M}), 
the matter density perturbation 
$\delta = \frac{\delta \rho_\mathrm{m}}{\rho_\mathrm{m}}$ 
satisfies the following equation:
\begin{equation}\label{gmp1}
\ddot \delta \, + \, 2 H \dot \delta \, - \, 4 \pi G_\mathrm{eff}(a,k) \rho_\mathrm{m} \delta = 0
\end{equation}
with $k$ being the comoving wavenumber and $G_\mathrm{eff}(a,k)$ being 
the effective gravitational ``constant'' given by 
\begin{equation}\label{gmp2}
G_\mathrm{eff}(a,k) = \frac{G}{F'(R)} \left[ 1 + \frac{\left( k^2/a^2 \right) \left( F''(R)/F'(R) \right)}{1 + 3 \left( k^2/a^2 \right) \left( F''(R)/F'(R) \right)} \right]\,.
\end{equation}
It is worth noting that the appearance of the comoving wavenumber $k$ in the effective gravitational constant makes the evolution of the matter density perturbations dependent on the comoving wavenumber $k$. It can be checked easily, by taking $F(R) = R$ in Eq.~(\ref{gmp2}), that the evolution of the matter density perturbation does not have this kind of dependence in the case of GR. 
In Fig.~\ref{EM_eff_grav_const}, we show the cosmological evolution 
as a function of the redshift $z$ and the scale dependence on the comoving wavenumber $k$ of this effective gravitational constant for the case of model $F_1(R)$ in Eq.~(\ref{F3exp}), while 
in Fig.~\ref{HS_eff_grav_const} we depict those for the case of model $F_2(R)$ 
in Eq.~(\ref{F3HS}). 
In both these cases, we have fixed $b=1$ and used $\gamma=1/1000$. 


Another important remark is to state that in deriving Eq.~(\ref{gmp1}), we have assumed the subhorizon approximation (see~\cite{EspositoFarese:2000ij}). 
Namely, 
comoving wavelengths $\lambda \equiv a/k$ are considered to be much shorter than the Hubble radius $H^{-1}$ as 
\begin{equation}\label{gmp3}
\frac{k^2}{a^2} \gg H^2\,.
\end{equation}
This means that we examine the scales of $\log k \geq -3$. 
On the other hand, as it was pointed out in Ref.~\cite{Cardone:2012xv}, for large $k$ we have to take into account deviations from the linear regime. Hence, we do not consider the scales of $\log k > -1$ 
and take the results obtained for $\log k$ close to $-1$. 

{}From Figs.~7 and 8, we see that $G_\mathrm{eff}$ measured today 
can significantly be different from the Newton's constant in the past. 
The Newton's constant should be normalized to the current one 
as $(G_\mathrm{eff}/G)$. 
This implies that the Newton's constant at the decoupling epoch 
must be much lower than what is implicitly assumed in CMB codes such 
as CAMB~\cite{Lewis:1999bs, CAMB-code}. 
This could significantly change the CMB power spectrum 
because it changes, for example, 
the relation between the gravitational interaction and the Thomson scattering 
rate. 
Since we use the CMB data when we examine whether the theoretical results 
are consistent with the observational ones analyzed in the framework of GR, 
it should be important for us to take into account this point. 
Therefore, strictly speaking, if we compare our results with 
the observations, we has to use the observational results obtained by 
analyzing the CMB data with using the present value of $G_\mathrm{eff}$ 
in our $F(R)$ gravity models instead of the Newton's constant $G$ 
in GR. 

Instead of solving Eq.~(\ref{gmp1}) for the matter density perturbation $\delta$, we now introduce the growth rate $f_\mathrm{g} \equiv d \ln{\delta}/d \ln{a}$ and solve the equivalent equation to Eq.~(\ref{gmp1}) for 
the growth rate in terms of the redshift $z$, given by 
\begin{equation}\label{gmp4}
\frac{df_\mathrm{g}(z)}{dz} \, + \, \left( \frac{1 + z}{H(z)} \frac{dH(z)}{dz} - 2 - f_\mathrm{g}(z) \right) \frac{f_\mathrm{g}(z)}{1 + z} + \frac{3}{2} \frac{\tilde{m}^2 (1 + z)^2}{H^2(z)} \frac{G_\mathrm{eff}(a(z),k)}{G} = 0\,. 
\end{equation}

\begin{figure}[!h]
\subfigure[]{\includegraphics[width=0.45\textwidth]{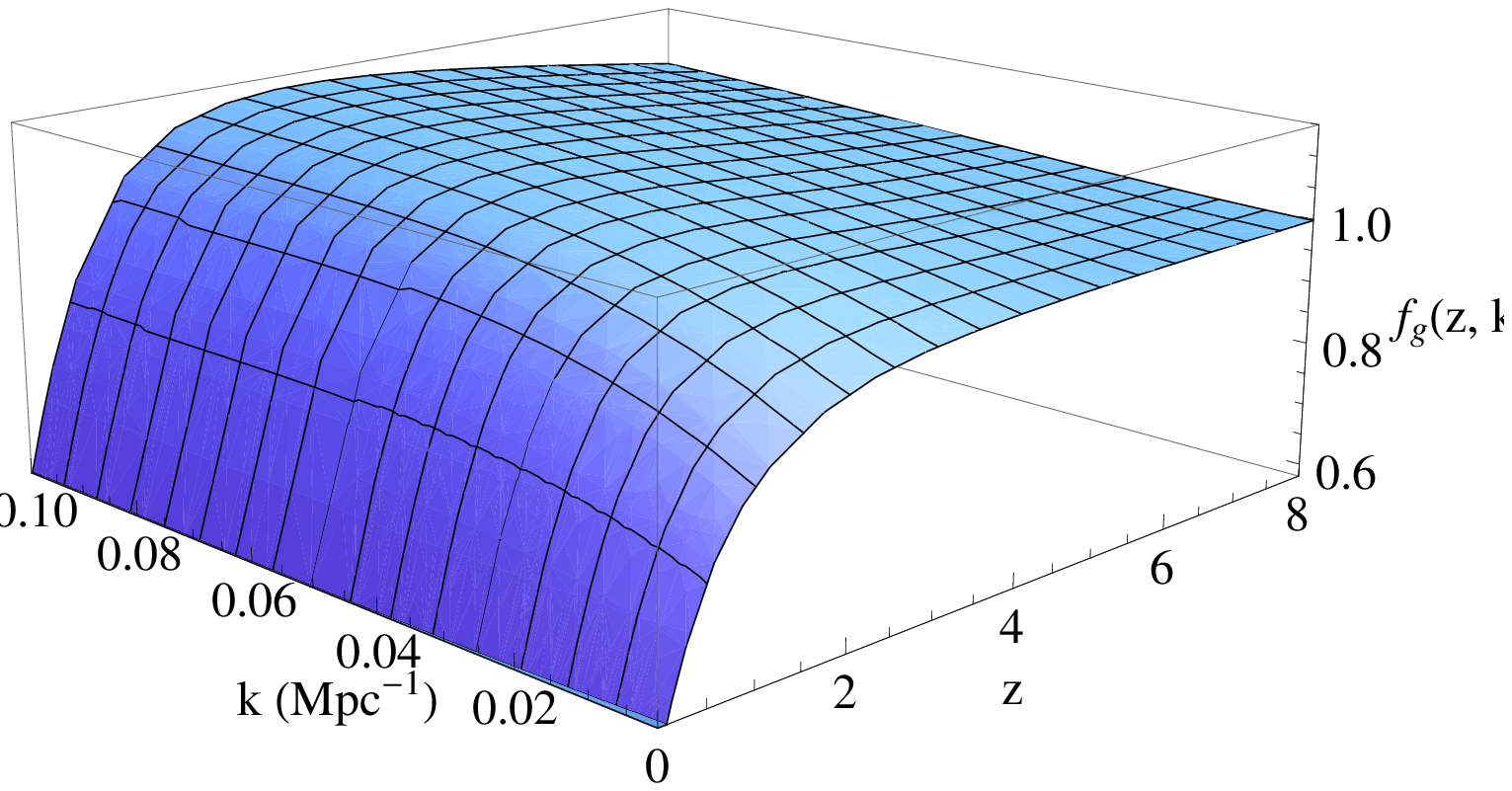}}
\qquad
\subfigure[]{\includegraphics[width=0.45\textwidth]{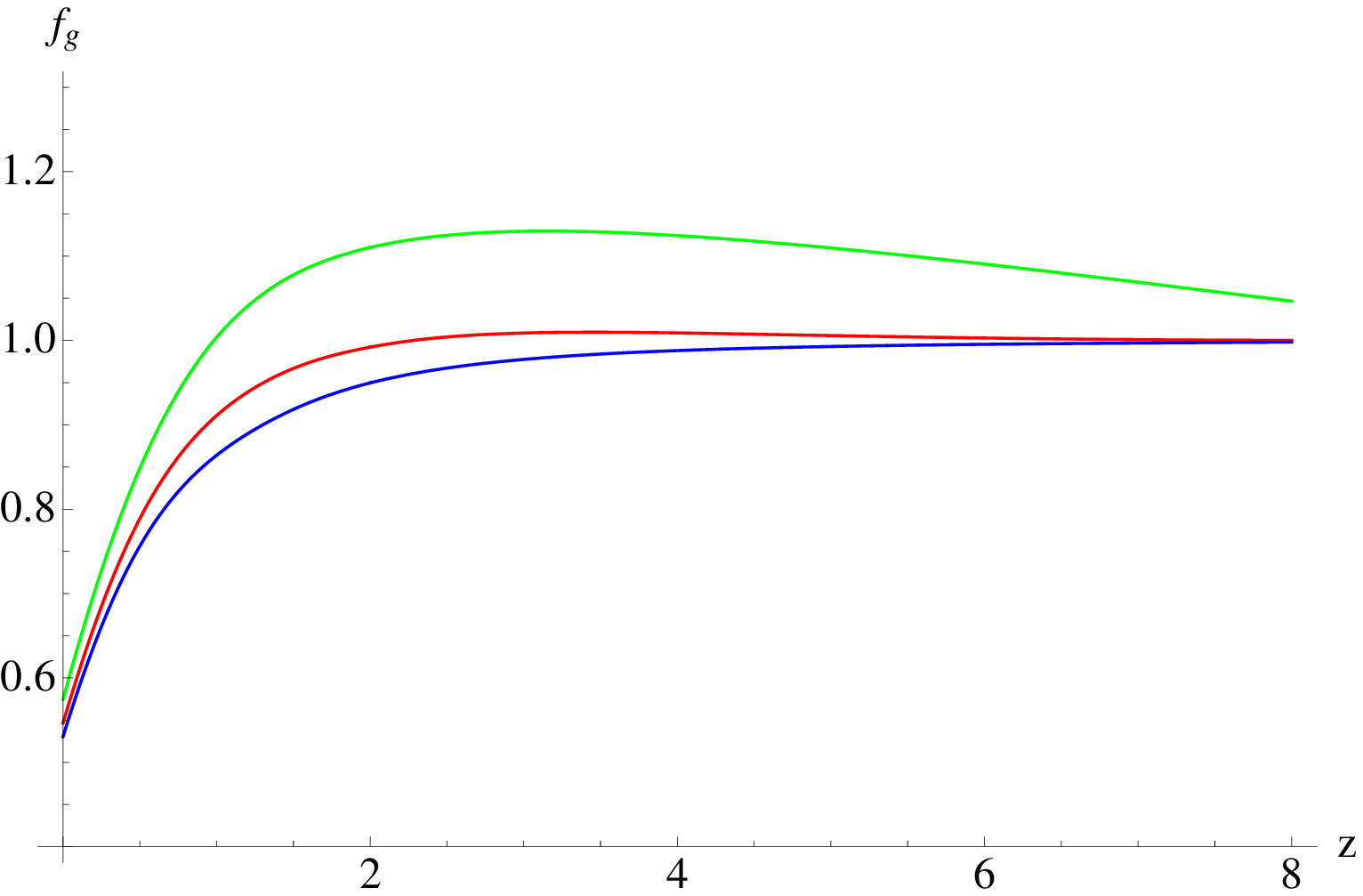}}
\caption{(a) Cosmological evolution as a function of the redshift $z$ and the scale dependence on the comoving wavenumber $k$ of the growth rate 
$f_\mathrm{g}$ for the model $F_1(R)$. (b) Cosmological evolution of the 
growth rate $f_\mathrm{g}$ as a function of $z$ in the model $F_1(R)$ for $k = 0.1 \mathrm{Mpc}^{-1}$ (green), $k = 0.01 \mathrm{Mpc}^{-1}$ (red) and $k = 0.001 \mathrm{Mpc}^{-1}$ (blue).}
\label{EM_growth_rate}
\end{figure}

\begin{figure}[!h]
\subfigure[]{\includegraphics[width=0.45\textwidth]{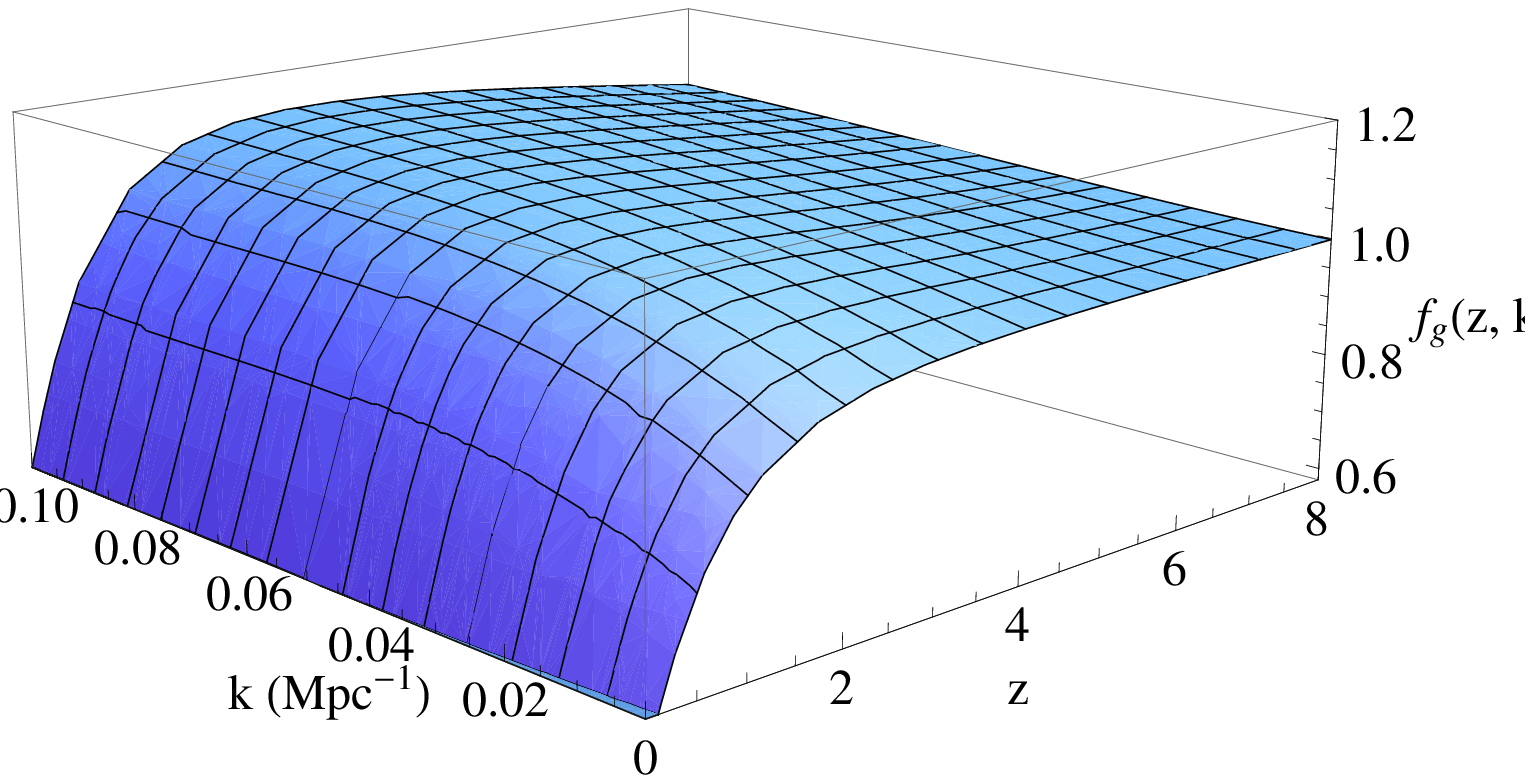}}
\qquad
\subfigure[]{\includegraphics[width=0.45\textwidth]{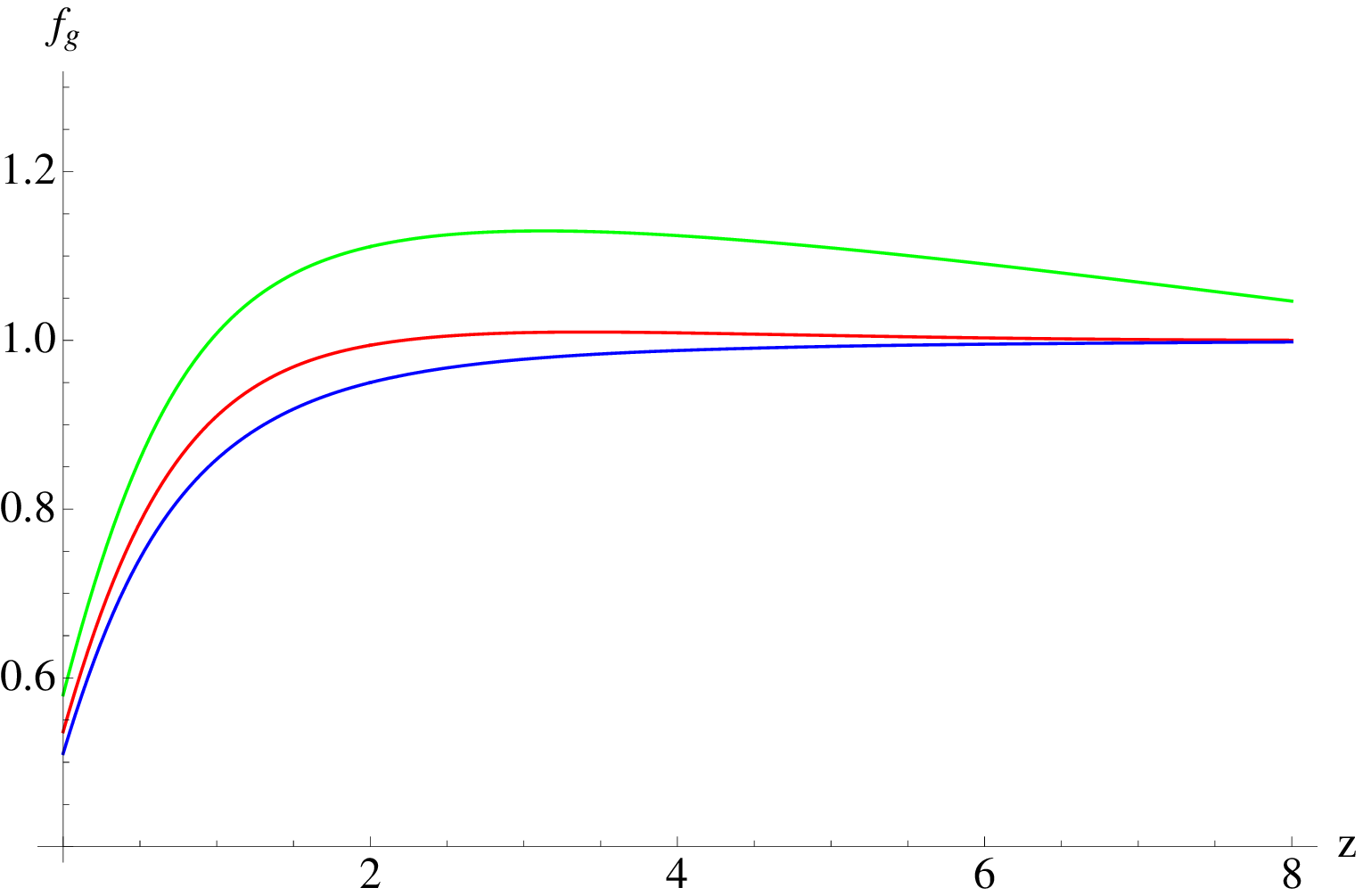}}
\caption{(a) Cosmological evolution as a function of the redshift $z$ and the scale dependence on the comoving wavenumber $k$ of the growth rate 
$f_\mathrm{g}$ for the model $F_2(R)$. (b) Cosmological evolution of 
the growth rate $f_\mathrm{g}$ as a function of $z$ for the model $F_2(R)$. 
Legend is the same as Fig.~\ref{EM_growth_rate}.}
\label{HS_growth_rate}
\end{figure}

Unfortunately, Eq.~(\ref{gmp4}) cannot be solved analytically for the models $F_1(R)$ and $F_2(R)$, but it can be solved numerically by imposing the initial conditions. Therefore, we execute the numerical calculations for both the model $F_1(R)$ and the model $F_2(R)$ with the condition that at a very high redshift the growth rate becomes that in the $\Lambda$CDM model. 
In Fig.~\ref{EM_growth_rate}, we illustrate 
the cosmological evolution 
as a function of the redshift $z$ and the scale dependence on the comoving wavenumber $k$ of the growth rate for the model $F_1(R)$, while we depict those of the growth rate for the model $F_2(R)$ in Fig.~\ref{HS_growth_rate}. 


One way of characterizing the growth of the matter density perturbations could be to use the so-called growth index $\gamma$, which is defined as the quantity satisfying the following equation: 
\begin{equation}\label{gi}
f_\mathrm{g}(z) = \Omega_\mathrm{m}(z)^{\gamma(z)}\,,
\end{equation}
with $\Omega_\mathrm{m}(z) = \frac{8 \pi G \rho_m}{3 H^2}$ being 
the matter density parameter. 

It is known that the growth index $\gamma$ in Eq.~(\ref{gi}) cannot be 
observed directly, but it can be determined from the observational data of both the growth factor $f_\mathrm{g}(z)$ and the matter density parameter $\Omega_\mathrm{m}(z)$ at the same redshift $z$. 
Even if the growth index is not directly observable quantity, it could have a 
fundamental importance in discriminating among the different cosmological models. One of the reasons is that in general, the growth factor $f_\mathrm{g}(z)$, which can be estimated from redshift space distortions in the galaxy power spectra at different $z$~\cite{Kaiser:1987qv, Hamilton:1997zq}, may not be expressed in terms of elementary functions and this fact makes the comparison among the different models difficult. If Eq.~(\ref{gi}) is satisfied with any ansatz for the growth index $\gamma$, then its determination could provide an easy and fast way to distinguish between cosmological models.

Various parameterizations for the growth index $\gamma$ have been proposed in the literature. In the first stage works on this topic, $\gamma$ was taken constant (see~\cite{Peebles:1984}). In the case of dark fluids with the constant EoS 
$\omega_0$ in GR, it is $\gamma = 3 \left(\omega_0 - 1\right)/\left(6 \omega_0 - 5\right)$ (for the $\Lambda$CDM model, the growth index is $\gamma \approx 0.545$). Although taking $\gamma$ constant is very appropriated for a wide class of dark energy models in the framework of GR (for which $\left| \gamma'(0) \right| < 0.02$), for modified gravity theories $\gamma$ is not constant in general (the cases of some viable $F(R)$ gravity models have been investigated in Refs.~\cite{Gannouji:2008wt, Cardone:2012xv}) and the measurement of $\left| \gamma'(0) \right|$ could be very important in order to discriminate between different theories. 
For this reason, another parameterizations has been proposed. The case of a linear dependence $\gamma(z) = \gamma_0 + \gamma'_0 z$ was treated in Ref.~\cite{Polarski:2007rr}. Recently, an ansatz of the type $\gamma(z) = \gamma_0 + \gamma_1 z/(1 + z)$ with $\gamma_0$ and $\gamma_1$ being constants was explored in Ref.~\cite{Belloso:2011ms} and a generalization given by $\gamma(z) = \gamma_0 + \gamma_1 z/(1 + z)^\alpha$ with $\alpha$ being a constant in Ref.~\cite{Cardone:2012xv}.
In the following, we study some of these parameterizations of the growth index for the case of the models $F_1(R)$ and $F_2(R)$.

\subsubsection{$\gamma = \gamma_0$}

We consider the ansatz for the growth index given by 
\begin{equation} 
\gamma = \gamma_0\,, \nonumber 
\end{equation}
where $\gamma_0$ is a constant.

\begin{figure}
\subfigure[]{\includegraphics[width=0.45 \textwidth]{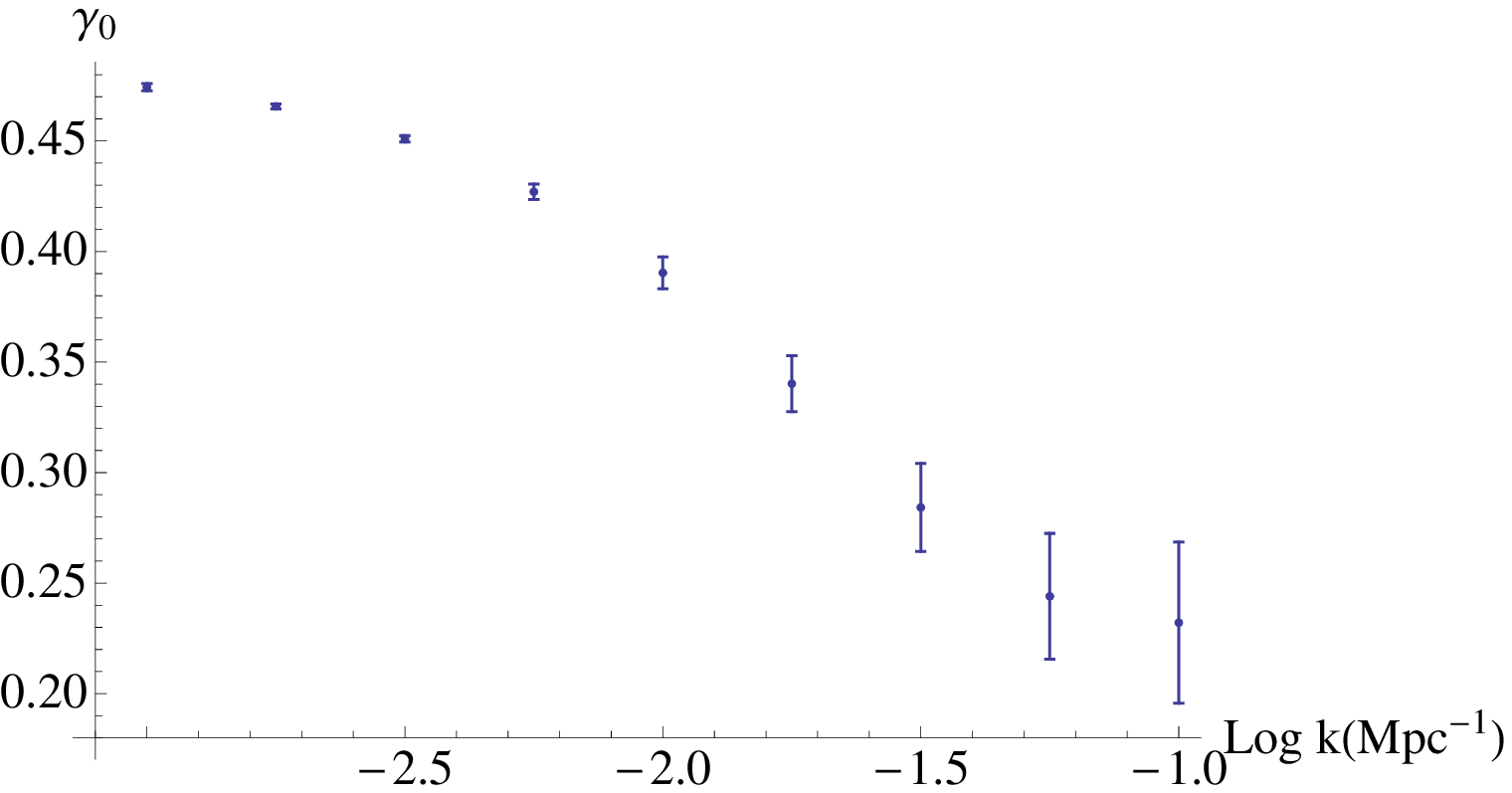}}
\quad
\subfigure[]{\includegraphics[width=0.45 \textwidth]{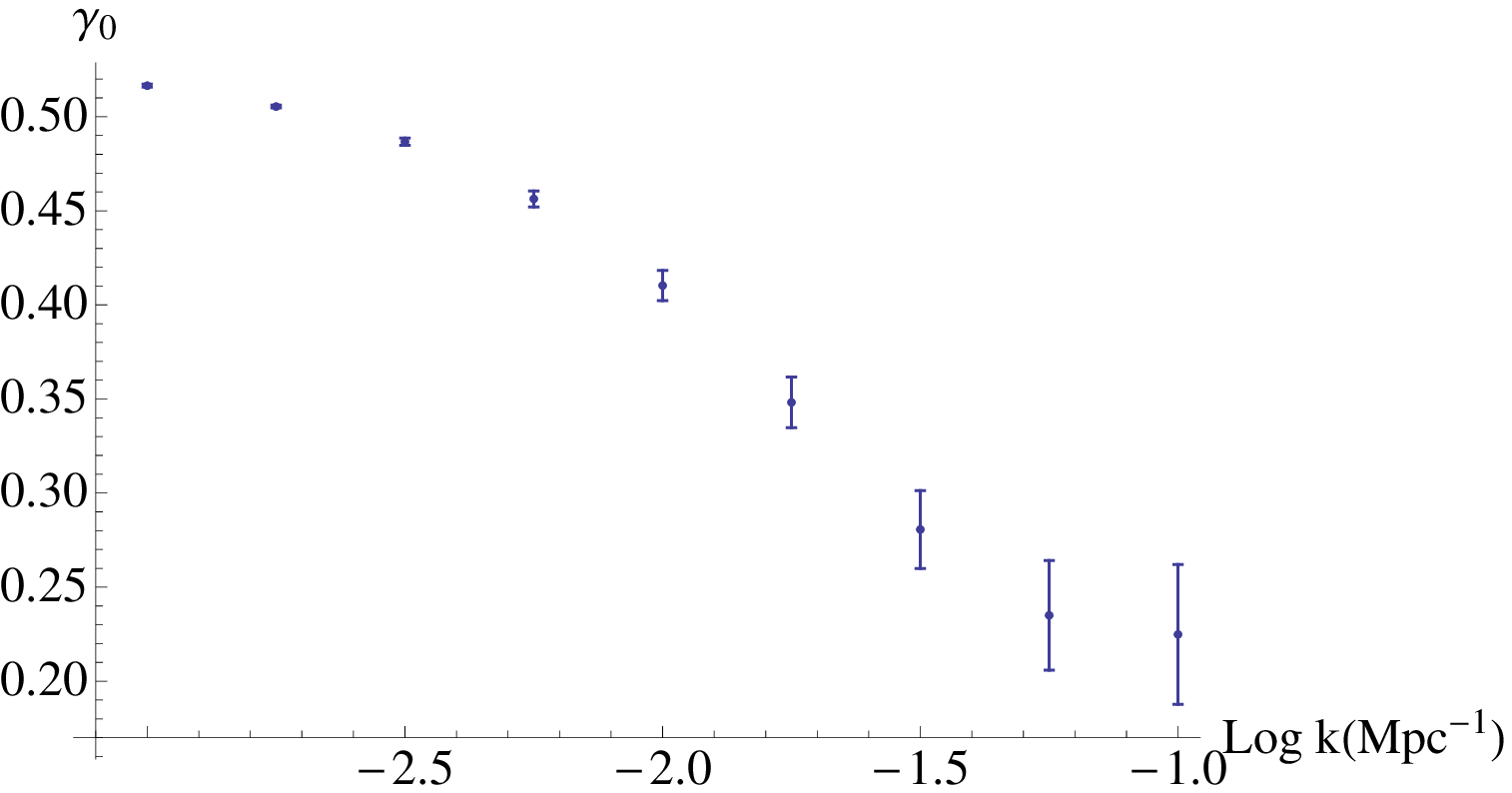}}
\caption{Constant growth index as a function of $\log k$ for the model $F_1(R)$ (a) and for the model $F_2(R)$ (b). The bars express the $68\%$ CL.}
\label{constant_growth_index_vs_logk}
\end{figure}

In Fig.~\ref{constant_growth_index_vs_logk}, we display the results obtained by fitting Eq.~(\ref{gi}) to the solution of Eq.~(\ref{gmp4}) for different values of the comoving wavenumber $k$ for the two models $F_1(R)$ and $F_2(R)$. We note that in these and following plots, the bars express the $68\%$ confidence 
level (CL) and the point denotes the median value. The first important result for both models is that the value of the growth index has a strong dependence with $\log k$. This scale dependence seems to be quite similar in both models. 

\begin{figure}[!h]
\subfigure[]{\includegraphics[width=0.3\textwidth]{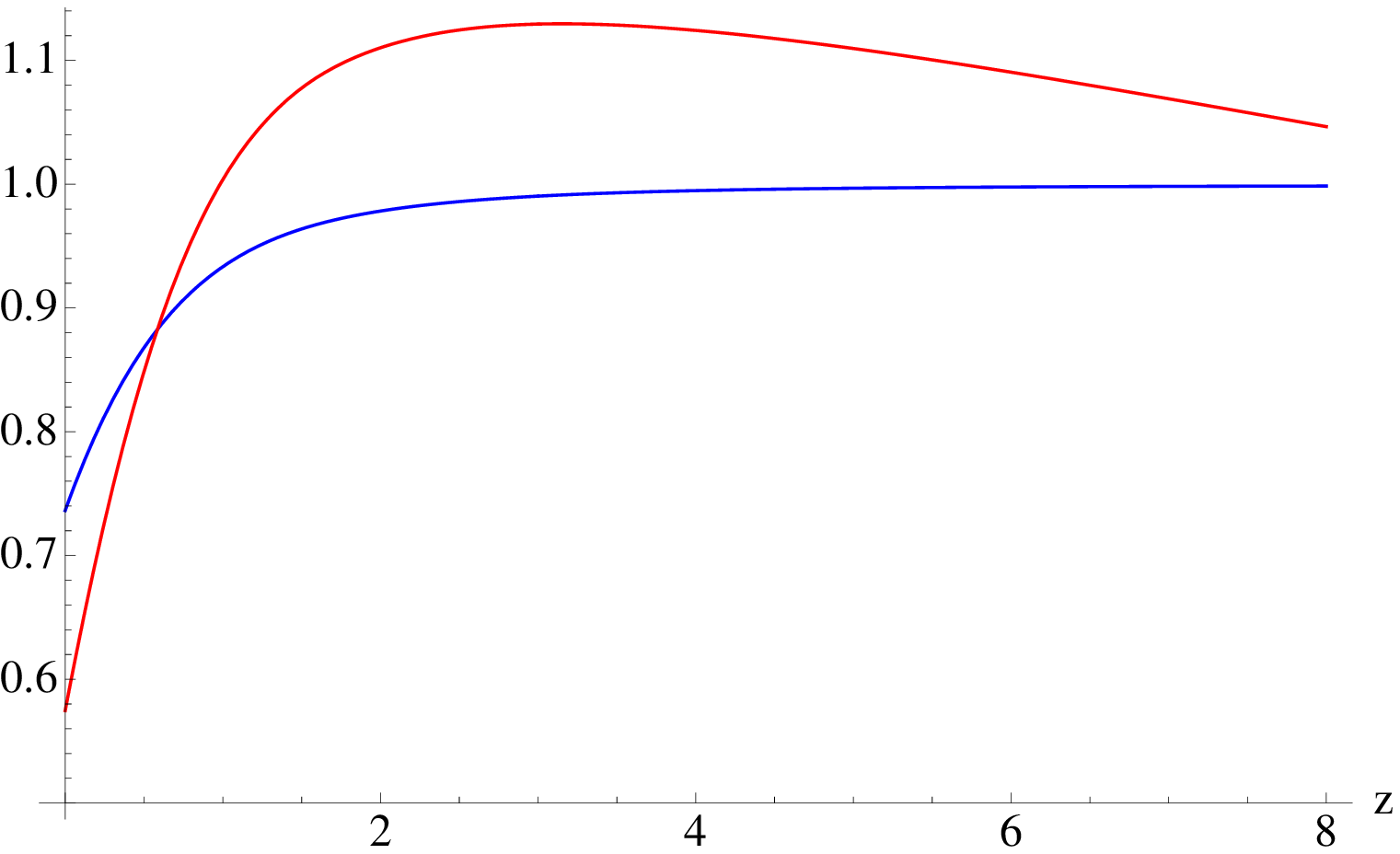}}
\quad
\subfigure[]{\includegraphics[width=0.3\textwidth]{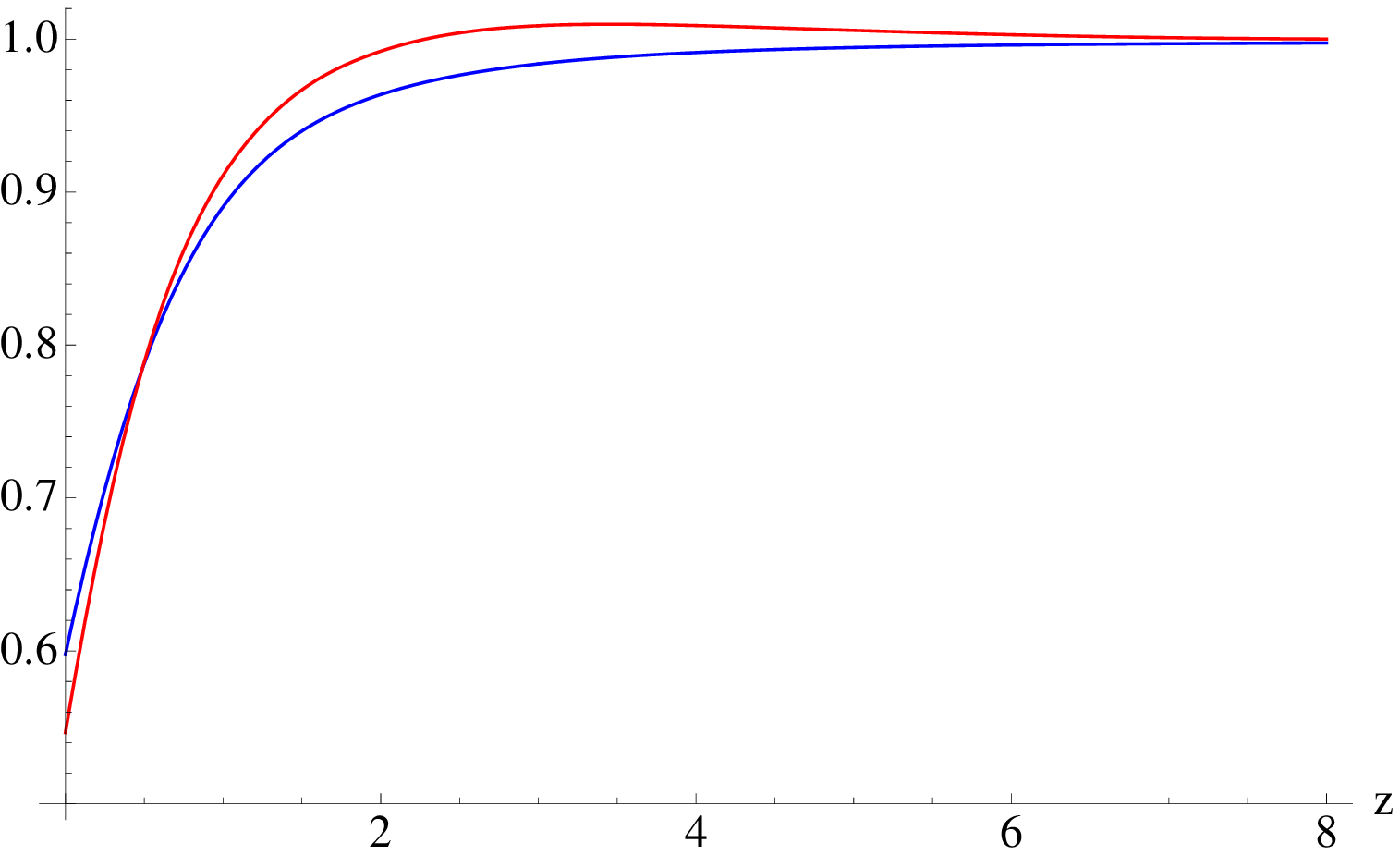}}
\quad
\subfigure[]{\includegraphics[width=0.3\textwidth]{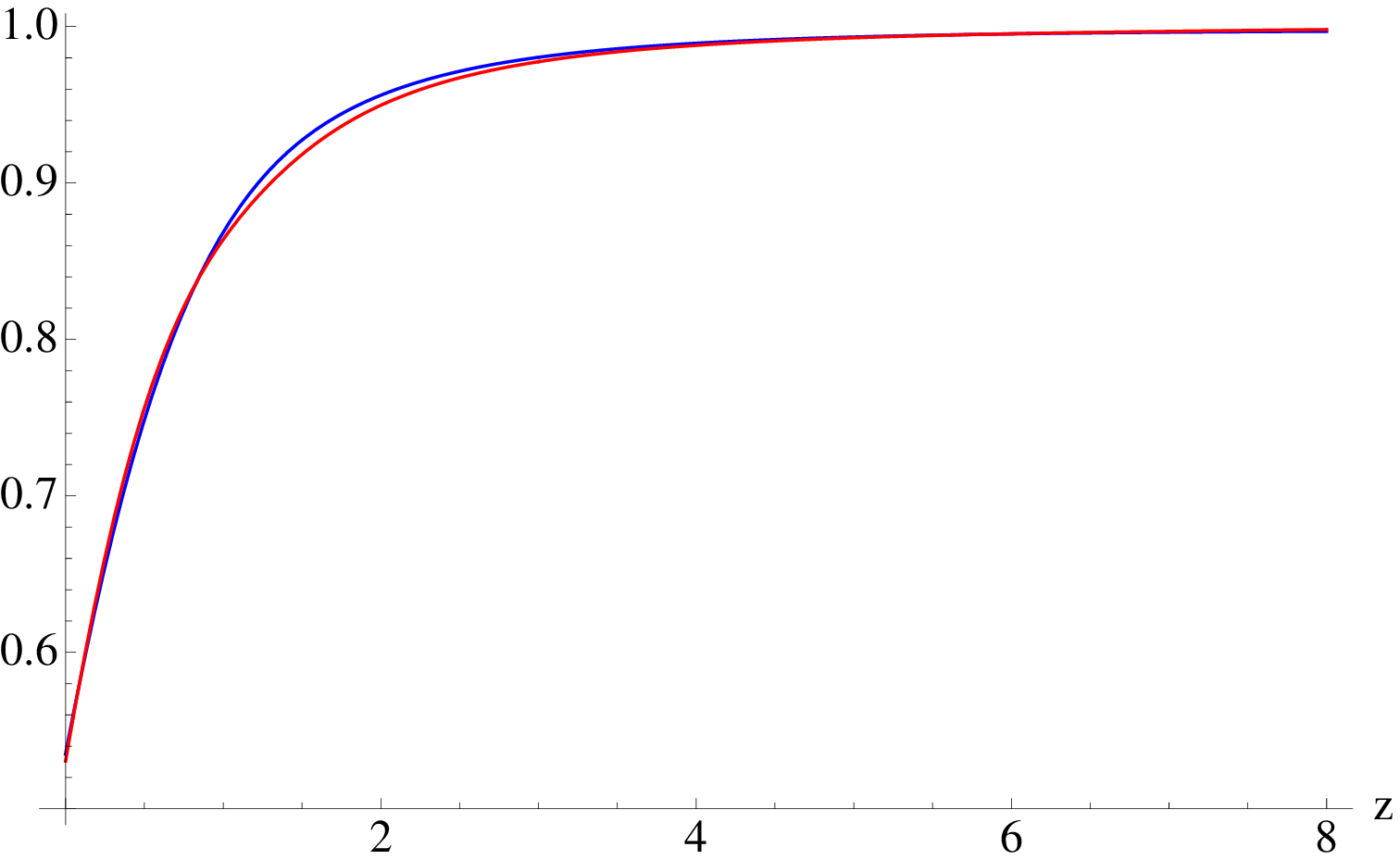}}
\quad
\subfigure[]{\includegraphics[width=0.3\textwidth]{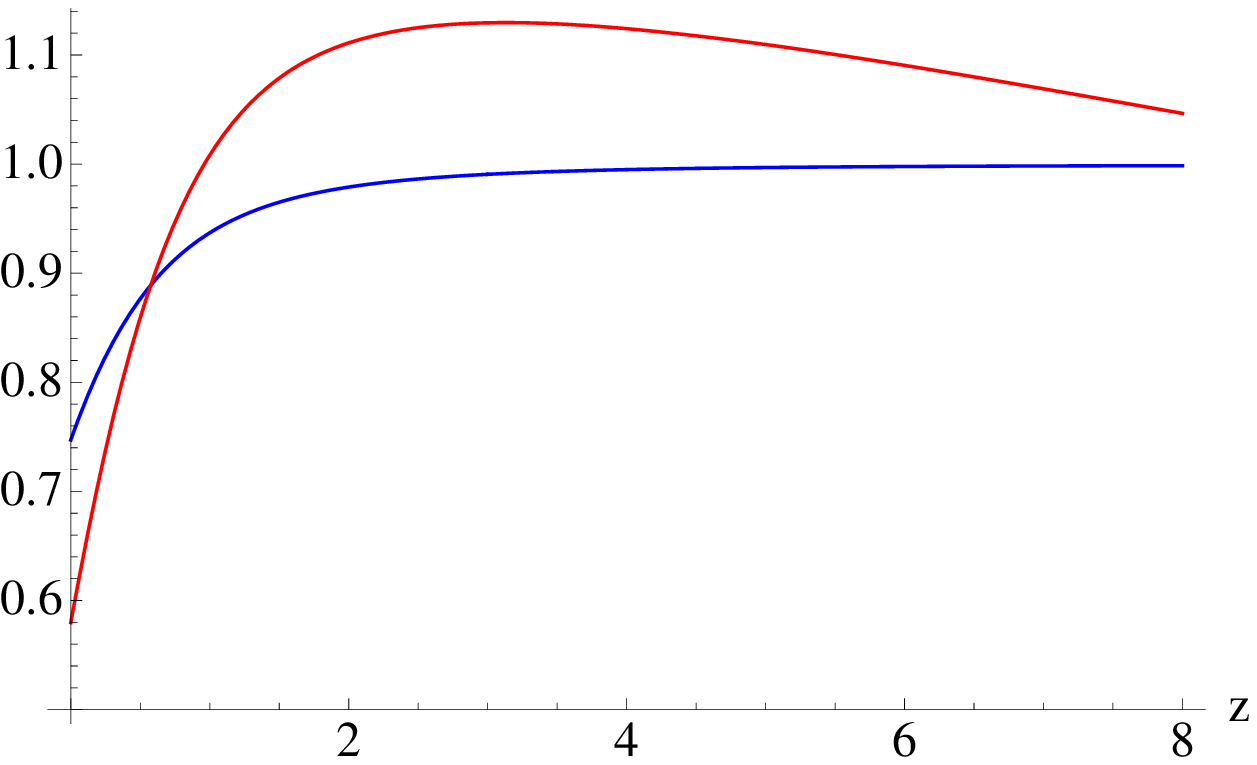}}
\quad
\subfigure[]{\includegraphics[width=0.3\textwidth]{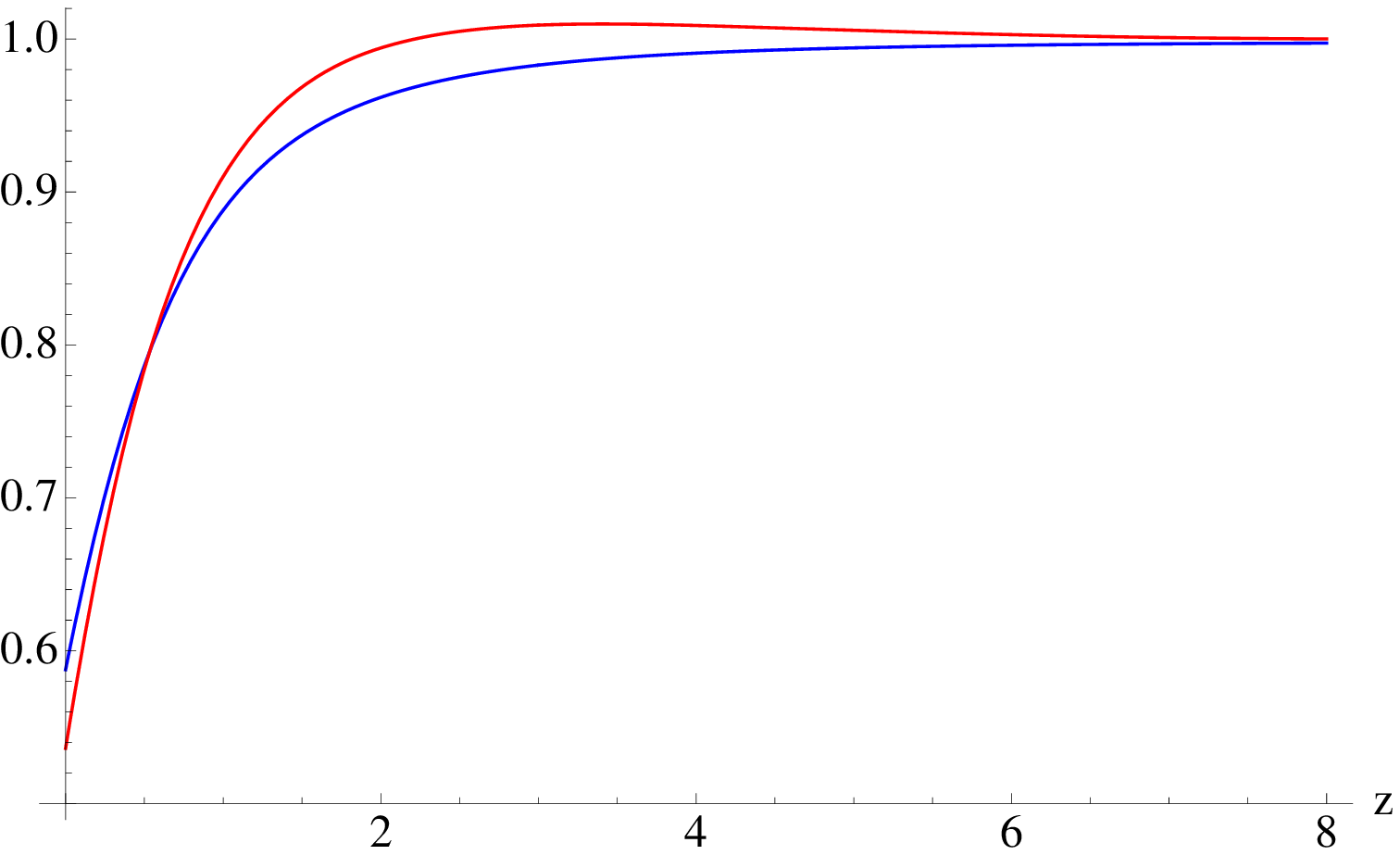}}
\quad
\subfigure[]{\includegraphics[width=0.3\textwidth]{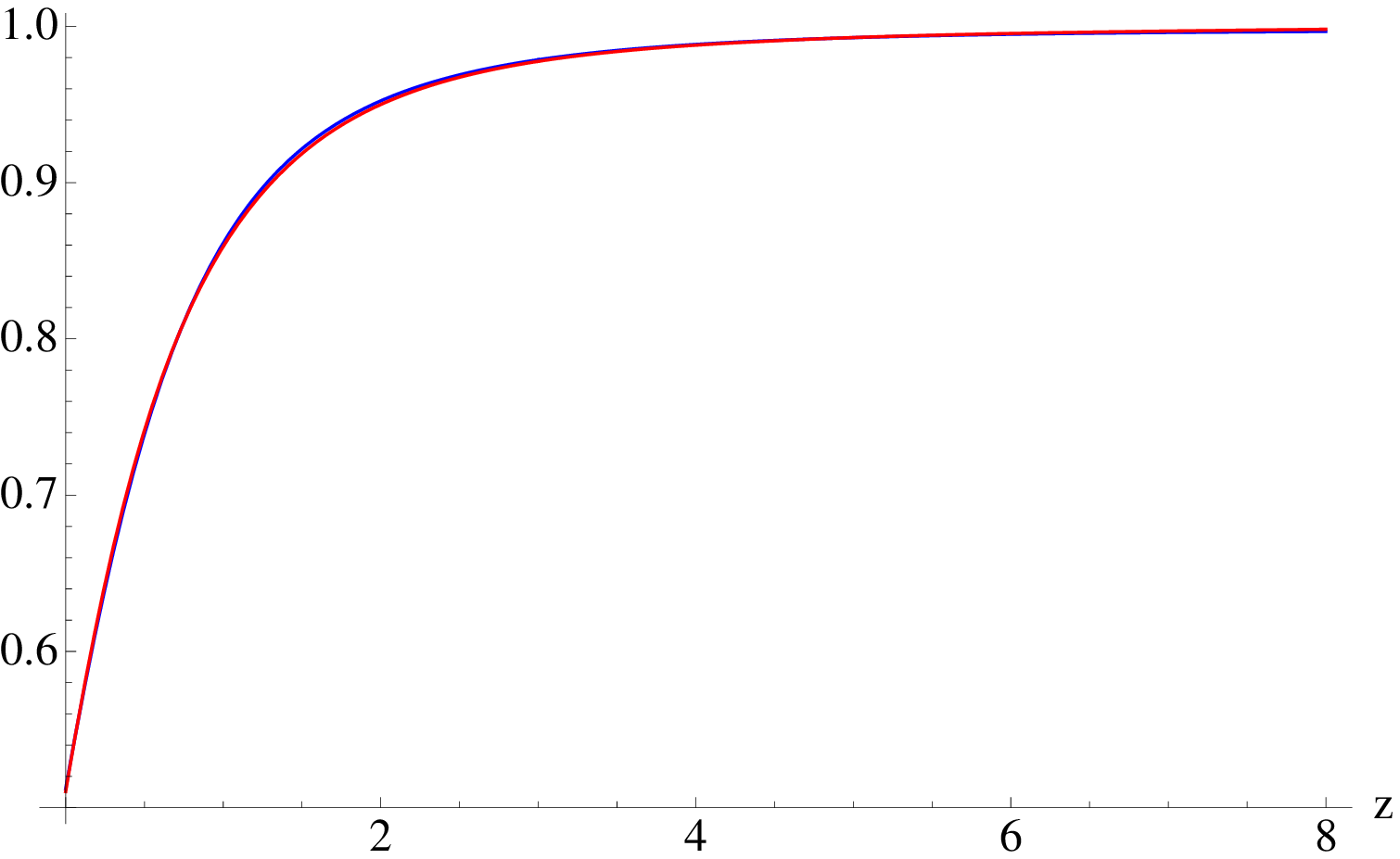}}
\caption{Cosmological evolutions of the 
growth rate $f_\mathrm{g}$ (red) and $\Omega_\mathrm{m}^\gamma$ (blue) with $\gamma = \gamma_0$ as functions of the redshift $z$ in the model $F_1(R)$ for $k = 0.1 \mathrm{Mpc}^{-1}$ (a), $k = 0.01 \mathrm{Mpc}^{-1}$ (b) and $k = 0.001 \mathrm{Mpc}^{-1}$ (c), and those in the model $F_2(R)$ for $k = 0.1 \mathrm{Mpc}^{-1}$ (d), $k = 0.01 \mathrm{Mpc}^{-1}$ (e) and $k = 0.001 \mathrm{Mpc}^{-1}$ (f).}
\label{HS_figure_constant_growth_index}
\end{figure}

\begin{figure}[!h]
\subfigure[]{\includegraphics[width=0.45\textwidth]{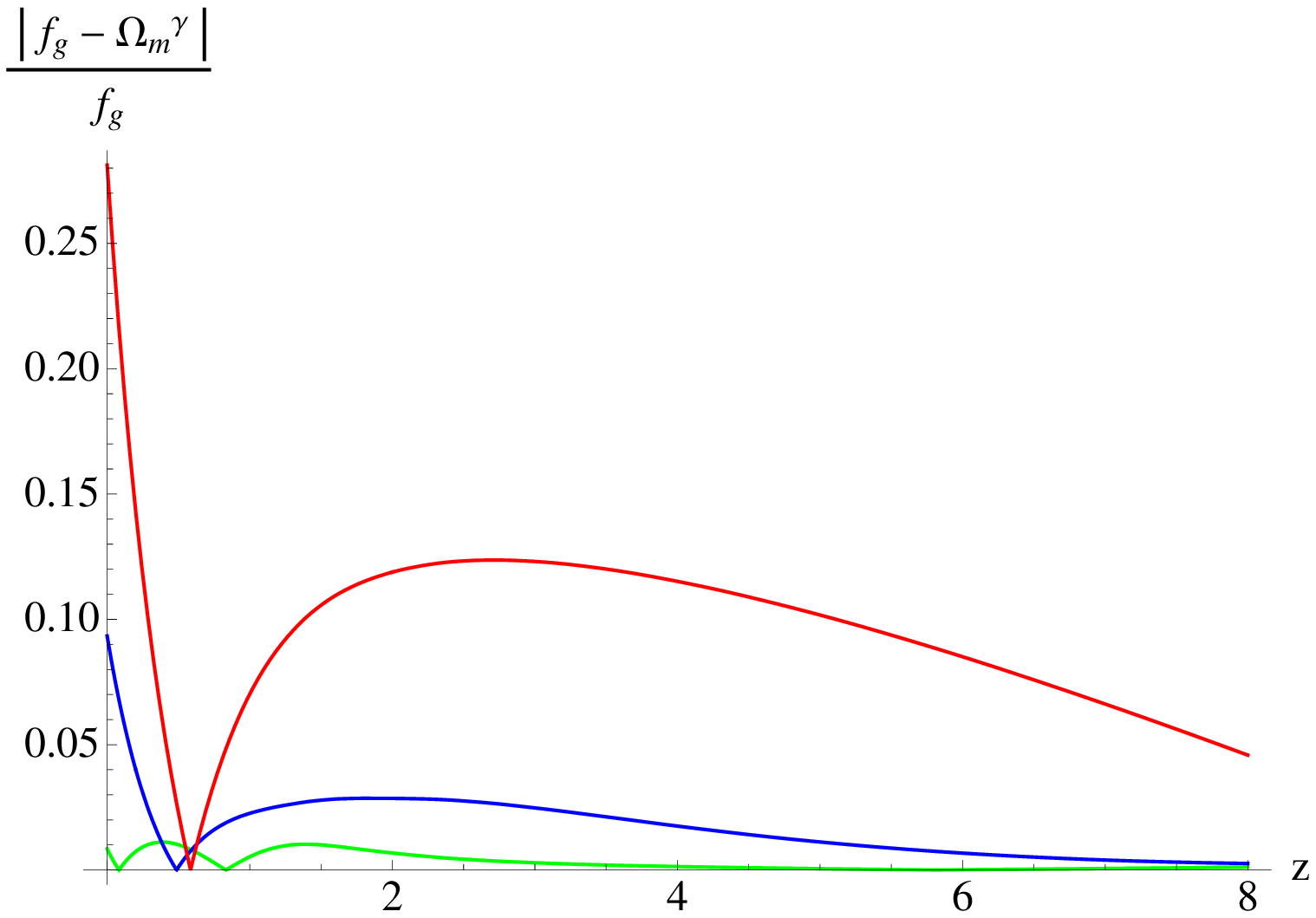}}
\quad
\subfigure[]{\includegraphics[width=0.45\textwidth]{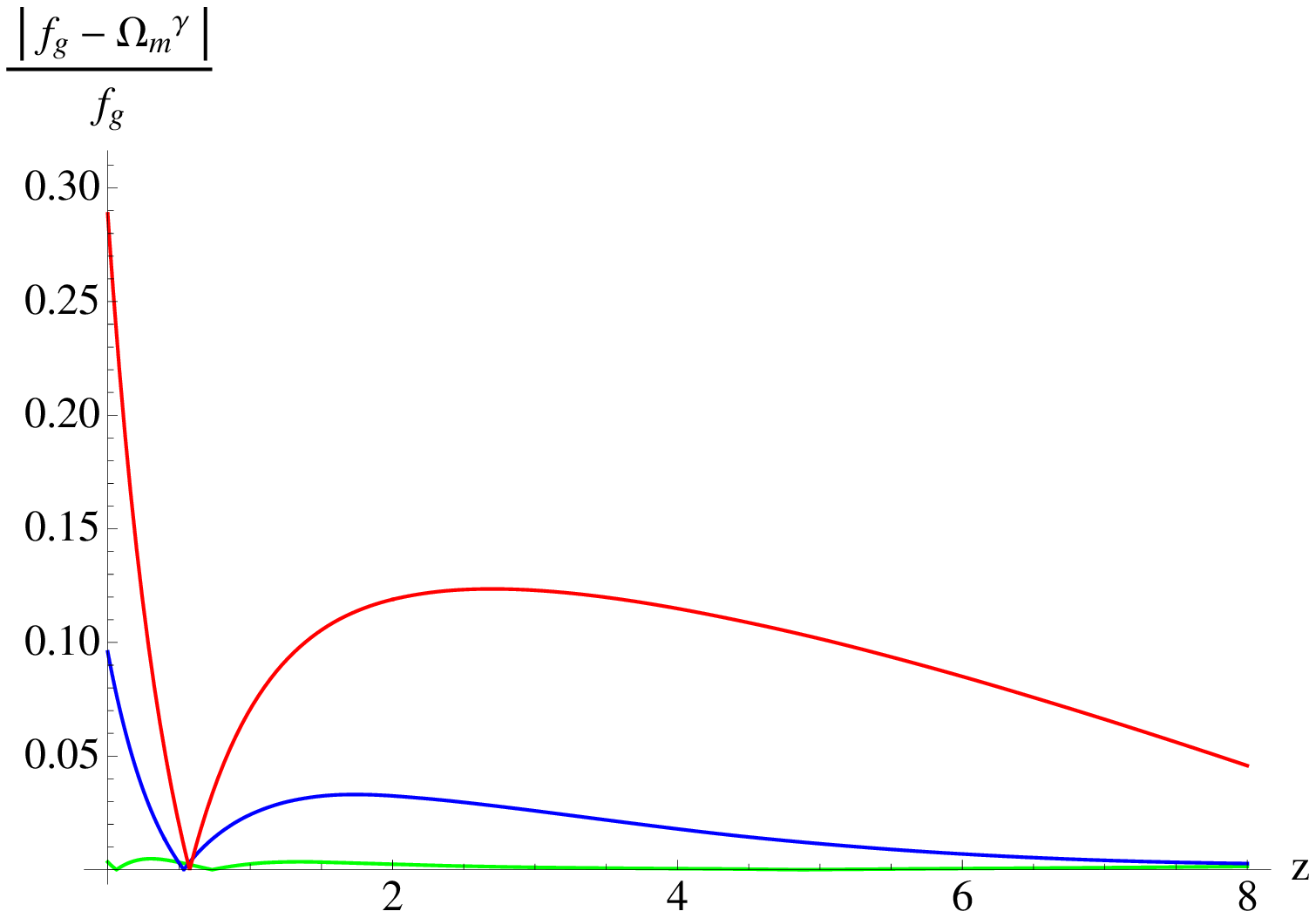}}
\caption{Cosmological evolution of the 
relative difference $\frac{\left| f_\mathrm{g} - \Omega_\mathrm{m}^\gamma \right|}{f_\mathrm{g}}$ with $\gamma = \gamma_0$ for $k = 0.1 \mathrm{Mpc}^{-1}$ (red), $k = 0.01 \mathrm{Mpc}^{-1}$ (blue) and $k = 0.001 \mathrm{Mpc}^{-1}$ (green) in the model $F_1(R)$ (a) and the model $F_2(R)$ (b).}
\label{const_rel_dif}
\end{figure}

In order to check the goodness of our fits, in Fig.~\ref{HS_figure_constant_growth_index} we show cosmological evolutions of 
the growth rate $f_\mathrm{g}(z)$ and $\Omega_\mathrm{m}(z)^{\gamma_0}$ as functions of the redshift $z$ together for several values of the comoving wavenumber $k$ for the models $F_1(R)$ and $F_2(R)$. 
To clarify these results, in Fig.~\ref{const_rel_dif} we also illustrate the 
cosmological evolution of the relative difference between $f_\mathrm{g}(z)$ and $\Omega_\mathrm{m}(z)^{\gamma_0}$ as a function of $z$ 
for the same values of $k$ in these models. The first remarkable thing is that for both models the function $\Omega_\mathrm{m}(z)^{\gamma_0}$ fits the growth rate for large scales (i.e., lower $k$) very well, but this is not anymore the case for larger values of $k$. In fact, if we do not consider lower values for $z$ (i.e., $z < 0.2$), for $\log k = -2$ the relative difference is smaller than $3\%$ for both models, while for $\log k = -1$ can arrive up to almost $13\%$. For $\log k = -3$, we see that the relative difference is always smaller than $1.5\%$ for the model $F_1(R)$ and smaller than $1\%$ for the model $F_2(R)$.


\subsubsection{$\gamma = \gamma_0 + \gamma_1 z$}

With the same procedure used in the previous subsection, we explore a linear dependence for the growth index 
\begin{equation}
\gamma = \gamma_0 + \gamma_1 z\,, 
\end{equation}
where $\gamma_1$ is a constant. 

\begin{figure}[!h]
\subfigure[]{\includegraphics[width=0.45\textwidth]{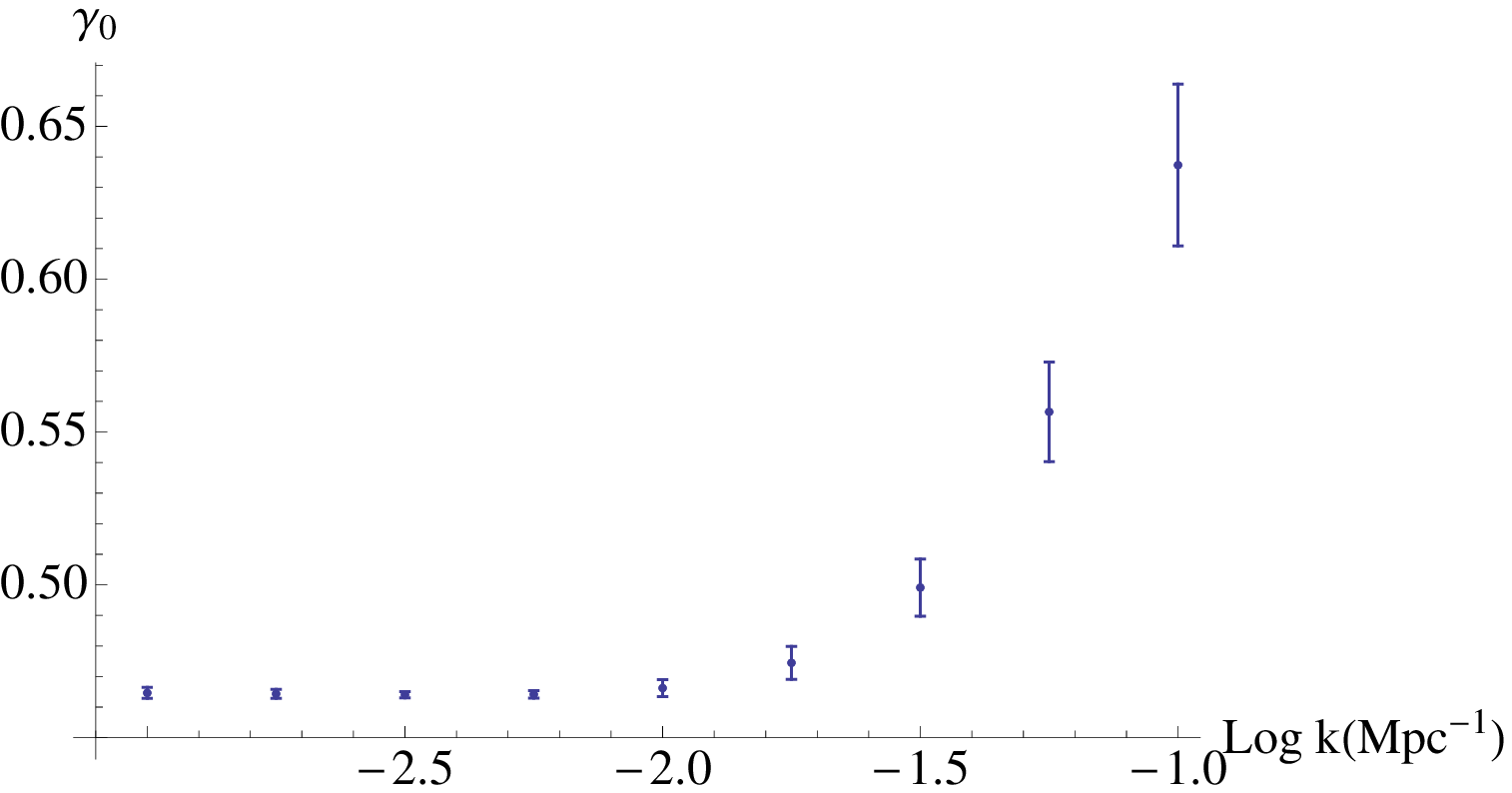}}
\quad
\subfigure[]{\includegraphics[width=0.45\textwidth]{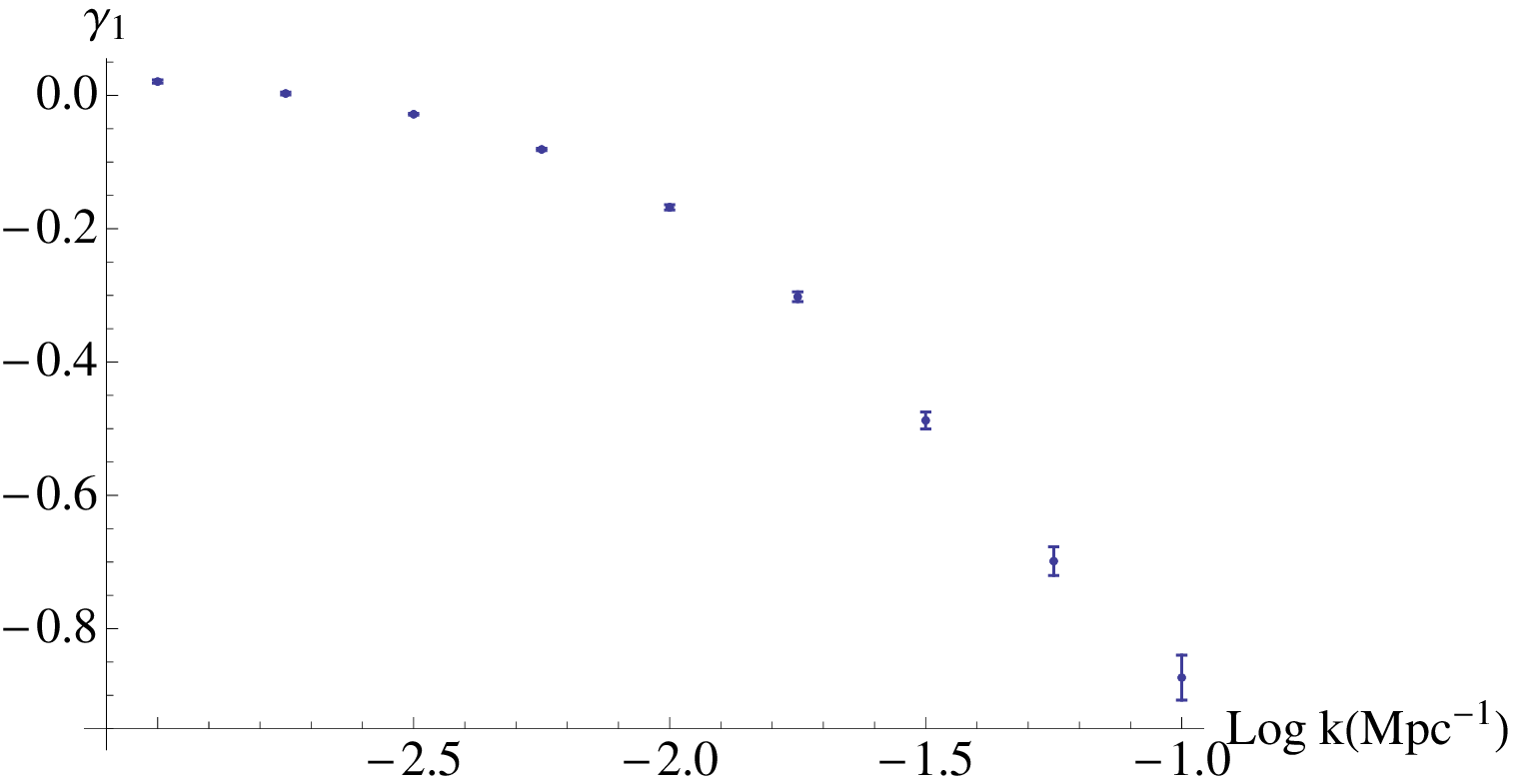}}
\quad
\subfigure[]{\includegraphics[width=0.45\textwidth]{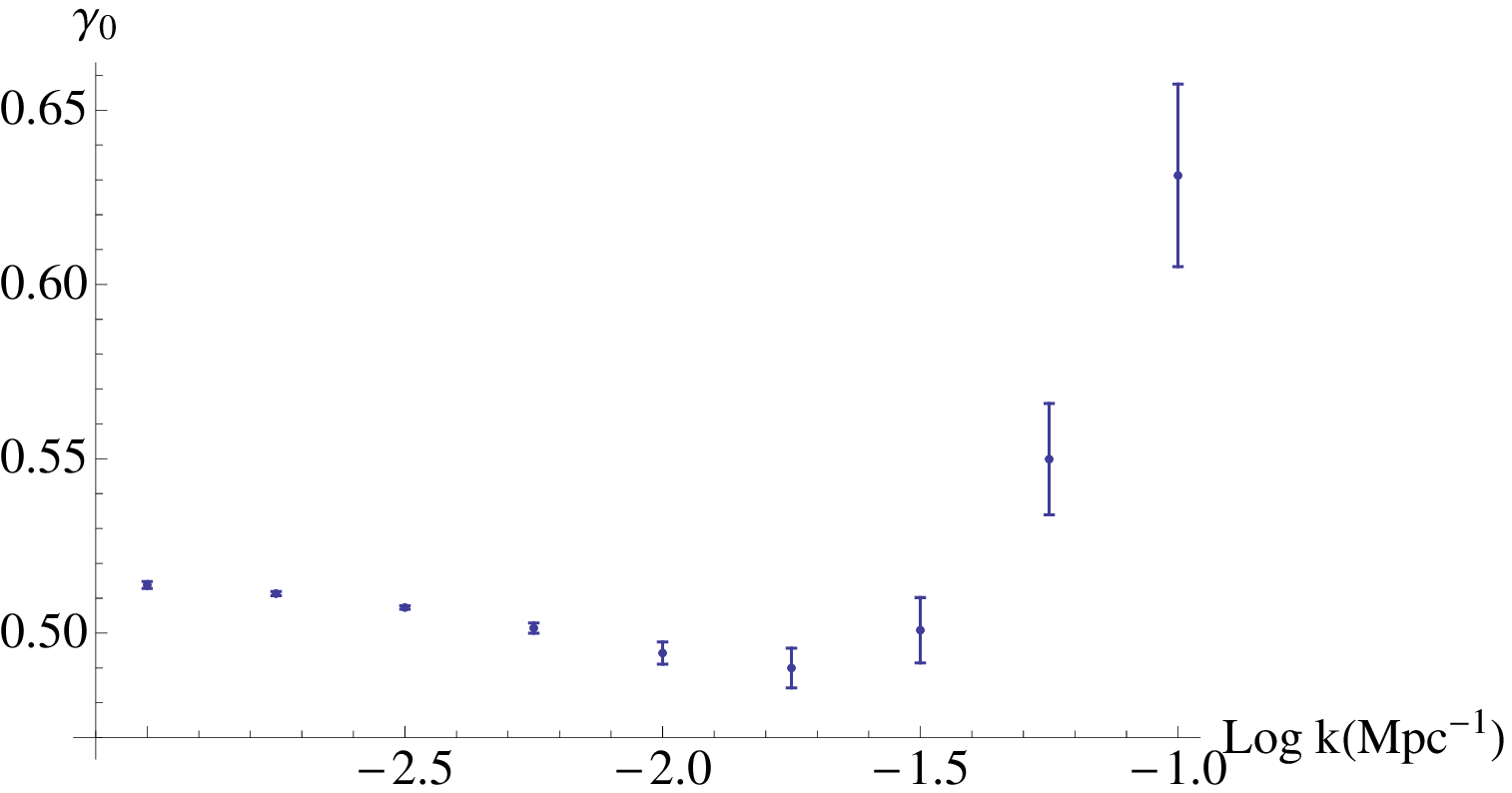}}
\quad
\subfigure[]{\includegraphics[width=0.45\textwidth]{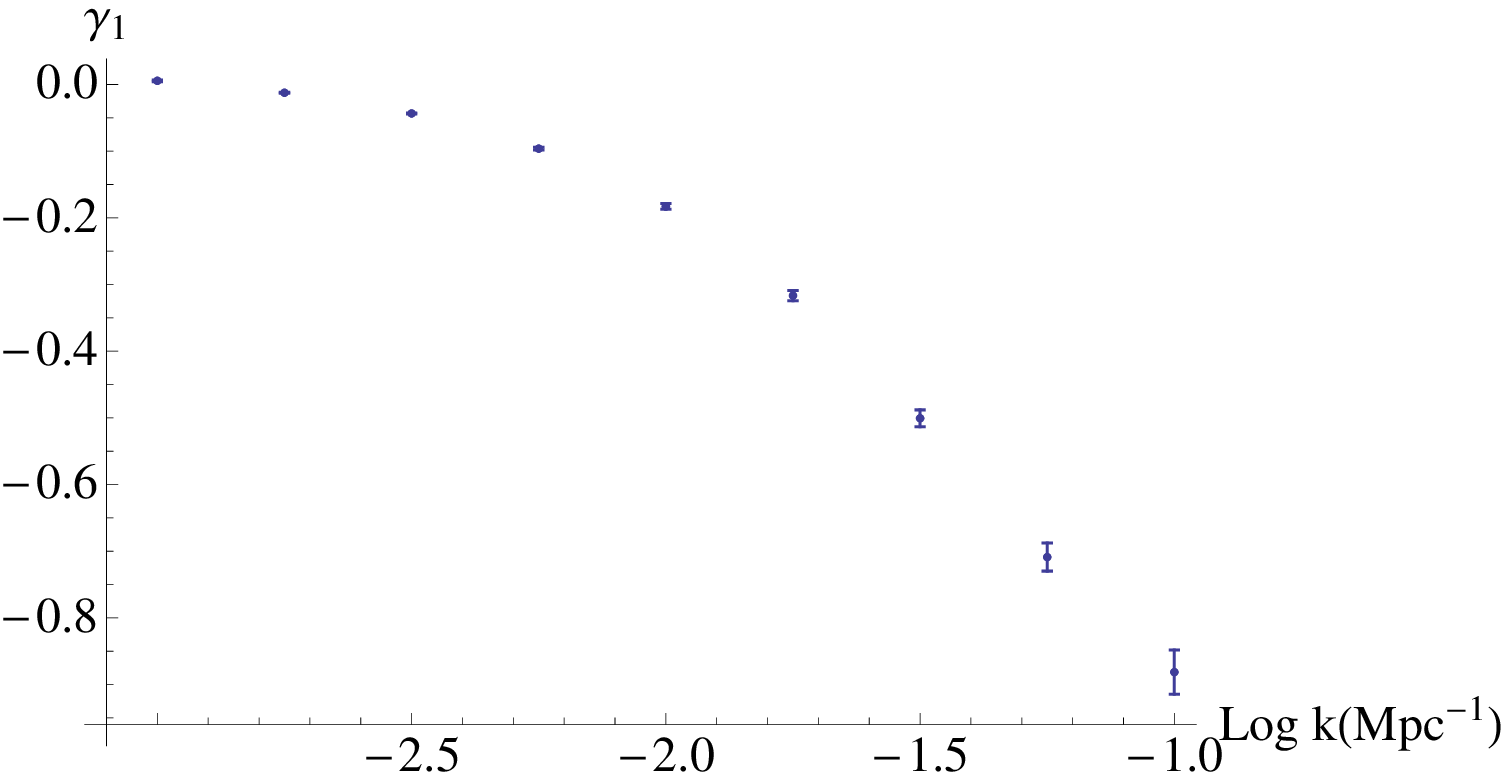}}
\caption{
Growth index fitting parameters in the case $\gamma = \gamma_0 + \gamma_1 z$ 
as a function of $\log k$ for the model $F_1(R)$ [(a) and (b)] 
and the model $F_2(R)$ [(c) and (d)]. 
Legend is the same as Fig.~\ref{constant_growth_index_vs_logk}. 
}
\label{HS_lineal_growth_index_vs_logk}
\end{figure}

In Fig.~\ref{HS_lineal_growth_index_vs_logk},  
we depict the parameters $\gamma_0$ and $\gamma_1$ for several values of 
$\log k$ in both the models. 
As is the same as the case $\gamma = \gamma_0$, it can easily be seen that 
the scale dependence of the parameters $\gamma_0$ and $\gamma_1$ is similar 
in these models. We can also find that $\gamma_0 \sim 0.46$ for the model $F_1(R)$ when $\log k \leq -2$, whereas $\gamma_0 \sim 0.51$ for the model $F_2(R)$ when $\log k \leq -2.5$. 
For both these models, 
the value of $\gamma_1$ has 
a strong dependence on $k$ in the range of $\log k > -2.25$, but in the range 
of $\log k < -2.25$ this dependence becomes weaker. 

\begin{figure}[!h]
\subfigure[]{\includegraphics[width=0.3\textwidth]{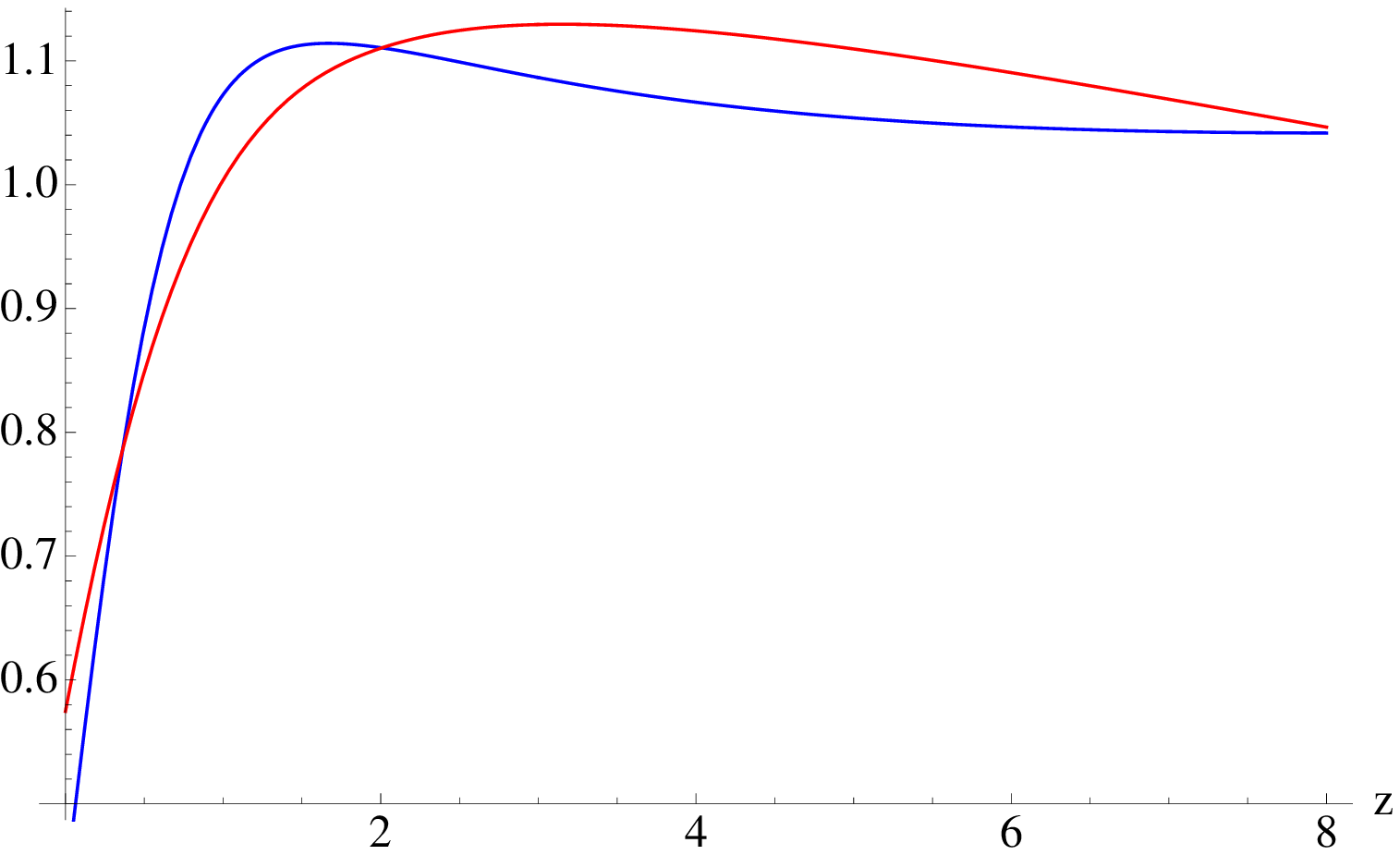}}
\quad
\subfigure[]{\includegraphics[width=0.3\textwidth]{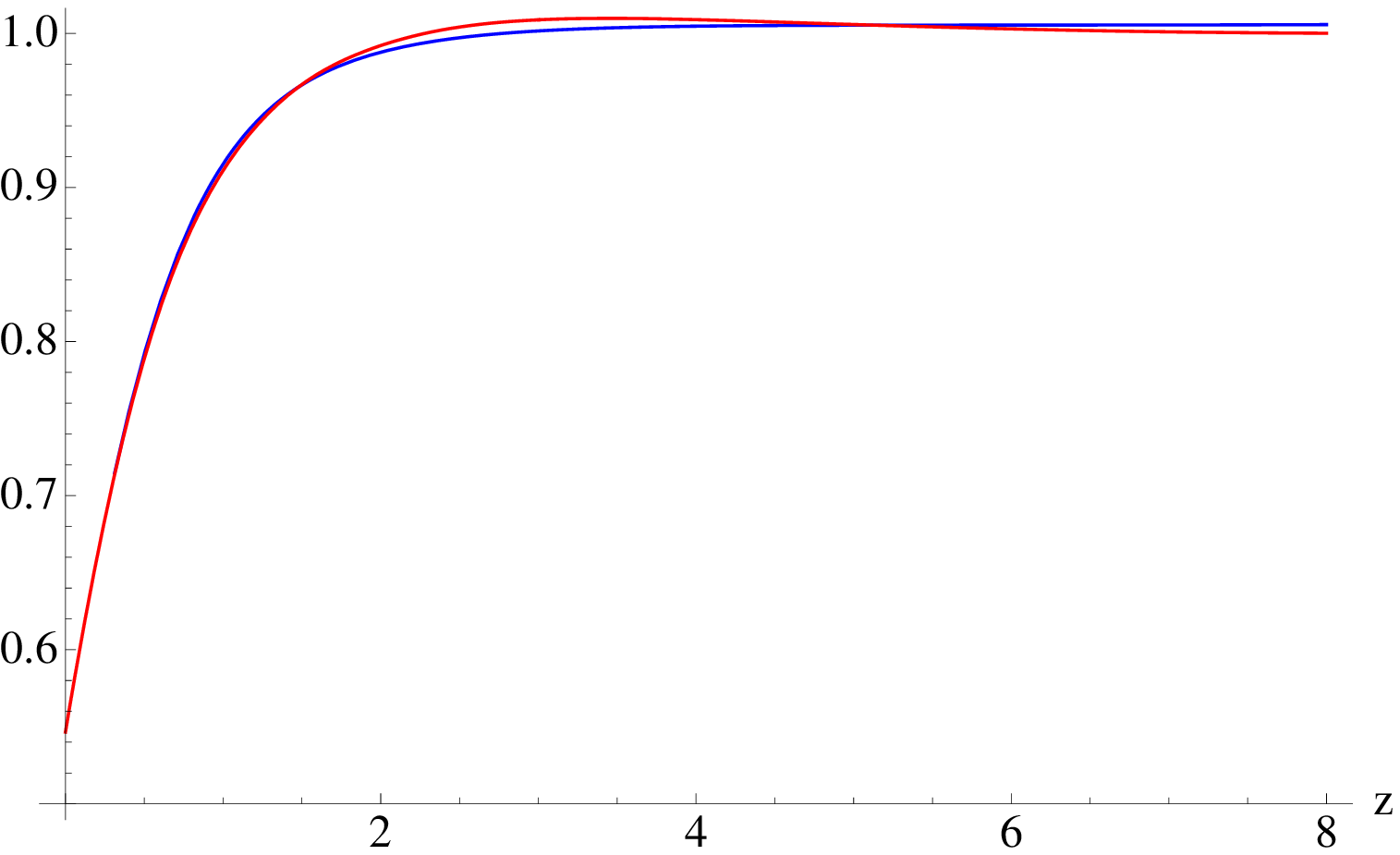}}
\quad
\subfigure[]{\includegraphics[width=0.3\textwidth]{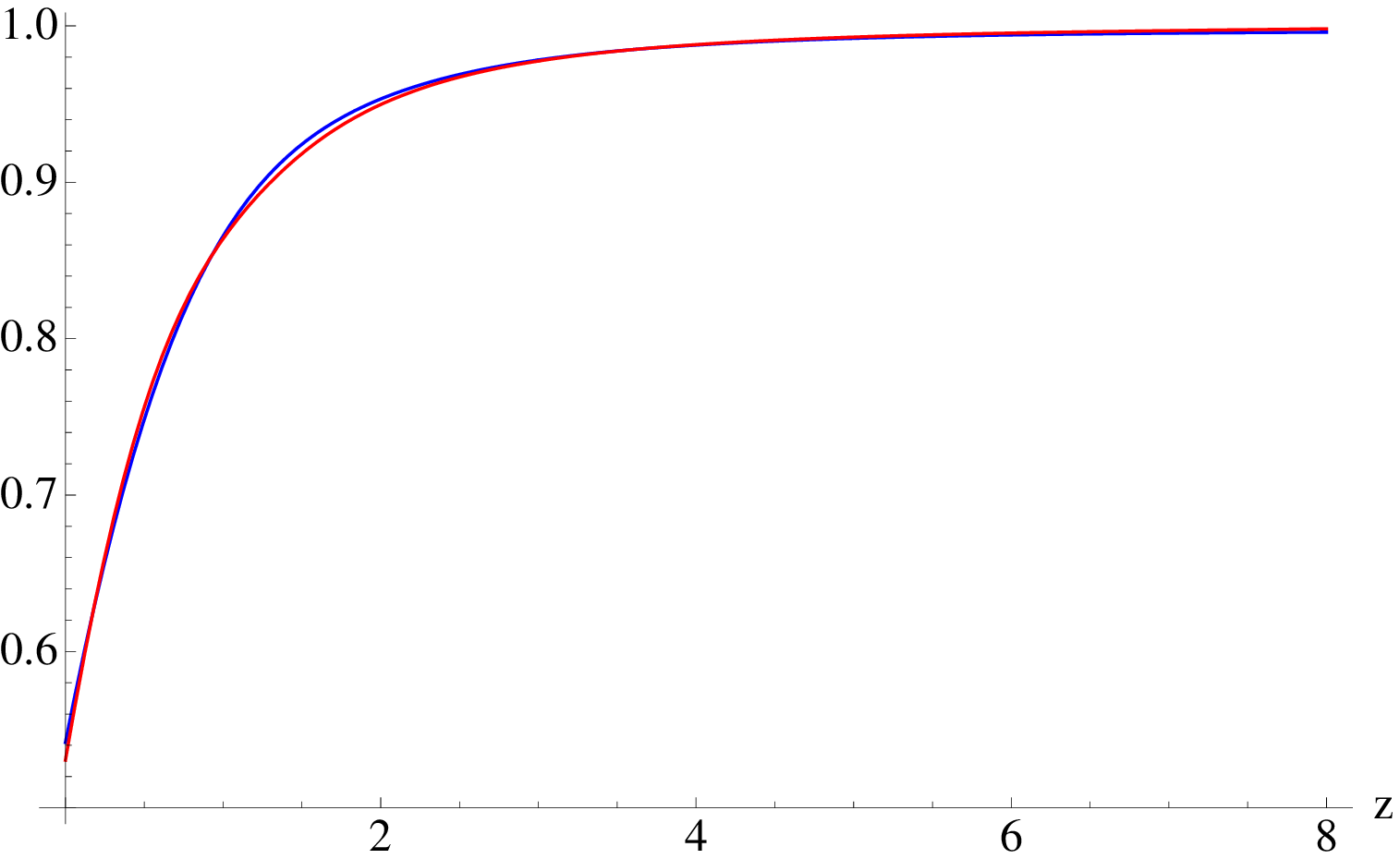}}
\quad
\subfigure[]{\includegraphics[width=0.3\textwidth]{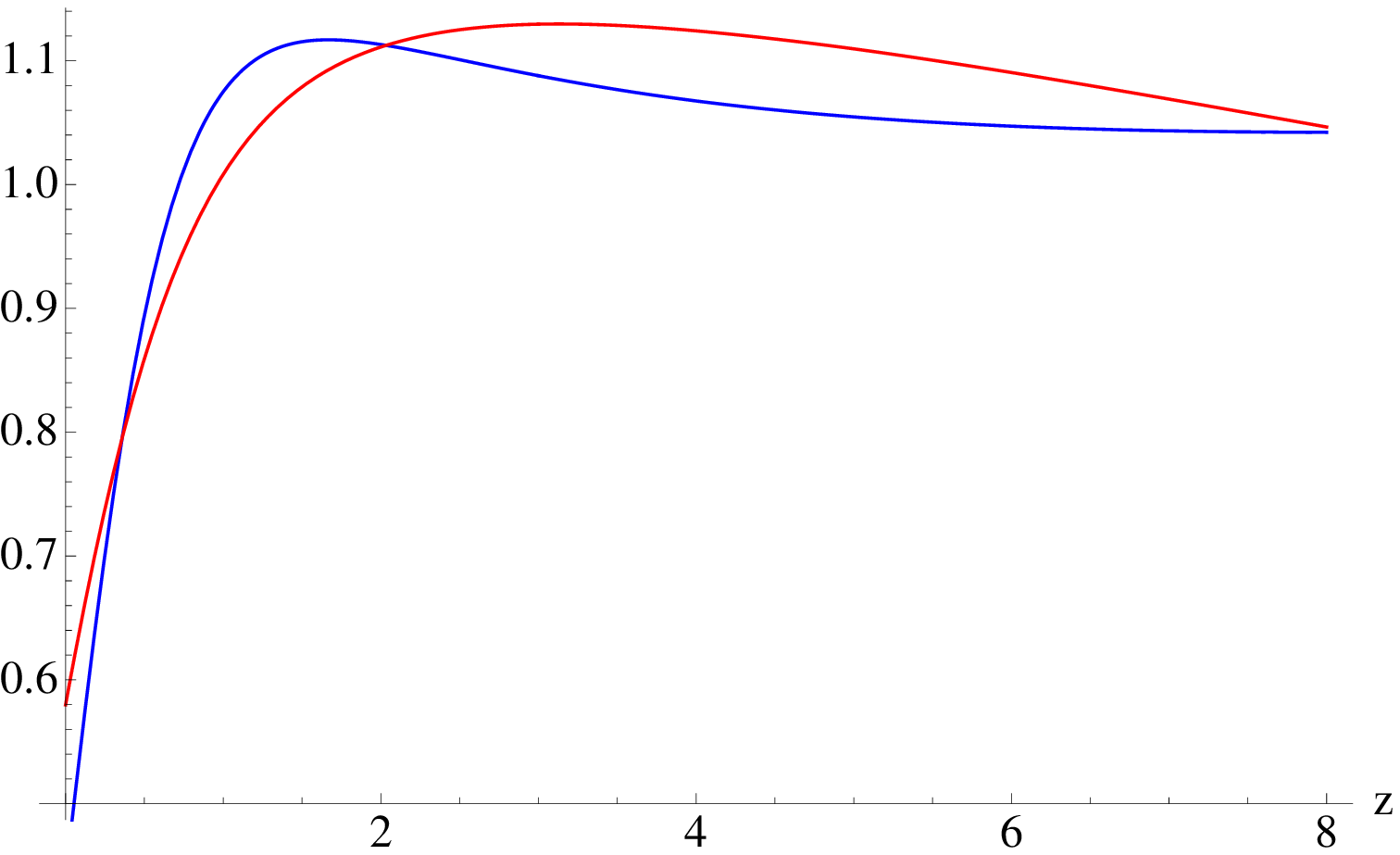}}
\quad
\subfigure[]{\includegraphics[width=0.3\textwidth]{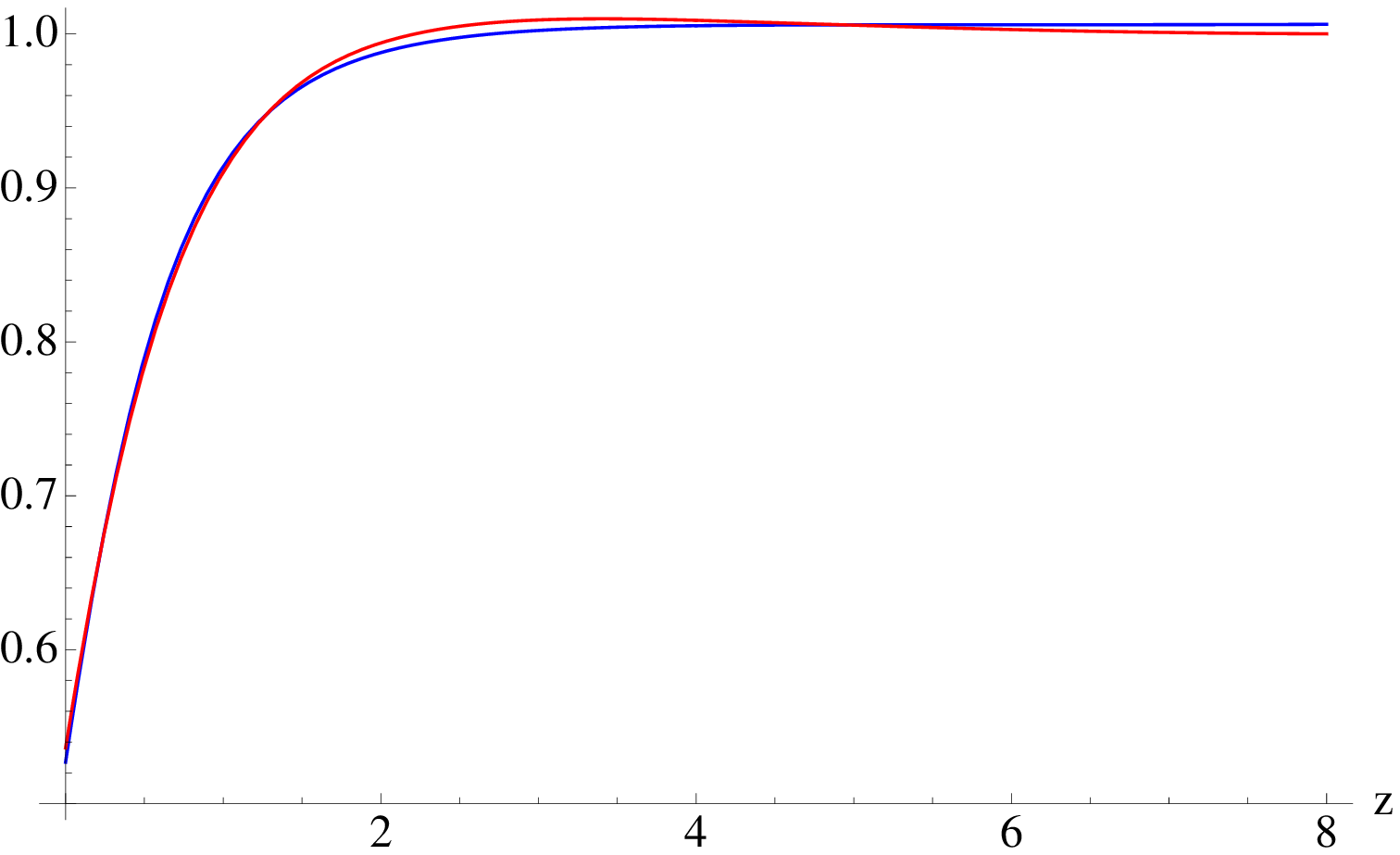}}
\quad
\subfigure[]{\includegraphics[width=0.3\textwidth]{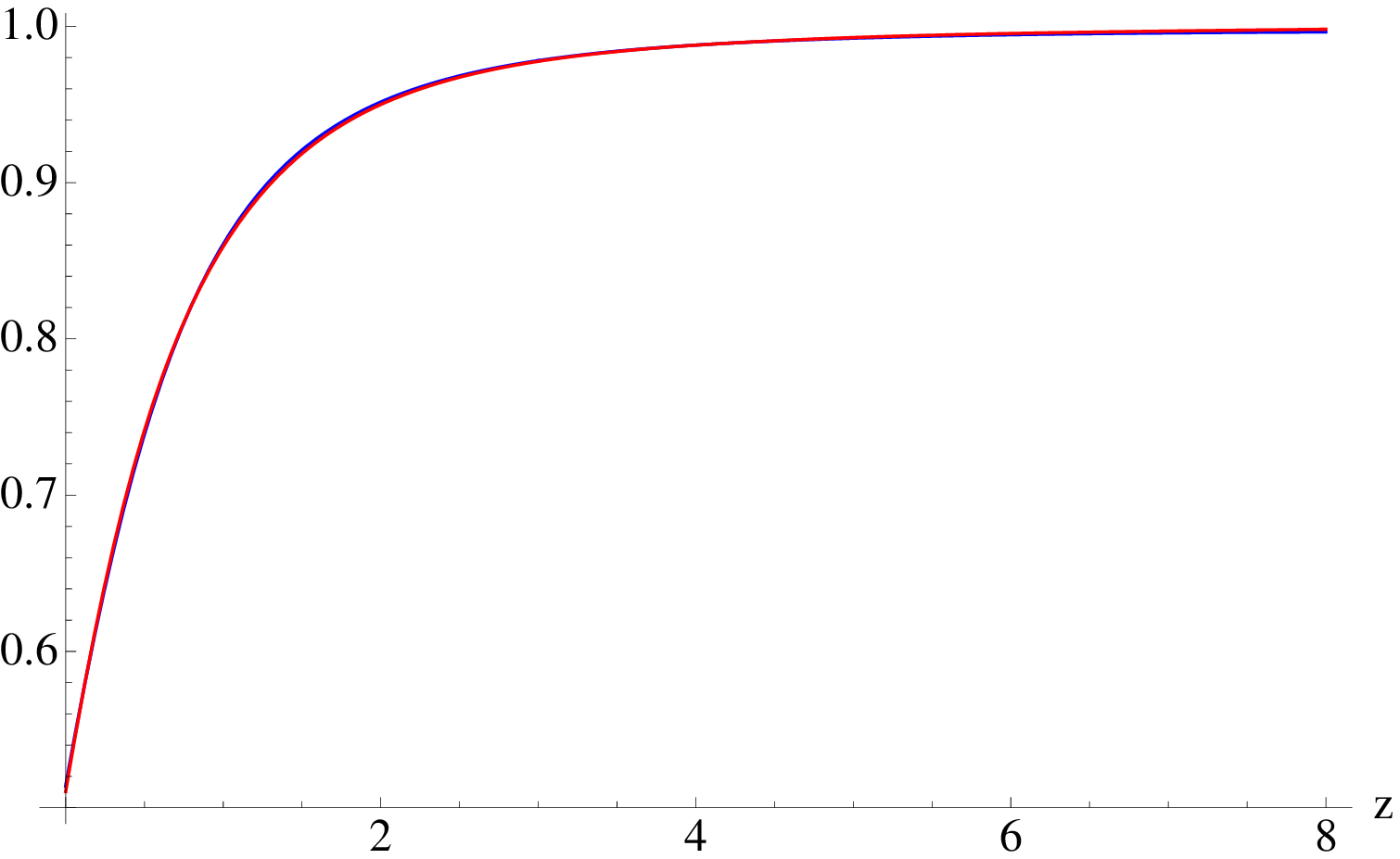}}
\caption{Cosmological evolutions of the 
growth rate $f_\mathrm{g}$ (red) and $\Omega_\mathrm{m}^\gamma$ (blue) with $\gamma = \gamma_0 + \gamma_1 z$ as functions of the redshift $z$ in the model $F_1(R)$ for $k = 0.1 \mathrm{Mpc}^{-1}$ (a), $k = 0.01 \mathrm{Mpc}^{-1}$ (b) and $k = 0.001 \mathrm{Mpc}^{-1}$ (c), and 
those in the model $F_2(R)$ for $k = 0.1 \mathrm{Mpc}^{-1}$ (d), $k = 0.01 \mathrm{Mpc}^{-1}$ (e) and $k = 0.001 \mathrm{Mpc}^{-1}$ (f).}
\label{HS_figure_lineal_growth_index}
\end{figure}

\begin{figure}[!h]
\subfigure[]{\includegraphics[width=0.45\textwidth]{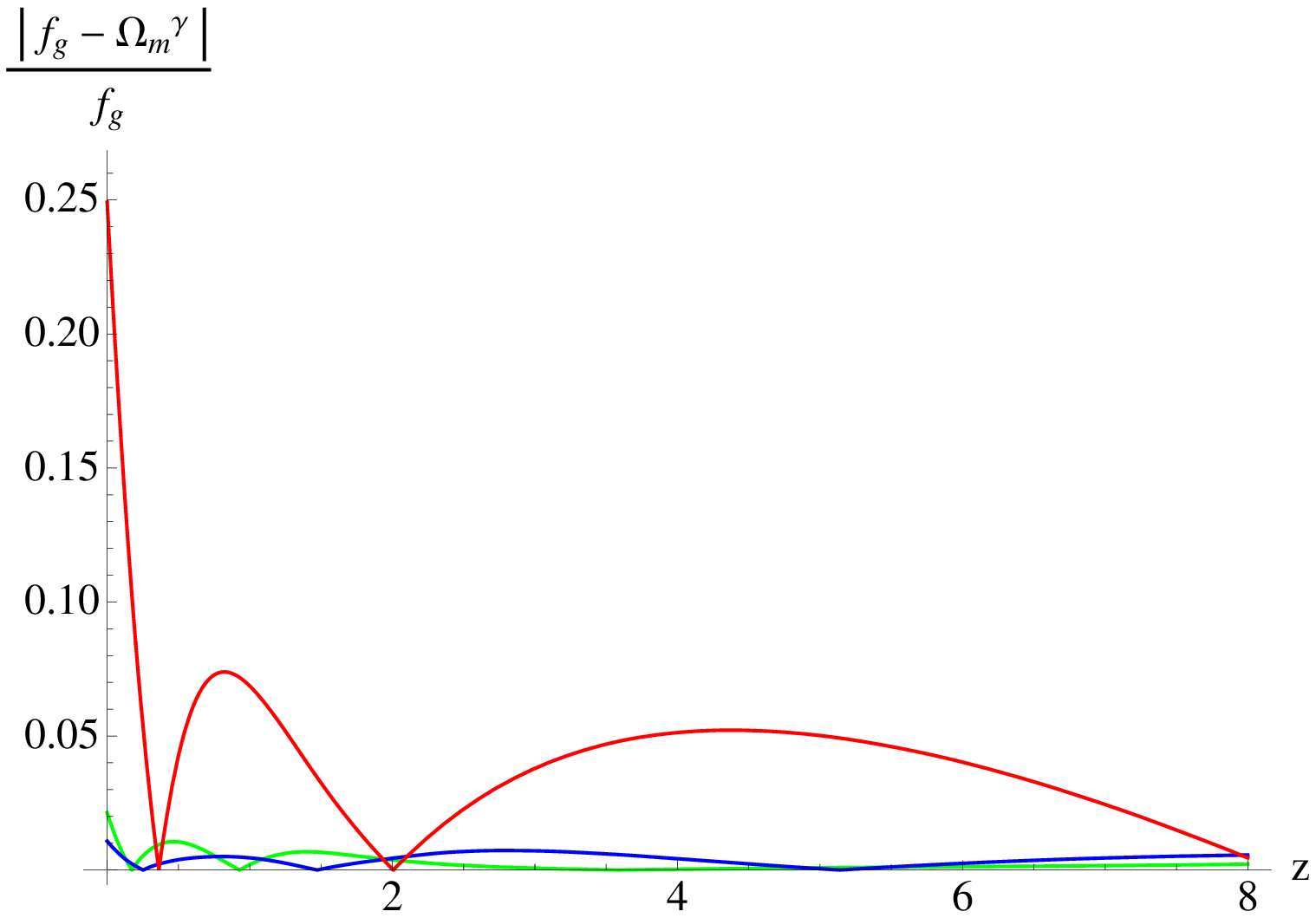}}
\quad
\subfigure[]{\includegraphics[width=0.45\textwidth]{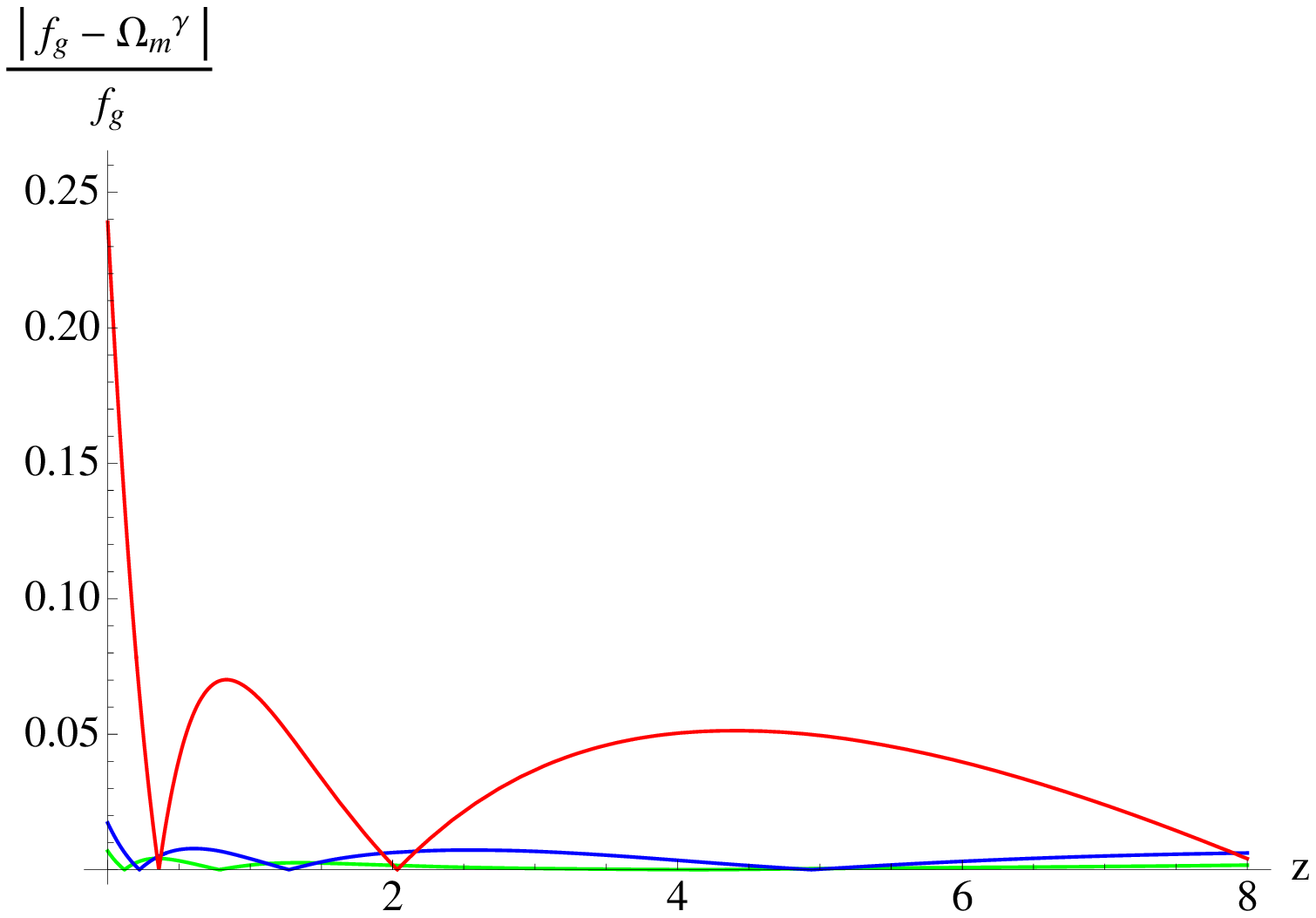}}
\caption{Cosmological evolution of the 
relative difference $\frac{\left| f_\mathrm{g} - \Omega_\mathrm{m}^\gamma \right|}{f_\mathrm{g}}$ with $\gamma = \gamma_0 + \gamma_1 z$ for $k = 0.1 \mathrm{Mpc}^{-1}$ (red), $k = 0.01 \mathrm{Mpc}^{-1}$ (blue) and $k = 0.001 \mathrm{Mpc}^{-1}$ (green) in the model $F_1(R)$ (a) and the model $F_2(R)$ (b).}
\label{lin_rel_dif}
\end{figure}

In Fig.~\ref{HS_figure_lineal_growth_index}, we illustrate cosmological evolutions of the growth rate $f_\mathrm{g}(z)$ and $\Omega_\mathrm{m}(z)^{\gamma(z)}$ as functions of the redshift $z$ together for 
the models $F_1(R)$ and $F_2(R)$. 
We can see that the fits for $\log k = 0.1$ have been improved in comparison 
with the same fits as the case with a constant growth index. 
Also, for $\log k < 0.1$ the fits continue to be quite good. 
In order to demonstrate these facts quantitatively, 
in Fig.~\ref{lin_rel_dif} we plot the cosmological evolution of 
the relative difference between $f_\mathrm{g}(z)$ and $\Omega_\mathrm{m}(z)^{\gamma(z)}$ as a function of $z$ 
for several values of $k$ in the models $F_1(R)$ and $F_2(R)$. 
In this case, for $\log k = -1$ the relative difference is smaller than $7.5\%$ in both the models if we do not consider lower values for $z$ (i.e., $z < 0.2$). We also see that the linear growth index improves the fits in both the models for $\log k = -2$ in comparison with those for a constant growth index. 
In this case, the relative difference for the model $F_1(R)$ is always smaller than $1\%$, whereas that for model $F_2(R)$ is smaller than $2\%$. 
Finally, for $\log k = -3$ the results obtained for a constant growth index are quite similar to those for a linear dependence on $z$. 


\subsubsection{$\gamma = \gamma_0 + \gamma_1 \frac{z}{1 + z}$}

Next, we examine the following ansatz for the growth index:
\begin{equation} 
\gamma = \gamma_0 + \gamma_1 \frac{z}{1 + z}\,. 
\end{equation}

\begin{figure}[!h]
\subfigure[]{\includegraphics[width=0.45\textwidth]{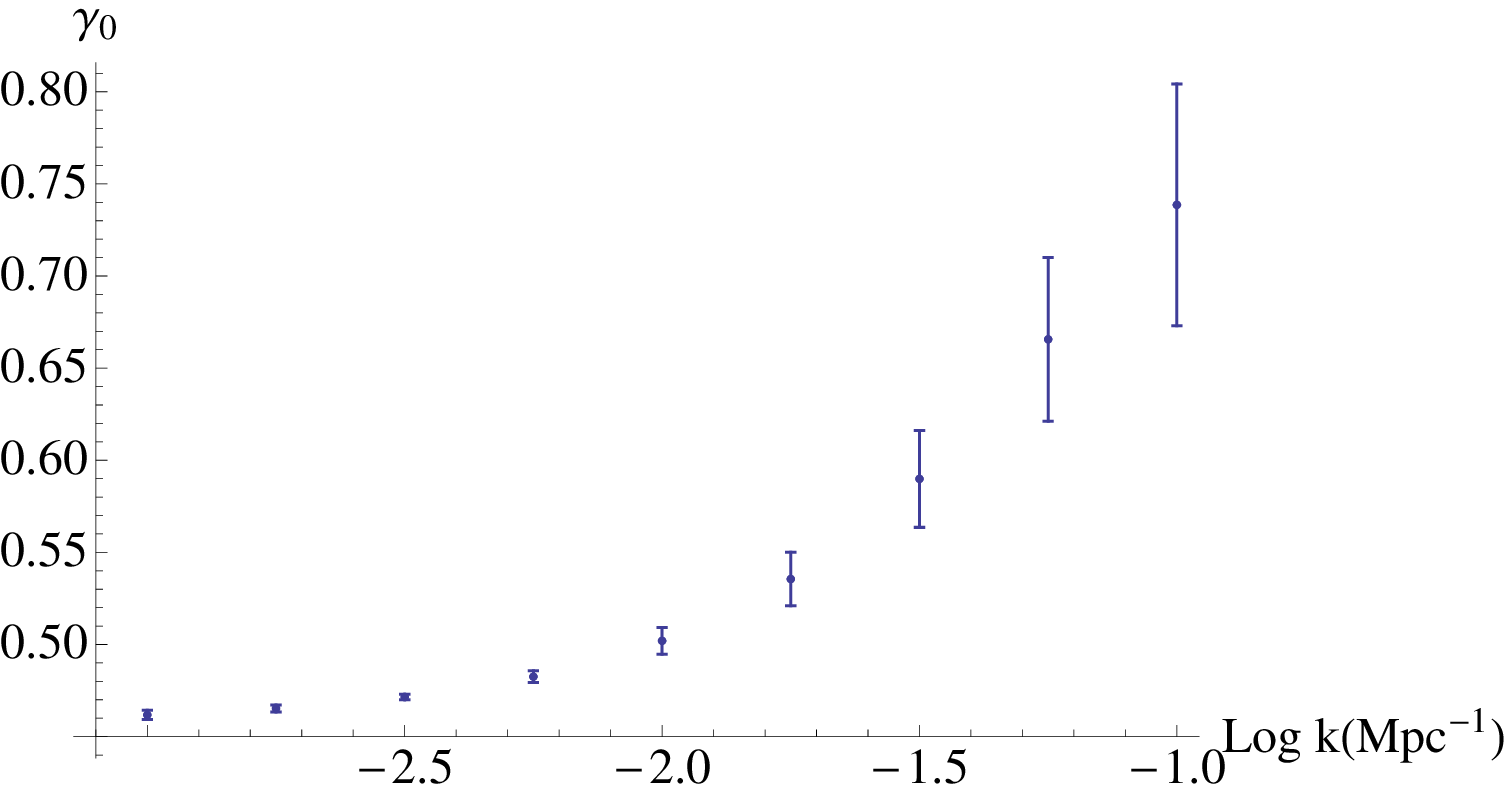}}
\quad
\subfigure[]{\includegraphics[width=0.45\textwidth]{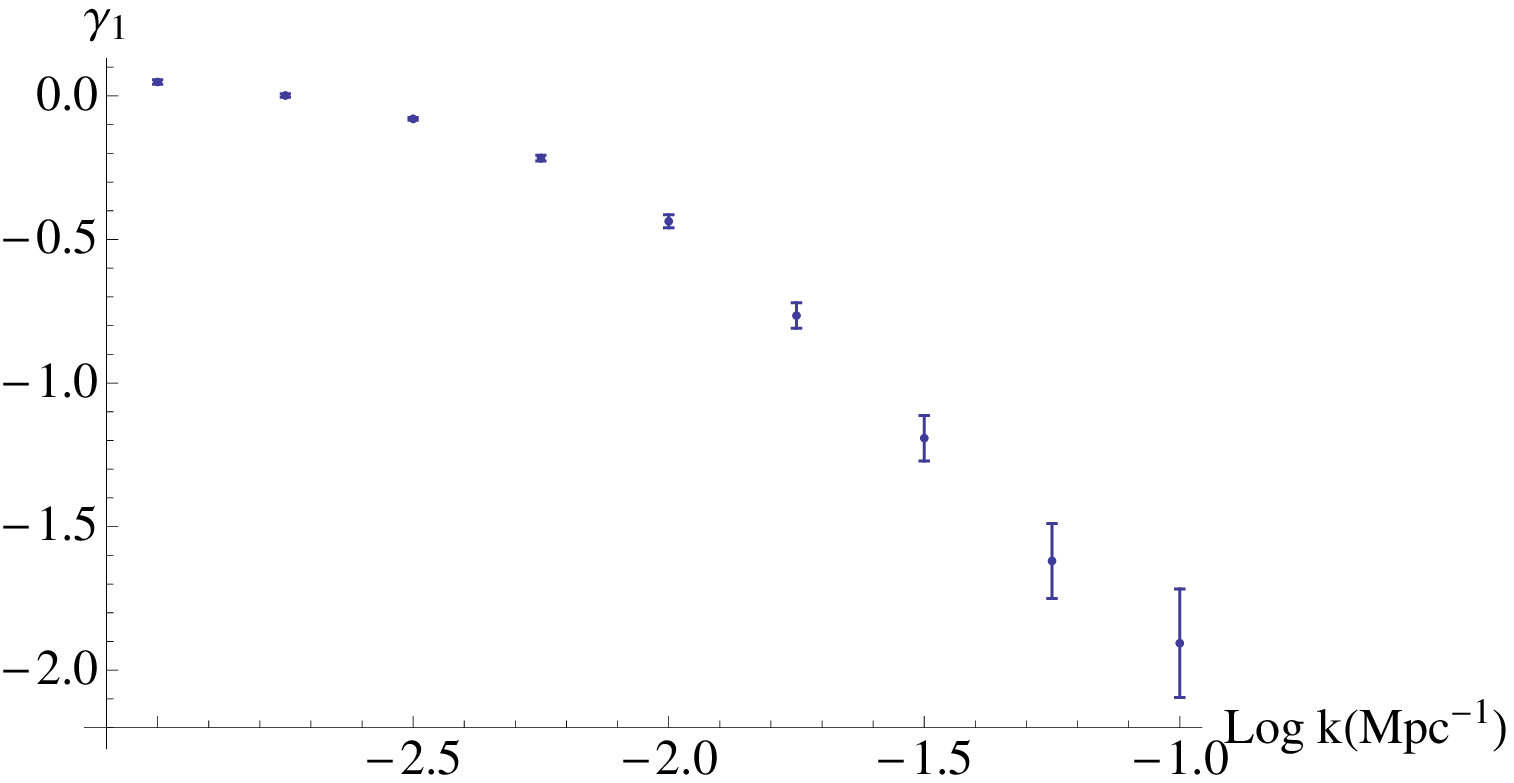}}
\subfigure[]{\includegraphics[width=0.45\textwidth]{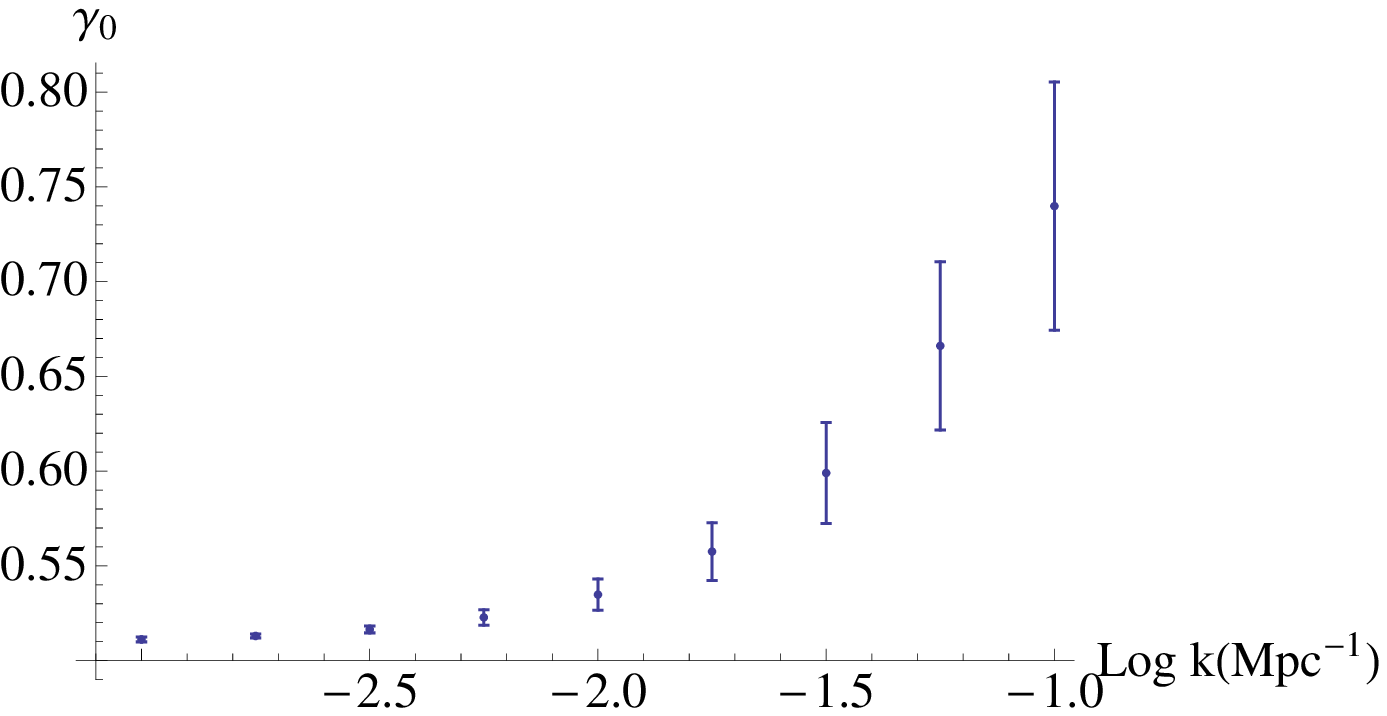}}
\quad
\subfigure[]{\includegraphics[width=0.45\textwidth]{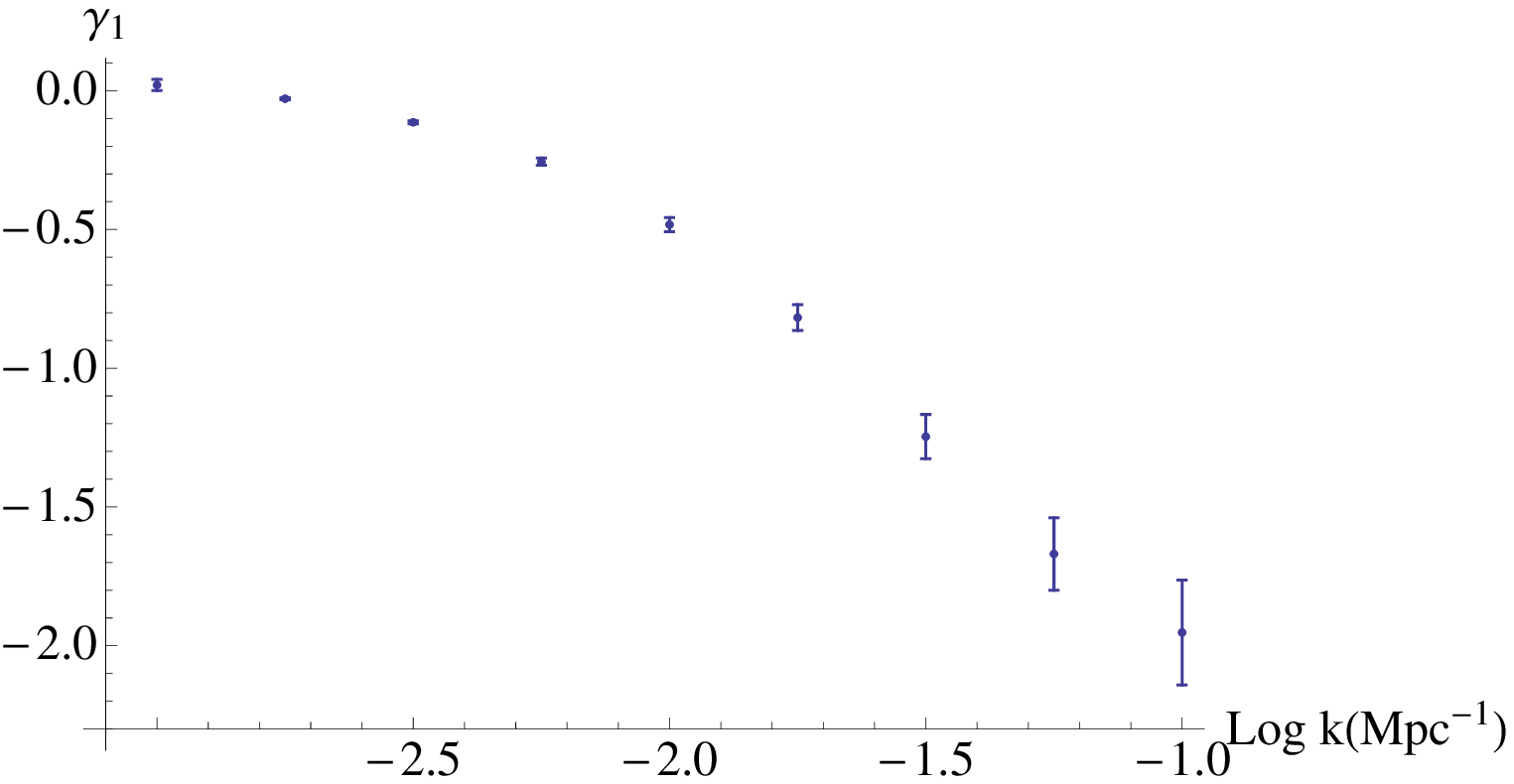}}
\caption{
Growth index fitting parameters in the case $\gamma = \gamma_0 + \gamma_1 \frac{z}{1 + z}$ as a function of $\log k$ for the model $F_1(R)$ [(a) and (b)] 
and the model $F_2(R)$ [(c) and (d)]. 
Legend is the same as Fig.~\ref{constant_growth_index_vs_logk}.
}
\label{HS_rational_growth_index_vs_logk}
\end{figure}

In Fig.~\ref{HS_rational_growth_index_vs_logk}, we depict 
the parameters $\gamma_0$ and $\gamma_1$ for several values of $\log k$ for both the models. 
The scale dependence of these parameters on $k$ is shown. 
The behavior of the parameter $\gamma_1$ seems to be quite similar to 
that for the previous case $\gamma = \gamma_0 + \gamma_1 z$, but it is worth cautioning that the scale of the figures are different from each other, and that for the present ansatz the scale dependence of $\gamma_1$ is stronger than that for the previous case. It can also be seen that $\gamma_0 \sim 0.465$ for the model $F_1(R)$ and $\gamma_0 \sim 0.513$ for the model $F_2(R)$ in the scale 
$\log k < -2.5$. 

\begin{figure}[!h]
\subfigure[]{\includegraphics[width=0.3\textwidth]{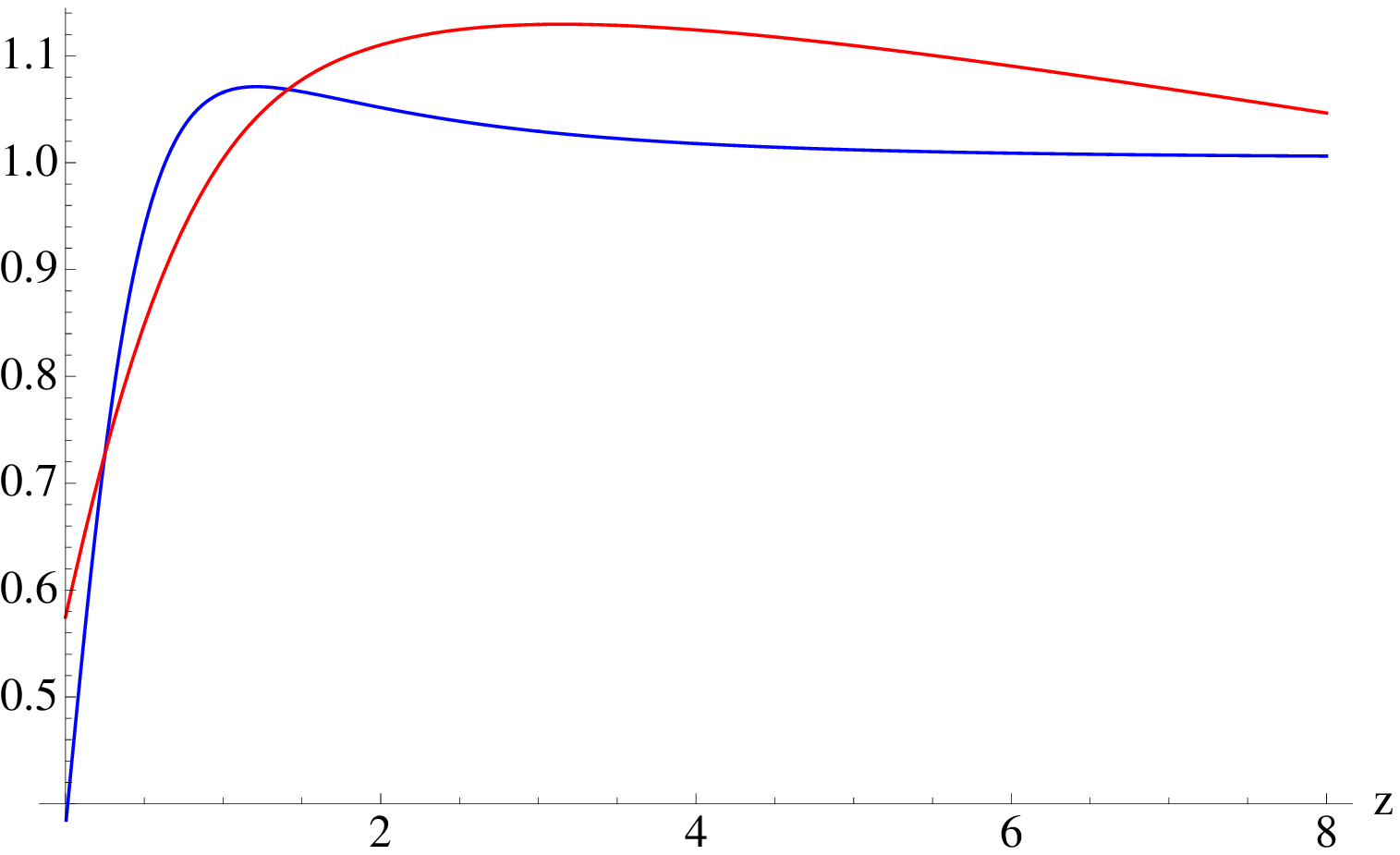}}
\quad
\subfigure[]{\includegraphics[width=0.3\textwidth]{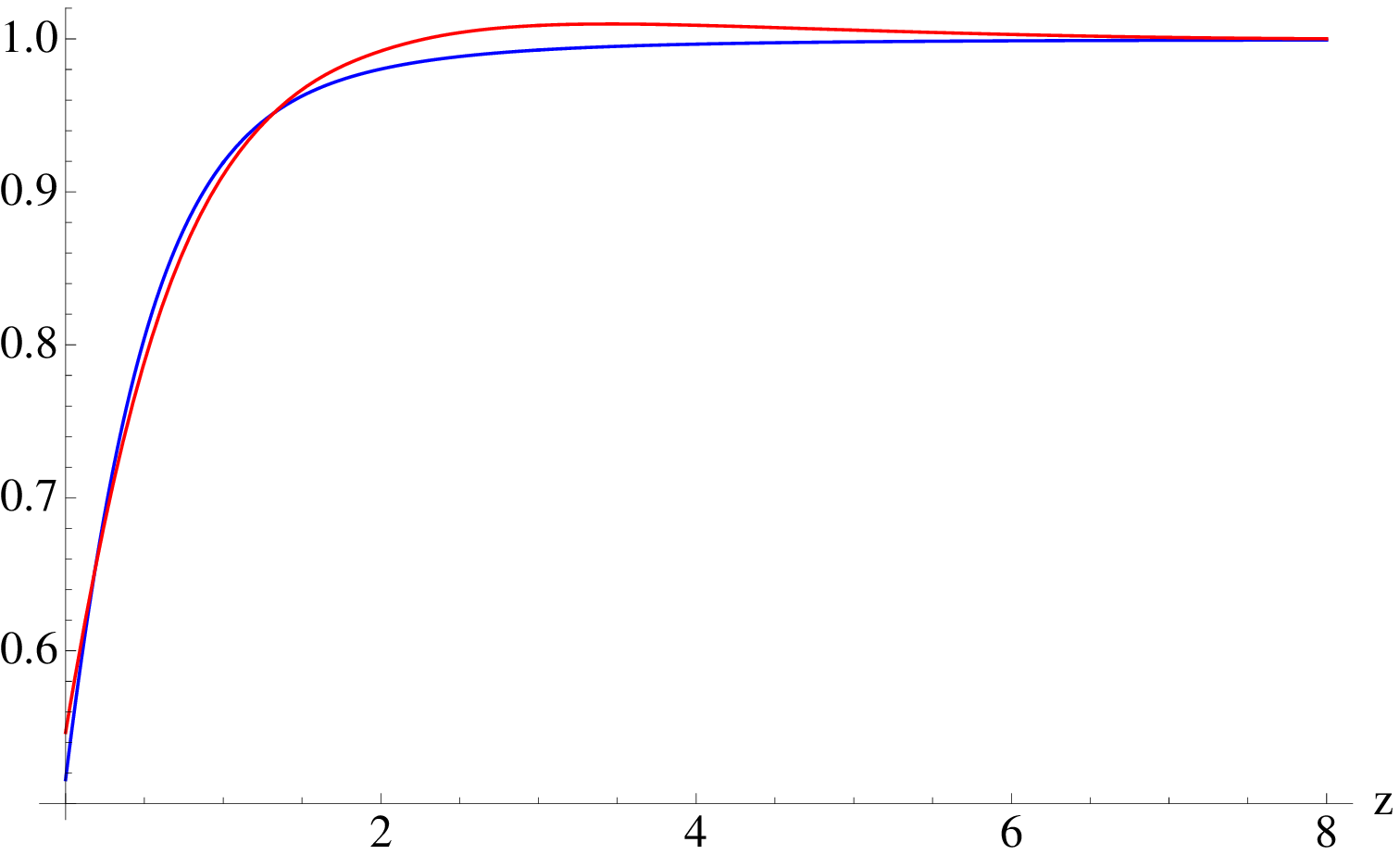}}
\quad
\subfigure[]{\includegraphics[width=0.3\textwidth]{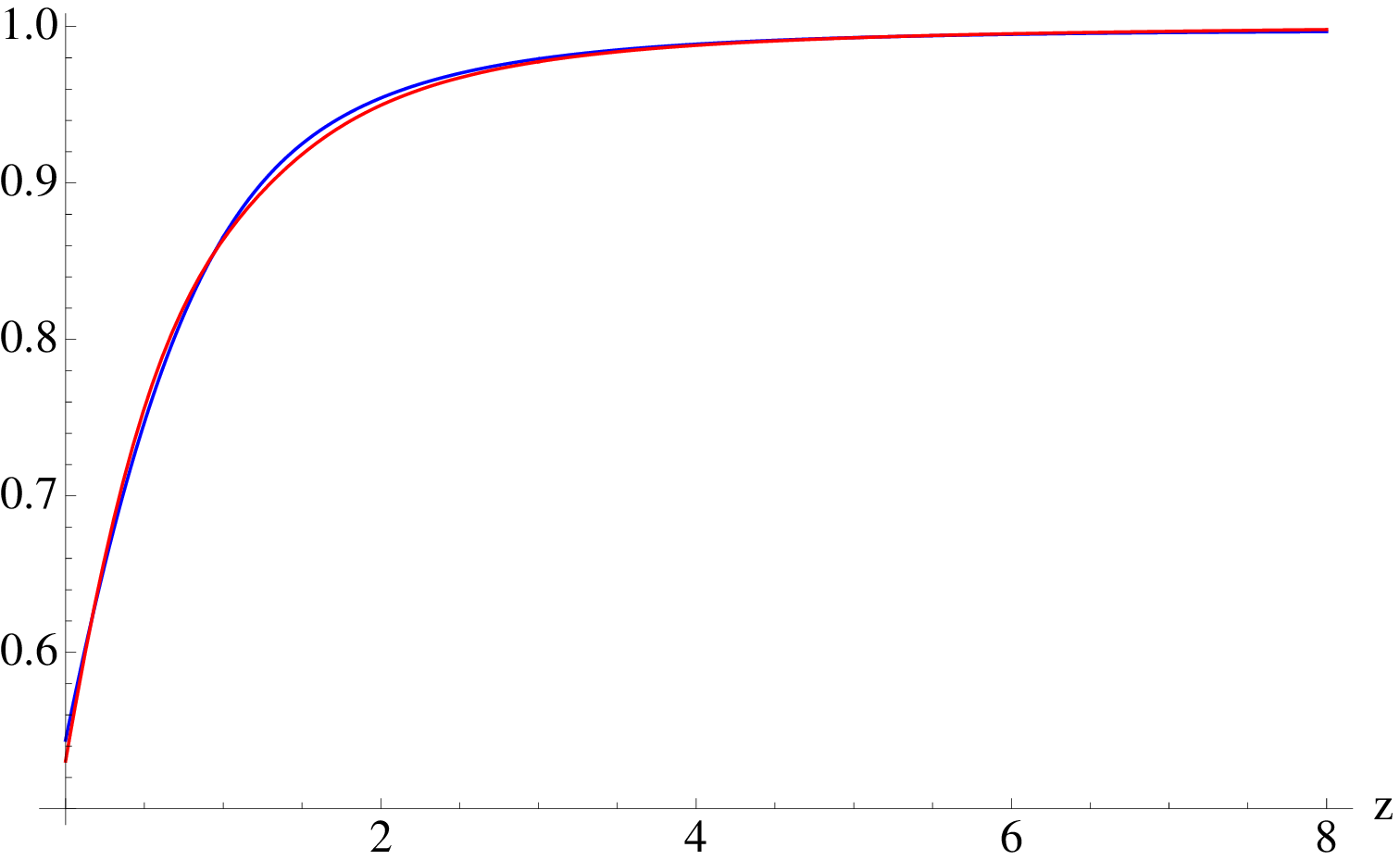}}
\quad
\subfigure[]{\includegraphics[width=0.3\textwidth]{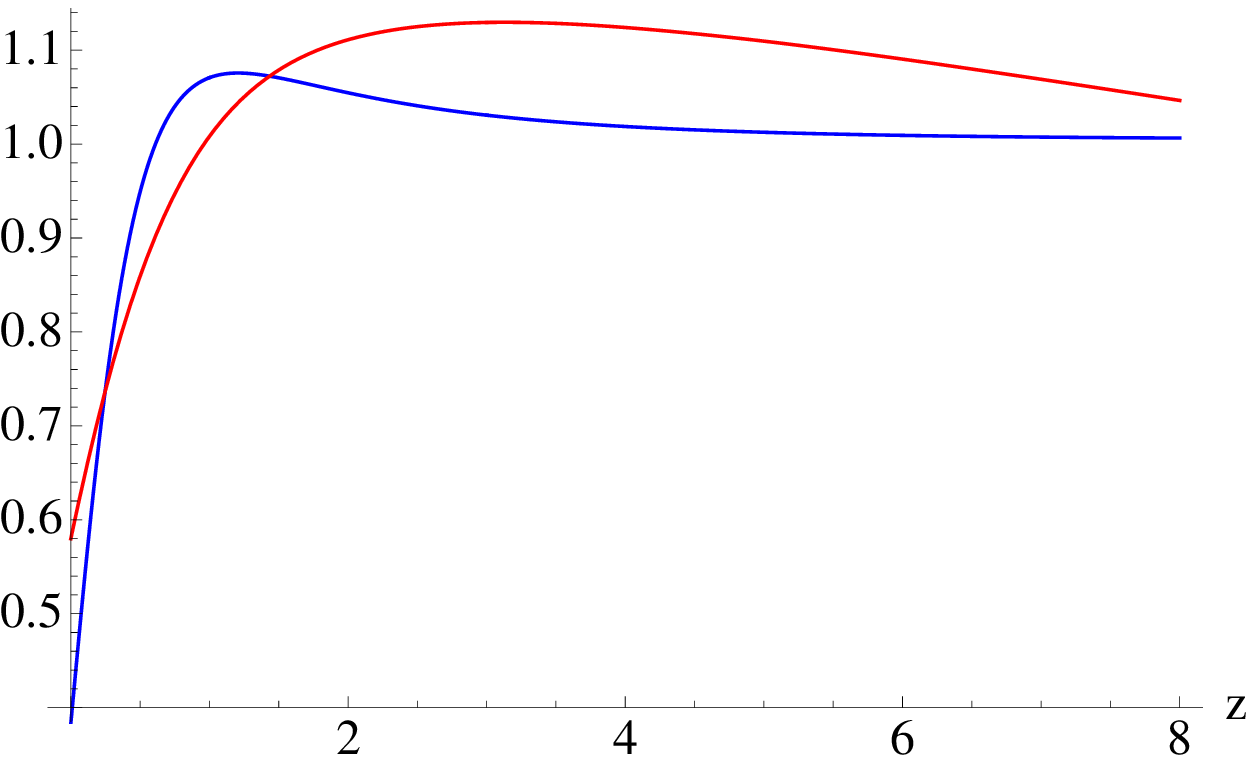}}
\quad
\subfigure[]{\includegraphics[width=0.3\textwidth]{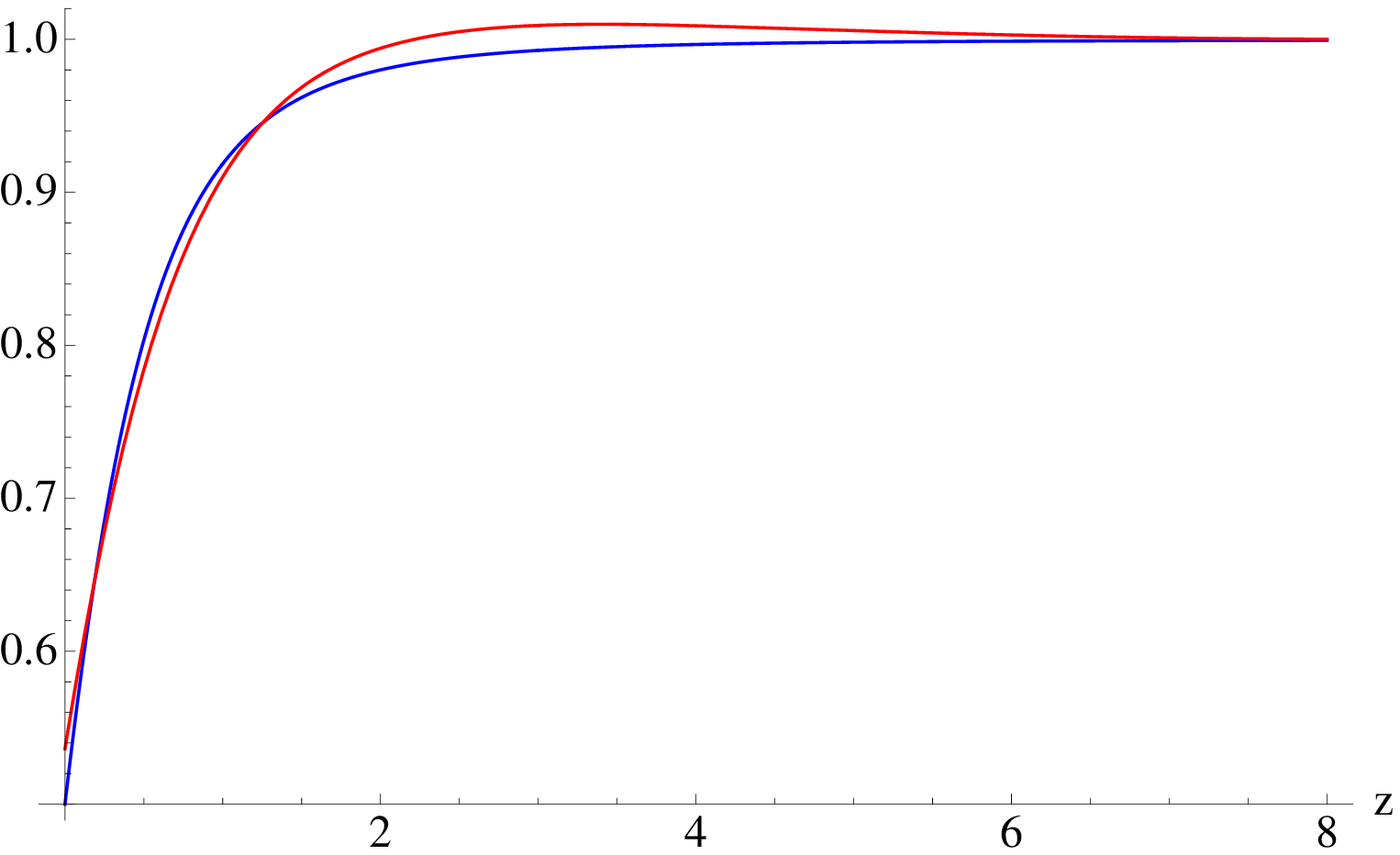}}
\quad
\subfigure[]{\includegraphics[width=0.3\textwidth]{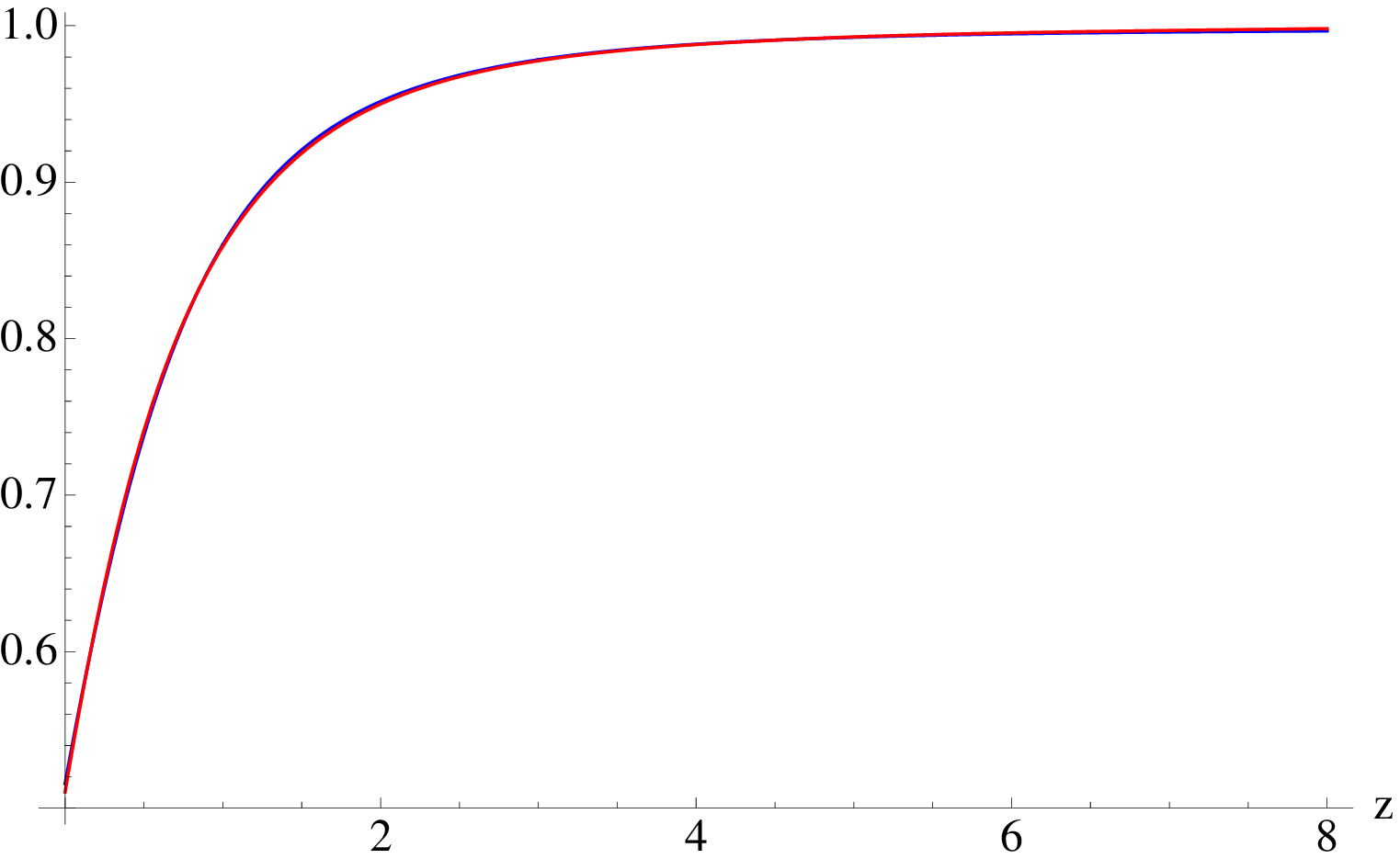}}
\caption{Cosmological evolutions of the 
growth rate $f_\mathrm{g}$ (red) and $\Omega_\mathrm{m}^\gamma$ (blue) with $\gamma = \gamma_0 + \gamma_1 \frac{z}{1 + z}$ as functions of the redshift $z$ in 
the model $F_1(R)$ for $k = 0.1 \mathrm{Mpc}^{-1}$ (a), $k = 0.01 \mathrm{Mpc}^{-1}$ (b) and $k = 0.001 \mathrm{Mpc}^{-1}$ (c), and 
those in the 
model $F_2(R)$ for $k = 0.1 \mathrm{Mpc}^{-1}$ (d), $k = 0.01 \mathrm{Mpc}^{-1}$ (e) and $k = 0.001 \mathrm{Mpc}^{-1}$ (f).}
\label{HS_figure_rational_growth_index}
\end{figure}

In Fig.~\ref{HS_figure_rational_growth_index}, we plot cosmological evolutions of the growth rate $f_\mathrm{g}(z)$ and $\Omega_\mathrm{m}(z)^{\gamma(z)}$ in the models $F_1(R)$ and $F_2(R)$ for several values of $k$, as demonstrated in the previous subsections. We can see the fits for $\log k \leq -2$ are quite good, as those in the previous ansatz for the growth index. 
In the case of higher values of $\log k$, it seems that the fits are similar to those for a constant growth rate 
and these fits do not reach the goodness of those for the case of $\gamma = \gamma_0 + \gamma_1 z$. 

\begin{figure}[!h]
\subfigure[]{\includegraphics[width=0.45\textwidth]{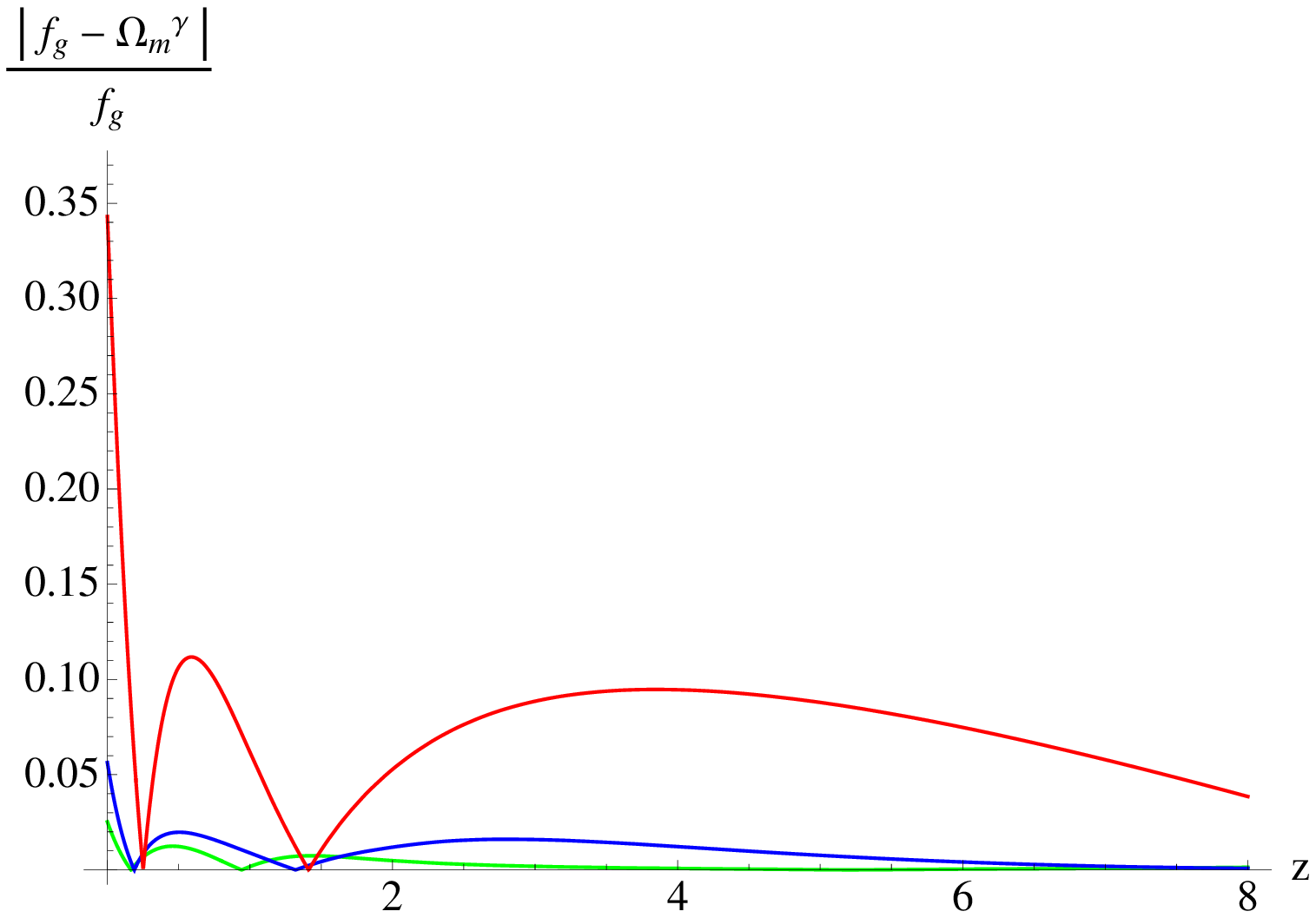}}
\quad
\subfigure[]{\includegraphics[width=0.45\textwidth]{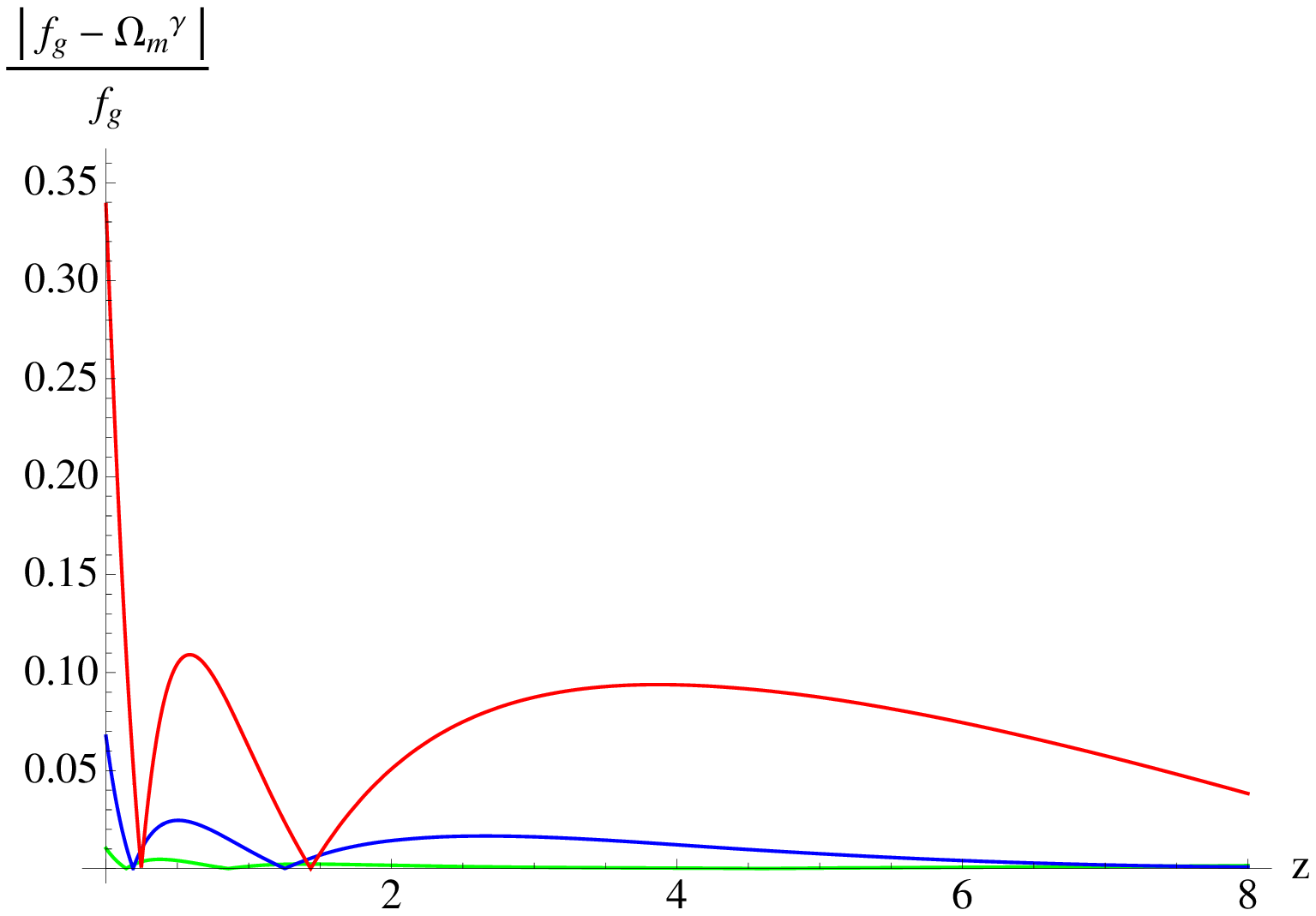}}
\caption{Cosmological evolution of the 
relative difference $\frac{\left| f_\mathrm{g} - \Omega_\mathrm{m}^\gamma \right|}{f_\mathrm{g}}$ with $\gamma = \gamma_0 + \gamma_1 \frac{z}{1 + z}$ for $k = 0.1 \mathrm{Mpc}^{-1}$ (red), $k = 0.01 \mathrm{Mpc}^{-1}$ (blue) and $k = 0.001 \mathrm{Mpc}^{-1}$ (green) in the model $F_1(R)$ (a) and the model $F_2(R)$ (b).}
\label{rat_rel_dif}
\end{figure}

In order to analyze the fits quantitatively, in Fig.~\ref{rat_rel_dif} 
we display the cosmological evolution of the relative difference between $f_\mathrm{g}(z)$ and $\Omega_\mathrm{m}(z)^{\gamma(z)}$ for several values of $k$ in the models $F_1(R)$ and $F_2(R)$. 
We see that the relative difference for $\log k = -1$ is smaller than $12\%$ 
(if we do not consider $z < 0.2$) for both the models. 
Thus, it is confirmed that these fits are better than those for the constant growth rate, but these are worse than those for $\gamma = \gamma_0 + \gamma_1 z$. 
For lower values of $\log k$, the relative difference is smaller than $2\%$ 
in $z > 0.2$. 

As a consequence, 
through the investigations of these different ansatz for the growth index, 
it is concluded that $\gamma = \gamma_0 + \gamma_1 z$ is the 
parameterization that can fit Eq.~(\ref{gi}) to the solution of 
Eq.~(\ref{gmp4}) better in a wide range of values for $k$. 
Even though the behavior of the parameters $\gamma_0$ and $\gamma_1$ in the 
models $F_1(R)$ and $F_2(R)$ is quite similar to each other, 
in order to distinguish between these models 
in 
Fig.~\ref{HS_lineal_growth_index_vs_logk}
we can see that 
the more differences between these models come from 
the values of $\gamma_0$ for $\log k \leq -2$. 
In fact, as remarked before, for $\log k \leq -2.5$ we have 
$\gamma_0 \sim 0.46$ for the model $F_1(R)$ and $\gamma_0 \sim 0.51$ 
for the model $F_2(R)$.

\section{Unified models for early and late-time cosmic acceleration\label{generalinflation}}

The reason why we also study inflation in $F(R)$ gravity is that 
one of the most important goals on the study of modified gravity theories 
is to describe the consistent evolution history of the universe from 
inflation in the early universe to the dark energy dominated stage at 
the present time. 
Namely, 
the universe starts with an inflationary epoch, 
followed by the radiation dominated era and 
the matter dominated universe, and finally 
the late cosmic acceleration epoch is actually achieved 
without invoking the presence of dark components in the universe~\cite{Review-Nojiri-Odintsov}
(for a first $F(R)$ theory unifying inflation with dark energy, 
see Ref.~\cite{DEinflation}). 
In Secs.~III and IV, it has been demonstrated that the exponential gravity with the correction terms can be a realistic $F(R)$ gravity model. 
Therefore, in this section we investigate the possibility that in 
such exponential gravity with additional correction terms, 
inflation as well as the late-time cosmic acceleration can be 
realized. 
Since we examine exponential gravity among various models of $F(R)$ gravity, 
we consider the unification model between inflation and the late-time cosmic acceleration in this work. 

Models of the type~(\ref{model}) may be combined in a natural way 
to obtain the phenomenological description of the inflationary epoch. 
For example, a `two-steps' model may be the smooth version, given by 
%
\begin{equation}
F(R)=R-2\Lambda \left[1-\mathrm{e}^{-R/(b\,R)}\right]-\Lambda_{\rm i} \,\theta(R-R_\mathrm{i})\,.
\label{toymodel2}
\end{equation}
%
Here, $\theta(R-R_0)$ is the Heaviside's step distribution, 
$R_\mathrm{i}$ is the transition scalar curvature at inflationary scale and 
$\Lambda_{\rm i}$ is a suitable cosmological constant 
producing inflation, when $R\gg R_\mathrm{i}$. 
The main problem associated with this sharp model is the appearance of a 
possible antigravity regime in a region around the transition point between 
inflation and the universe described by the $\Lambda$CDM model. 
The antigravity in a past epoch is not phenomenologically acceptable. 
Furthermore, adding some terms would be necessary in order for 
inflation to end. 

In this section, we study two applications of exponential gravity to achieve an unified description of the early-time inflation and the late-time cosmic acceleration. In particular, we show how it is possible to obtain 
inflationary universes with different numbers of $e$-folds 
by choosing different models parameters in the presence of 
ultrarelativistic matter in the early universe. 

Following the first proposal of Ref.~\cite{twostep}, 
we start with the form of $F(R)$ with a natural possibility of 
a unified description of our universe 
\begin{equation}
F(R)=R-2\Lambda\left(1-\mathrm{e}^{-\frac{R}{b R}}\right)
-\Lambda_\mathrm{i}\left[1-\mathrm{e}^{-\left(\frac{R}{R_\mathrm{i}}\right)^n}\right]
+\bar{\gamma} \left(\frac{1}{\tilde R_\mathrm{i}^{\alpha-1}}\right)R^\alpha\,, 
\label{total}
\end{equation}
where $R_\mathrm{i}$ and $\Lambda_\mathrm{i}$ are the typical values of 
transition curvature and expected cosmological constant during inflation, 
respectively, and $n$ is a natural number larger than unity 
(here, we do not write the correction term for the stability of 
oscillations in the matter dominated era). 
In Eq.~(\ref{total}), 
the last term 
$\bar{\gamma}(1/\tilde R_\mathrm{i}^{\alpha-1}) R^\alpha$, 
where $\bar{\gamma}$ 
is a positive dimensional constant and $\alpha$ is a real number, 
works at the inflation scale $\tilde R_\mathrm{i}$ and 
is actually necessary in order to realize an exit from inflation. 

We also propose another nice inflation model based on the good behavior of exponential function described as 
\begin{equation}
F(R)=R-2\Lambda\left(1-\mathrm{e}^{-\frac{R}{b R}}\right)-\Lambda_\mathrm{i}\frac{\sin\left(\pi\,\mathrm{e}^{-\left(\frac{R}{R_\mathrm{i}}\right)^n}\right)}{\pi\,\mathrm{e}^{-\left(\frac{R}{R_\mathrm{i}}\right)^n}}+\bar{\gamma} \left(\frac{1}{\tilde R_\mathrm{i}^{\alpha-1}}\right)R^\alpha\,.
\label{sin}
\end{equation}
Here, the parameters have the same roles of the corresponding ones 
in the model in Eq.~(\ref{total}). 
We note that the second term of the model vanishes when $R\ll R_i$ and 
tends to $\Lambda_\mathrm{i}$ when $R\gg R_\mathrm{i}$. 
We analyze these models, i.e., Model I in Eq.~(\ref{total}) and Model II 
in Eq.~(\ref{sin}), and explore the possibilities to reproduce 
the phenomenologically acceptable inflation.

\subsection{Inflation in exponential model (Model I)\label{Inflation}}

First, we investigate the model in Eq.~(\ref{total}). 
For simplicity, we describe a part of it as 
\begin{equation}
f_\mathrm{i}(R) \equiv 
-\Lambda_\mathrm{i}\left(1-\mathrm{e}^{-\left(\frac{R}{R_\mathrm{i}}\right)^n}\right)+\bar{\gamma}\left(\frac{1}{\tilde R_\mathrm{i}^{\alpha-1}}\right)R^\alpha\,.
\end{equation}
We note that 
if $n>1$ and $\alpha>1$, 
when $R\ll R_{\mathrm i}(\sim\tilde R_\mathrm{i})$, we obtain 
\begin{equation}
R \gg \left|f_\mathrm{i}(R)\right| \simeq \left|-\frac{R^{n}}{R_\mathrm{i}^{n-1}}+\bar{\gamma}\frac{R^\alpha}{\tilde R_\mathrm{i}^{\alpha-1}}\right|\,,
\end{equation}
and the absence of the effects of inflation during the matter dominated era. 
We also find 
\begin{eqnarray}
f_\mathrm{i}'(R) \Eqn{=} -\frac{\Lambda_\mathrm{i} n
R^{n-1}}{R_\mathrm{i}^n}\mathrm{e}^{-\left(\frac{R}{R_\mathrm{i}}\right)^n}+\bar{\gamma}\alpha\,\left(\frac{R}{\tilde R_\mathrm{i}}\right)^{\alpha-1}\,, \\
f_\mathrm{i}''(R) \Eqn{=} -\frac{\Lambda_\mathrm{i} 
n(n-1)R^{n-2}}{R_\mathrm{i}^n}\mathrm{e}^{-\left(\frac{R}{R_\mathrm{i}}\right)^n}
+\Lambda_i\left(\frac{n R^{n-1}}{R_\mathrm{i}^n}\right)^2
\mathrm{e}^{-\left(\frac{R}{R_\mathrm{i}}\right)^n}+\bar{\gamma}\alpha(\alpha-1)\frac{R^{\alpha-2}}{\tilde R_\mathrm{i}^{\alpha-1}}\,.
\label{secondderivative}
\end{eqnarray}
Since when $R=R_\mathrm{i}\left[(n-1)/n\right]^{1/n}$ 
the negative term of $f_\mathrm{i}'(R)$ has a minimum, 
in order to avoid the anti-gravity effects 
(this means, $|f'_\mathrm{i}(R)|<1$), it is sufficient to require 
\begin{equation}
R_\mathrm{i}>\Lambda_\mathrm{i} n
\left(\frac{n-1}{n}\right)^{\frac{n-1}{n}}\mathrm{e}^{-\frac{n-1}{n}}\,.
\label{uno}
\end{equation}
It is necessary for the modification of gravity describing inflation 
not to have any influence on the stability of the matter dominated era 
in the small curvature limit. 
When $R \ll R_\mathrm{i}$, the second derivative of $f_\mathrm{i}''(R)$, 
given by 
\begin{equation}
f''_\mathrm{i}(R)\simeq
\frac{1}{R}\left[-n(n-1)\left(\frac{R}{R_\mathrm{i}}\right)^{n-1}
+\bar{\gamma} \alpha(\alpha-1)\left(\frac{R}{\tilde R_\mathrm{i}}\right)^{\alpha-1}\right]\,,
\label{zwei}
\end{equation}
must be positive, that is, 
\begin{equation}
n>\alpha\,.
\end{equation}
We require the existence of the de Sitter critical point $R_{\mathrm{dS}}$ 
which describes inflation in the high-curvature regime of $f_\mathrm{i}(R)$, 
so that 
$f_\mathrm{i}(R_{\mathrm{dS}} \gg R_\mathrm{i})\simeq -\Lambda_\mathrm{i}+\bar{\gamma}(1/R_\mathrm{i}^{\alpha-1})\,R^{\alpha}$. 
In this case, if we put $\tilde R_\mathrm{i}=R_{\mathrm{dS}}$, 
we may solve the trace of the field equation (\ref{Field equation}) in vacuum for a constant curvature, namely $2F(R)-R F' (0)=0$, and therefore we obtain
(in vacuum, namely, if the effective modified gravity energy density is dominant over matter), 
\begin{equation}
R_{\mathrm{dS}}=\frac{2\Lambda_\mathrm{i}}{\bar{\gamma}(2-\alpha)+1}\,,
\quad
\left(\frac{R_{\mathrm{dS}}}{R_\mathrm{i}}\right)^n \gg 1\,.
\label{due}
\end{equation}
The last two conditions have to be satisfied simultaneously. 
By using Eq.~(\ref{uno}), we also acquire 
\begin{equation}
\frac{2}{\bar{\gamma}(2-\alpha)+1}>n
\left(\frac{n-1}{n}\right)^{\frac{n-1}{n}}\mathrm{e}^{-\frac{n-1}{n}}\,.
\label{unouno}
\end{equation}
%

\subsubsection*{Instability and number of $e$-folds during inflation}

The well-known condition to have an instable de Sitter solution 
(see Sec.~{\ref{dS}}) is given by 
\begin{equation}
\frac{F'(R_\mathrm{dS})}{R_\mathrm{dS}\,F''(R_\mathrm{dS})}<1\,, 
\end{equation}
which leads to
\begin{equation}
\alpha\,\bar{\gamma}(\alpha-2)>1\,, 
\label{duedue}
\end{equation}
for our model. 
Here, we have considered $f_i(R_{\mathrm{dS}})\simeq -\Lambda_\mathrm{i}+\bar{\gamma}(1/R_\mathrm{i}^{\alpha-1})\,R^{\alpha}$. {}From Eqs.~(\ref{unouno})--(\ref{duedue}), we have to require 
\begin{equation}
2+1/\bar{\gamma}>\alpha>2\,.
\end{equation} 
Thus, we may evaluate the characteristic number of $e$-folds during inflation 
\begin{equation}
N=\log \frac{z_\mathrm{i}+1}{z_\mathrm{e}+1}\,,
\end{equation}
where $z_\mathrm{i}$ and $z_\mathrm{e}$ are the redshifts at the beginning and at the end of early time cosmic acceleration. 
Given a small cosmological perturbation $y_1(z_\mathrm{i})$ at the redshift 
$z_\mathrm{i}$, we have from Eq.~(\ref{result}) avoiding 
the matter contribution 
\begin{equation}
y_1(z_\mathrm{i})=C_0(z_\mathrm{i}+1)^x\,, 
\label{pertinfl}
\end{equation}
with
\begin{equation}
x=\frac{1}{2}\left(3-\sqrt{25-\frac{16F'(R_{\mathrm{dS}})}{R_{\mathrm{dS}}F''(\mathrm{R_{dS}})}}\right)\,,
\label{x}
\end{equation}
where $x<0$ if the de Sitter point is unstable. Thus, the perturbation $y_1(z)$ in Eq.~(\ref{result}) grows up in expanding universe as
\begin{equation}
y_1(z)=y_1(z_\mathrm{i})\left[\frac{(z+1)}{(z_\mathrm{i}+1)}\right]^x\,.
\end{equation}
Here, we have considered $C_0=y_1(z_\mathrm{i})/(z_\mathrm{i}+1)^x$. 
When $y_1(z)$ is on the same order of the effective modified gravity energy density $y_0$ of 
the de Sitter solution describing inflation 
(we remind, $y_0=R_{\mathrm{dS}}/(12\tilde{m}^2)$), the model exits from 
inflation. We can estimate the number of $e$-folds during inflation 
as
\begin{equation}
N\simeq \frac{1}{x}\log\left(\frac{y_1(z_\mathrm{i})}{y_0}\right)\,.
\label{Nfolding}
\end{equation}
A value demanded in most inflationary scenarios is at least 
$N = 50$--$60$.

A classical perturbation on the (vacuum) de Sitter solution may be given by 
the presence of ultrarelativistic matter in the early universe. 
The system gives rise to the de Sitter solution where the universe expands in 
an accelerating way but, suddenly, it exits from inflation and tends towards 
the minimal attractor at $R = 0$ (the trivial de Sitter point). 
In this way, the small curvature regime arises and the physics of the $\Lambda$CDM model is reproduced.

\subsection{Inflation in Model II\label{inflation2}}

Next, we study the inflation model in Eq.~(\ref{sin}). 
By performing a similar analysis to that in the previous subsection, 
we find that also in this case, if $\alpha>1$ and $n>1$, we avoid the effects 
of inflation at small curvatures and it does not influence the stability of 
the matter dominated era. 
The de Sitter point exists if $\tilde R_\mathrm{i}=R_{\mathrm{dS}}$ and 
it reads as in Eq.~(\ref{due}) under the condition $(R/R_\mathrm{i})^n\gg 1$. 
Thus, the inflation is unstable if the condition in Eq.~(\ref{duedue}) is 
satisfied. 
The bigger difference between the two models exists in those behaviors 
in the transition phase between the small curvature region 
(where the physics of the $\Lambda$CDM model emerges) 
and the high curvature region. 
This means that the no antigravity condition is different in the two models 
and such a condition becomes more critical in the transition region. 
Therefore, in the following we are able to 
make the different choices of parameters in the two models. 
We note that 
since dark energy sector of the above models only originates from 
exponential gravity, all qualitative results in terms of the behavior of the 
dark energy component in exponential gravity 
found in the previous sections remain to be valid.

\section{Analysis of inflation}

In this section, we perform the numerical analysis of the early time acceleration for the unified models in Eqs.~(\ref{total}) and (\ref{sin}), by choosing appropriate parameters according with the analysis in Sec.~\ref{generalinflation}. 
For this aim, it is worth rewriting Eq.~(\ref{superEq}) by introducing a suitable scale factor $M^2$ at the inflation. We can choose $M^2=\Lambda_\mathrm{i}$. The effective modified gravity energy density $y_H(z)$ is now defined as 
\begin{equation}
y_H (z)\equiv\frac{\rho_{\mathrm{DE}}}{M^2/\kappa^2}=\frac{3H^2}{M^2}
-\tilde\chi (z+1)^{4}\,.
\end{equation}
Here, we have neglected the contribution of standard matter and supposed the presence of ultrarelativistic matter/radiation in the hot universe scenario, whose energy density $\rho_{\mathrm{rad}}$ at the redshift equal to zero is related with the scale as 
\begin{equation}
\tilde\chi=\frac{\kappa^2\rho_{\mathrm{rad}}}{M^2}\,.
\end{equation}
Since the results are independent of the redshift scale, 
we set $z=0$ at some times around the end of inflation. 
Equation (\ref{superEq}) reads
\begin{eqnarray}
&&
y_H''(z)-\frac{y'(z)}{z+1}\left\{3+\frac{1-F'(R)}{2M^2F''(R)\left[y_H(z)+\tilde\chi(z+1)^4\right]}\right\} 
\nonumber\\ 
&&
{}+\frac{y_H(z)}{(z+1)^2}\frac{2-F'(R)}{M^2F''(R)\left[y_H(z)+\tilde\chi(z+1)^4\right]}
\nonumber\\ 
&&
{}
+\frac{(F'(R)-1)2\tilde\chi(z+1)^4+(F(R)-R)/M^2}{(z+1)^2\,2M^2F''(R)\left[y_H(z)+\tilde\chi(z+1)^4\right]}=0\,.
\label{inflsystem}
\end{eqnarray}
Moreover, the Ricci scalar is expressed as 
\begin{equation}
R=M^2\left[4y_H(z)-(z+1)\frac{d y_H(z)}{d z}\right]\,.
\end{equation}
Thus, 
it is easy to verify that in the de Sitter universe with $R=R_{\mathrm{dS}}$ the perturbation $y_1(z)$ on the solution $y_0=R_{\mathrm{dS}}/(4\,M^2)$ is effectively given by Eq.~(\ref{pertinfl}), i.e., 
$y_1=C_0(z+1)^x$, according with Eq.~(\ref{result}) 
if we neglect the contribution of standard matter. 
In this derivation, we have assumed the contribute of ultrarelativistic matter to be much smaller than $y_0$. However, as stated above, this small energy contribution may originate from the perturbation $y_1(z_\mathrm{i})$ at the beginning of inflation, which, if $x<0$, grows up in the expanding universe 
making inflation unstable.

\subsubsection*{Model I}

First, we explore the model in Eq.~(\ref{total}). 
We have to choose the parameters as $\Lambda_\mathrm{i}\simeq 10^{100-120}\Lambda$. The dynamics of the system is independent of this choice. 
Here, we summarize the conditions for inflation already stated in 
Sec.~\ref{Inflation}: 
\begin{eqnarray}
&&
R_\mathrm{i}>\Lambda_\mathrm{i} n
\left(\frac{n-1}{n}\right)^{\frac{n-1}{n}}\mathrm{e}^{-\frac{n-1}{n}}\,,\quad\text{(no antigravity effects)}\nonumber\\ 
&&
\tilde R_\mathrm{i}=R_{\mathrm{dS}}\,,\quad\alpha\bar{\gamma}(\alpha-2)>1\,,\quad\left(\frac{R_{\mathrm{dS}}}{R_\mathrm{i}}\right)^n\gg 1\,,\quad\text{(existence of unstable dS solution)} \nonumber
\end{eqnarray}
with $n>1$, $2+1/\bar{\gamma}>\alpha>2$ and $R_{\mathrm{dS}}=2\Lambda_\mathrm{i}/\left[\bar{\gamma}(2-\alpha)+1\right]$. Since $\bar{\gamma}$ and $\alpha$ are combined in $\gamma(\alpha-2)$, we can fix $\bar{\gamma}=1$, so that $R_{\mathrm{dS}}=2\Lambda_\mathrm{i}/(3-\alpha)$ and $3>\alpha>2$. 
The instability factor $x$ in Eq.~(\ref{x}) only depends on $R_{\mathrm{dS}}$. 
Hence, by studying the phenomenology of inflation, we examine the variation of $\alpha$ parameter (and, as a consequence, that of $\tilde R_\mathrm{i}$). 
We take $n=4$ and $R_\mathrm{i}=2\Lambda_\mathrm{i}$, which satisfy the condition for no antigravity well. 
We analyze three different cases of $\alpha = 5/2$, $8/3$, and $11/4$. 
In these cases, we have 
$R_{\mathrm{dS}}=4\Lambda_\mathrm{i}$, $6\Lambda_\mathrm{i}$, and $8\Lambda_\mathrm{i}$, respectively. 

\begin{figure}[!h]
\subfigure[]{\includegraphics[width=0.3\textwidth]{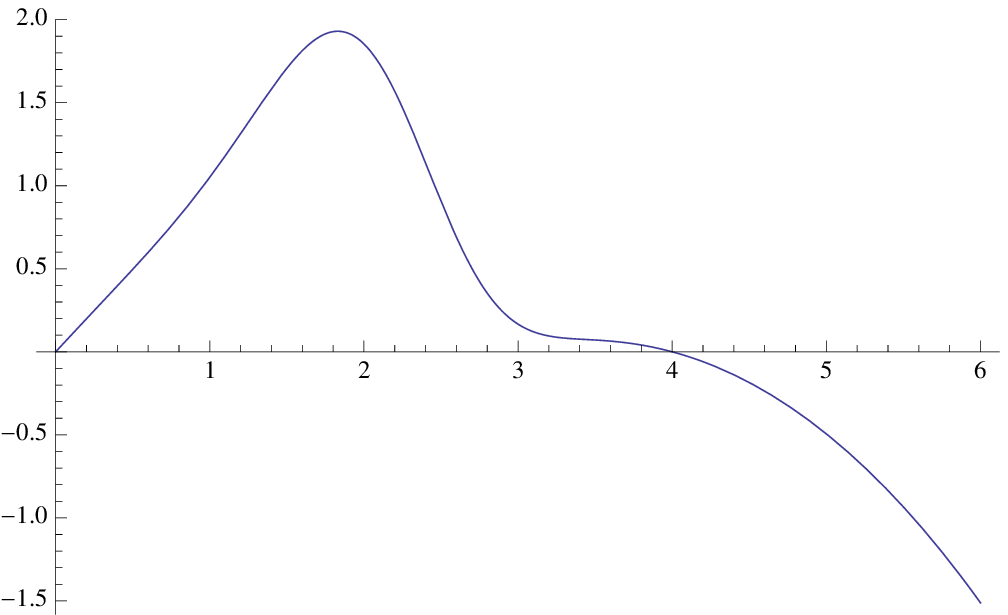}}
\centering
\subfigure[]{\includegraphics[width=0.3\textwidth]{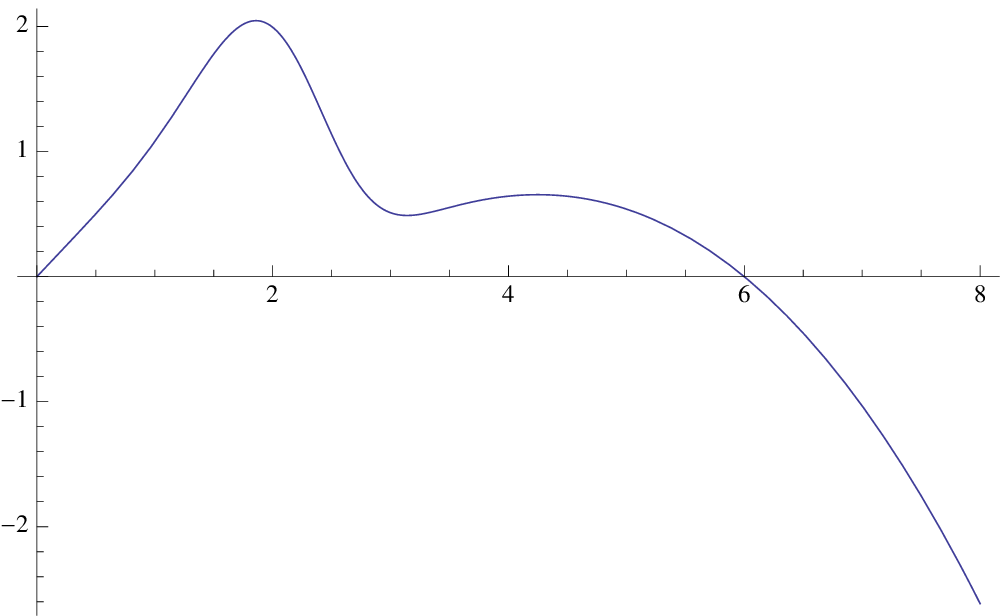}}
\centering
\subfigure[]{\includegraphics[width=0.3\textwidth]{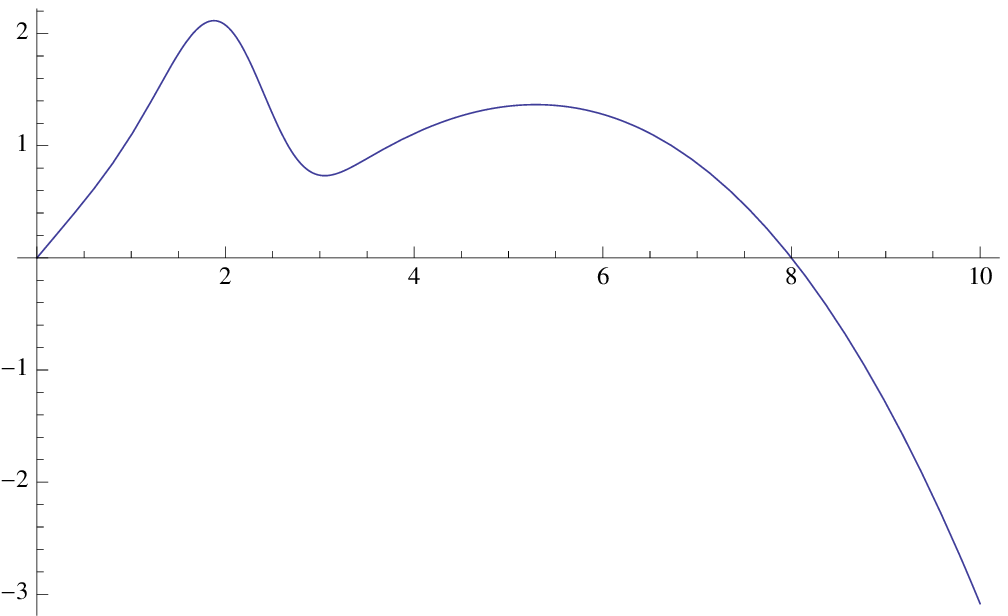}}
\caption{Cosmological evolution of the quantity $2F(R/\Lambda_\mathrm{i})-(R/\Lambda_\mathrm{i}) F'(R/\Lambda_\mathrm{i})$ as a function of the redshift $z$ 
for exponential model with $\alpha=5/2$ (a), $\alpha=8/3$ (b) and $\alpha=11/4$ (c). The ``zeros'' of these graphics indicate the de Sitter solutions of the model. 
\label{00}}
\end{figure}

In Fig.~\ref{00}, we plot the cosmological evolution of 
the quantity 
$2F(R/\Lambda_\mathrm{i})-(R/\Lambda_\mathrm{i}) F'(R/\Lambda_\mathrm{i})$ 
as a function of the redshift $z$ in the three cases. The value of ``zero'' of this quantity corresponds to the de Sitter points of the model. We can recognize the unstable de Sitter solutions of inflation and the attractor in zero (the de Sitter point of current acceleration is out of scale).

Despite the fact that the three considered values of $\alpha$ are very close 
each other, the values of $R_{\mathrm{dS}}$ and $x$ significantly change and the reactions of the system to small perturbations are completely different. 
By starting from Eq.~(\ref{Nfolding}), we may reconstruct the rate $y_1(z_\mathrm{i})/y_0$ between the abundances of ultrarelativistic matter/radiation and 
modified gravity energy at the beginning of inflation in order to obtain a determined 
number of $e$-folds during inflation in the three different cases, by taking into account that $x=-0.086$, $-0.218$, and $-0.270$ for $\alpha=5/2$, $8/3$, and $11/4$, respectively. 
For example, in order to have $N=70$, 
for $\alpha=5/2$, 
a perturbation of $y_1(z_\mathrm{i})/y_0\sim 10^{-3}$ is necessary; 
for $\alpha=8/3$, 
a perturbation of $y_1(z_\mathrm{i})/y_0\sim 10^{-7}$ is sufficient; 
whereas for $\alpha=11/4$, 
$y_1(z_\mathrm{i})/y_0\sim 10^{-9}$. 
The system becomes more unstable, as $\left(3-\alpha\right)$ 
is closer to zero. 

In studying the behavior of the cosmic evolution in 
Model I for the three different cases, 
we set $\tilde{\chi}=10^{-4}\,y_0/(z_\mathrm{i}+1)^4$ in Eq.~(\ref{inflsystem}) for the case $\alpha=5/2$ and $\tilde{\chi}=10^{-6}\,y_0/(z_\mathrm{i}+1)^4$ for the cases $\alpha=8/3, 11/4$. 
In these choices, 
the effective energy density originating from 
the modification of gravity is $10^4$ and $10^6$ times larger than 
that of ultrarelativistic matter/radiation during inflation. 
By using Eq.~(\ref{Nfolding}), we can predict the following numbers of 
$e$-folds: 
\begin{eqnarray}
N \Eqn{\simeq} 107 \quad (\mathrm{for} \,\,\, \alpha=5/2)\,,\nonumber\\
N \Eqn{\simeq} 64 \quad (\mathrm{for} \,\,\, \alpha=8/3)\,,\nonumber\\
N \Eqn{\simeq} 51 \quad (\mathrm{for} \,\,\, \alpha=11/4)\,.\label{stime}
\end{eqnarray}
In order to solve Eq.~(\ref{inflsystem}) numerically, 
we use the initial conditions
\begin{eqnarray}
\frac{d y_H(z)}{d (z)}\Big\vert_{z_i} \Eqn{=} 0\,,\nonumber\\ \nonumber\\
y_H(z)\Big\vert_{z_i} \Eqn{=} \frac{R_{\mathrm{dS}}}{4\Lambda_\mathrm{i}}\,,
\label{BC}
\end{eqnarray}
at the redshift $z_\mathrm{i}\gg 0$ when inflation starts. 
We put $z_\mathrm{i}=10^{46}$, $10^{27}$, and $10^{22}$ for $\alpha=5/2$, 
$8/3$, and $11/4$, respectively 
(just for a more comfortable reading of the graphics). 
We also remark that 
the initial conditions are subject to an artificial error that we can estimate to be in the order of $\exp\left[-\left(R_{\mathrm{dS}}/R_\mathrm{i}\right)^n\right]\sim 10^{-7}$. This is the reason for which we only consider $\tilde{\chi}>10^{-7}$. 

\begin{figure}[!h]
\subfigure[]{\includegraphics[width=0.3\textwidth]{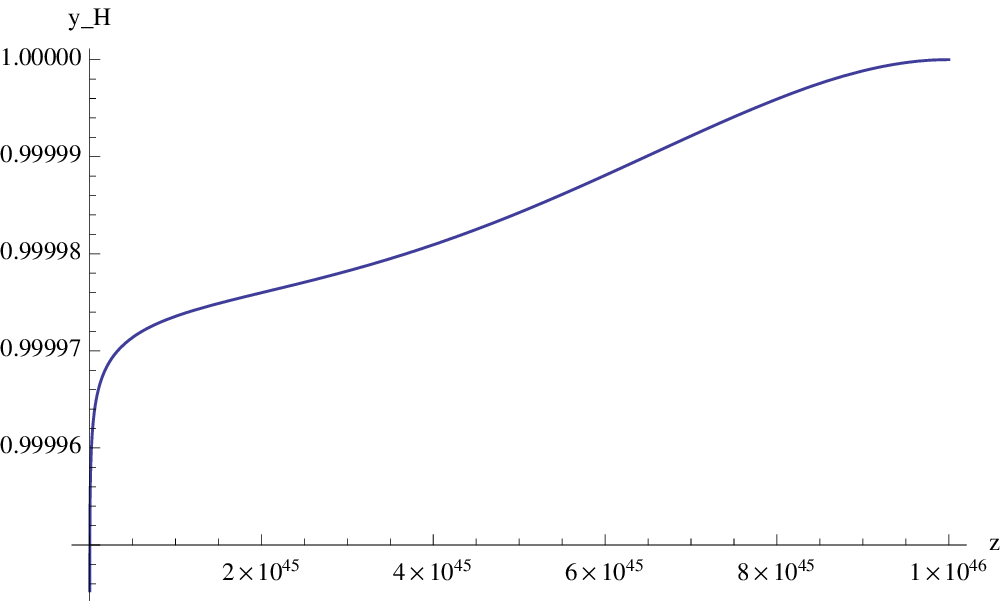}}
\centering
\subfigure[]{\includegraphics[width=0.3\textwidth]{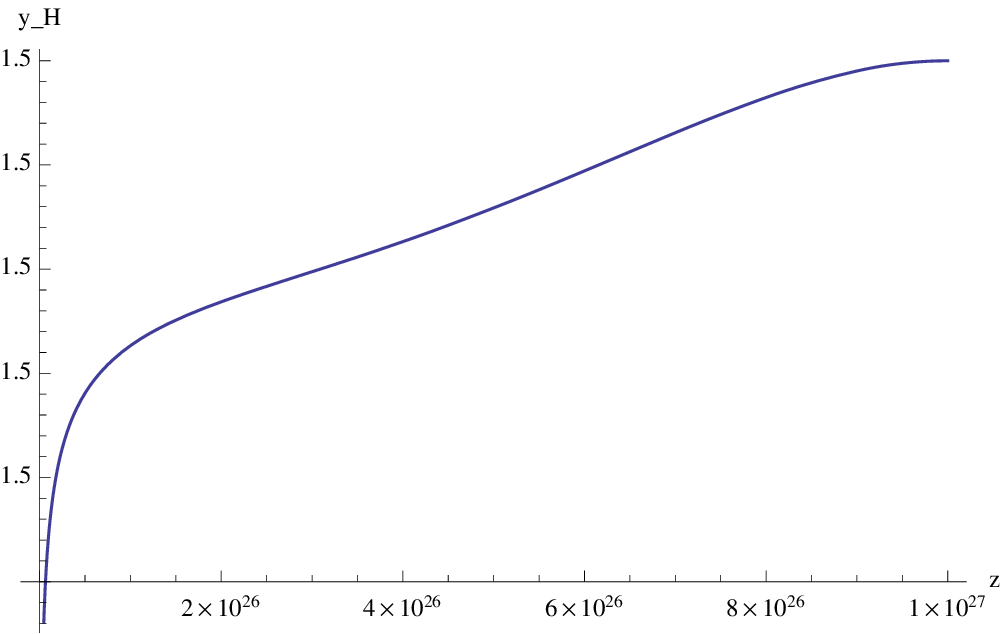}}
\centering
\subfigure[]{\includegraphics[width=0.3\textwidth]{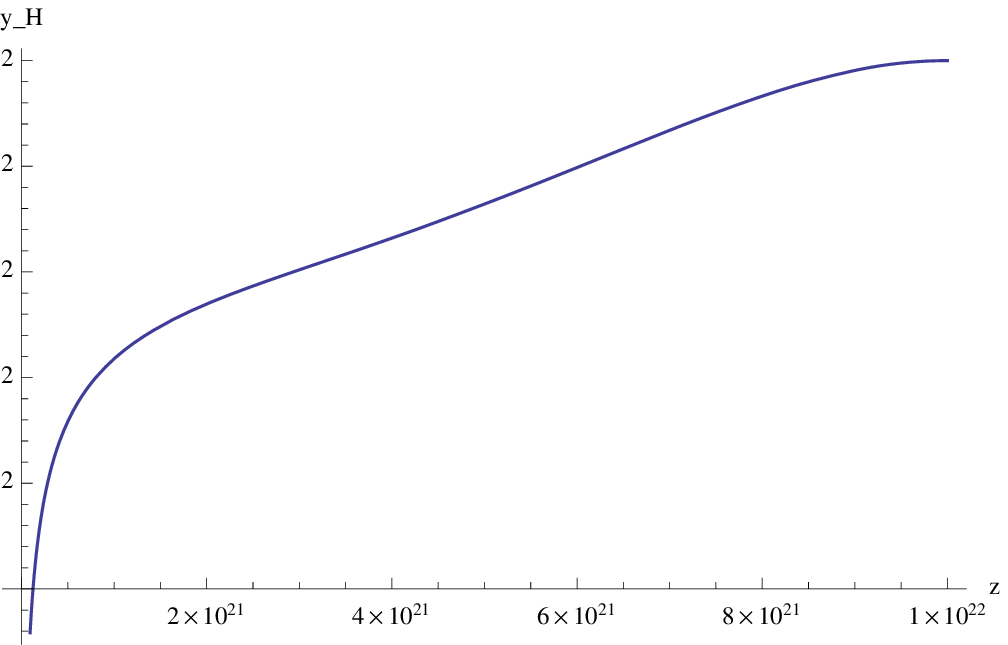}}
\quad
\subfigure[]{\includegraphics[width=0.3\textwidth]{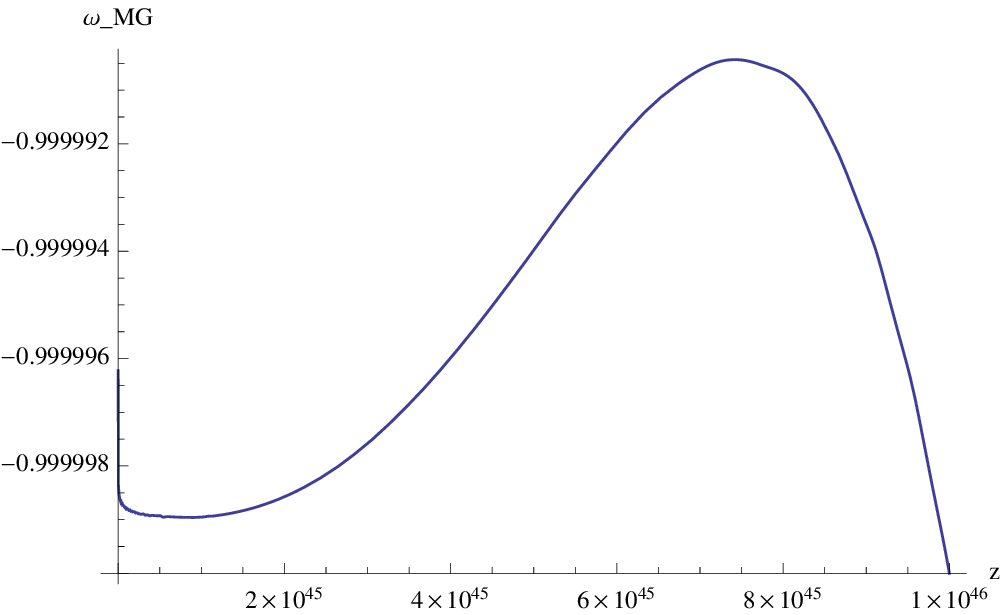}}
\centering
\subfigure[]{\includegraphics[width=0.3\textwidth]{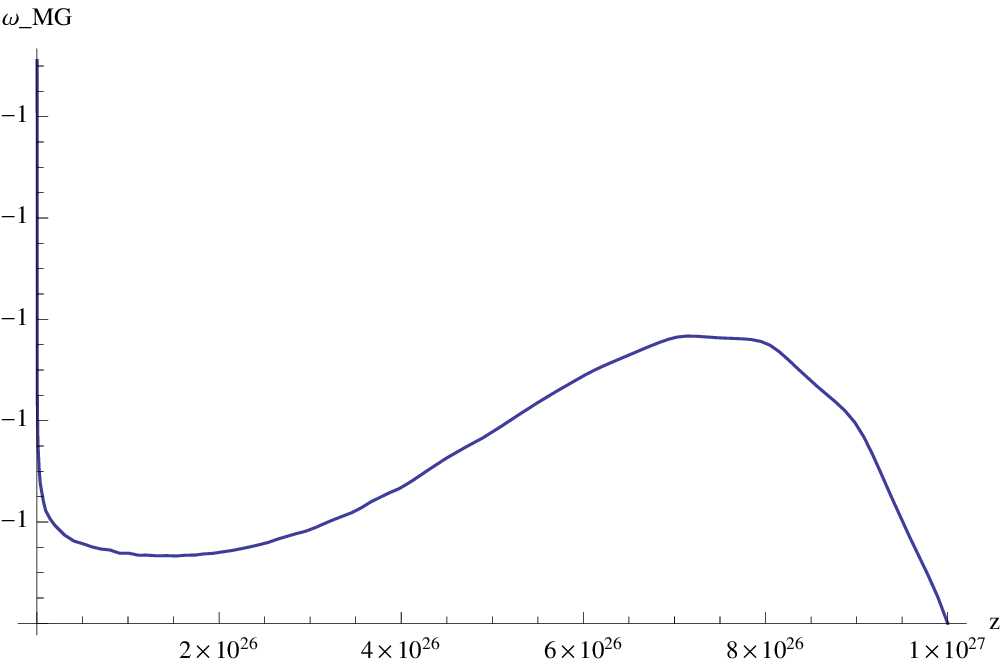}}
\centering
\subfigure[]{\includegraphics[width=0.3\textwidth]{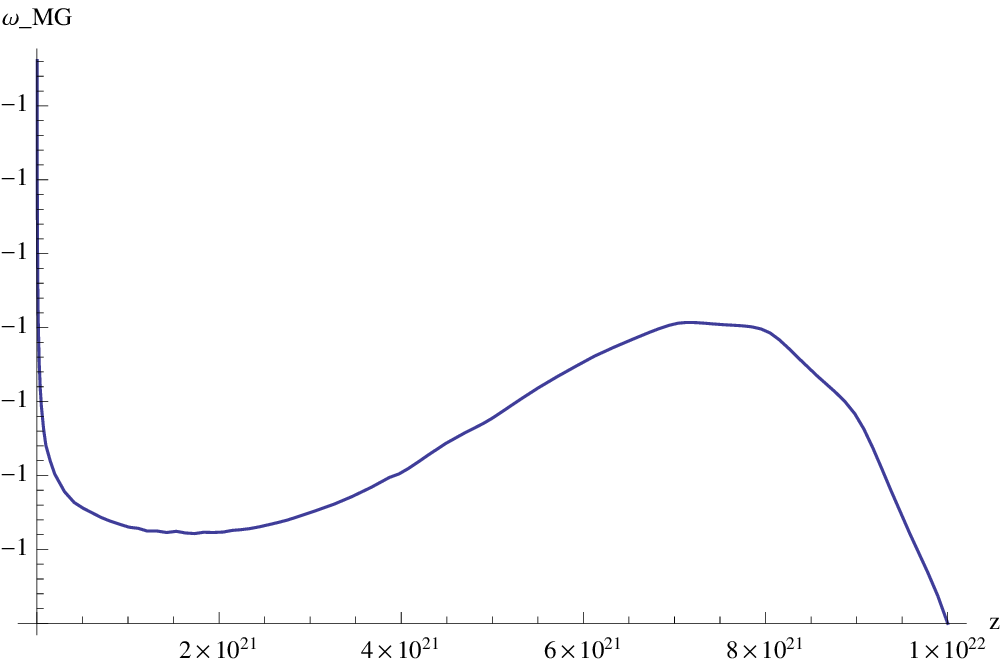}}
\caption{Plots of $y_H$ [a-c] and $\omega_{\mathrm{MG}}$ [d-f] as functions of the redshift $z$ for Model I with $\alpha=5/2$ [a-d], $\alpha=8/3$ [b-e] and $\alpha=11/4$ [c-f]. 
\label{001}}
\end{figure}

\begin{figure}[!h]
\subfigure[]{\includegraphics[width=0.3\textwidth]{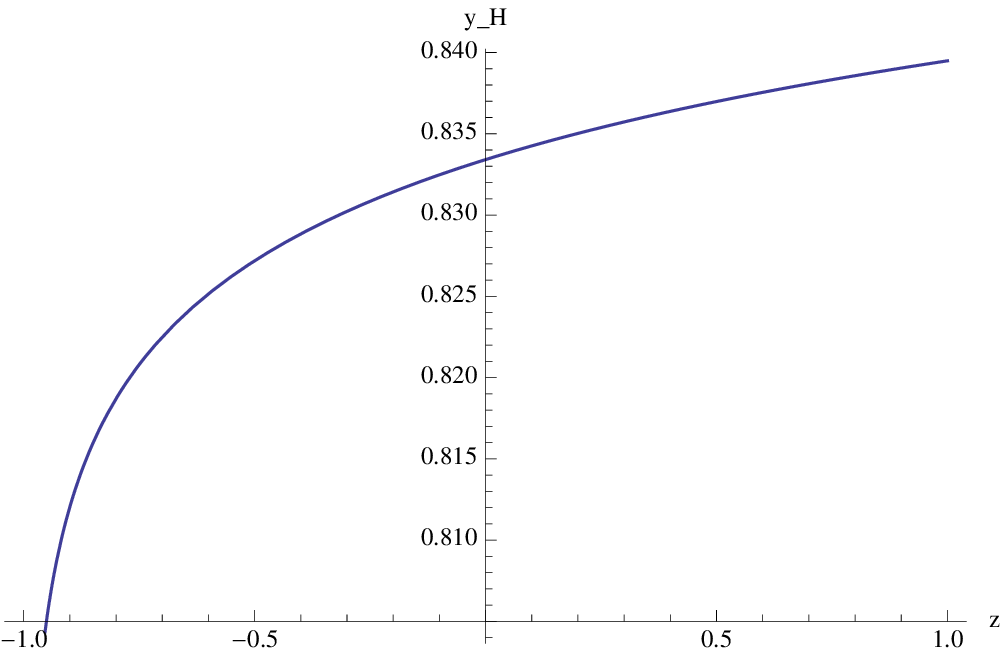}}
\centering
\subfigure[]{\includegraphics[width=0.3\textwidth]{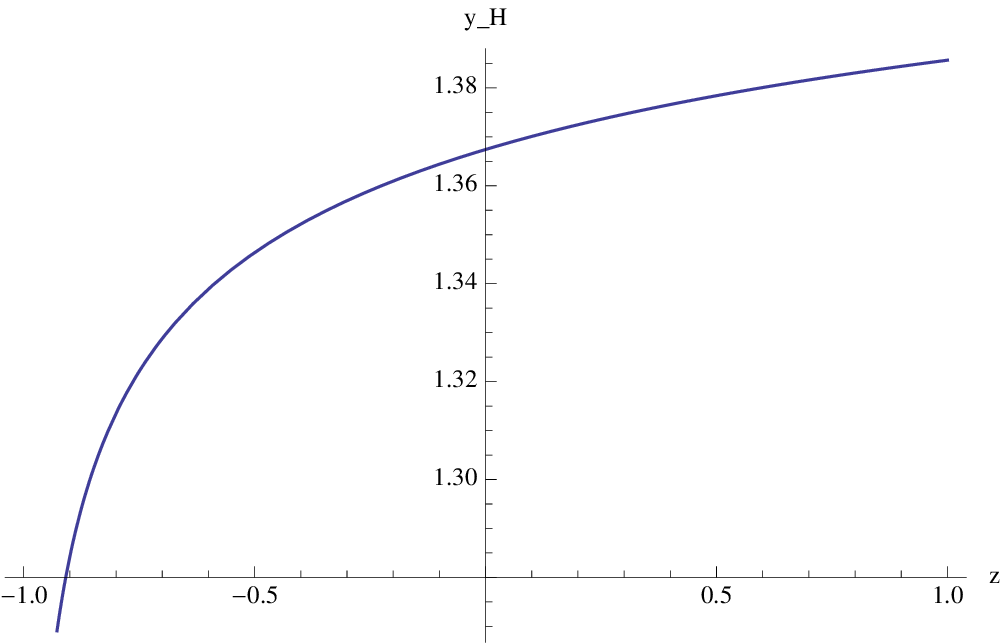}}
\centering
\subfigure[]{\includegraphics[width=0.3\textwidth]{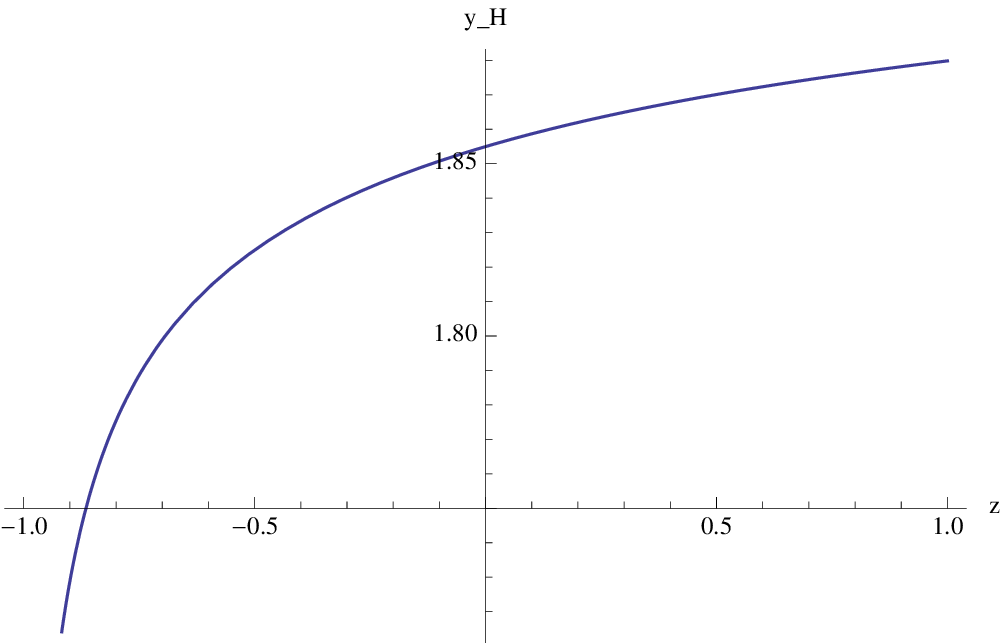}}
\quad
\subfigure[]{\includegraphics[width=0.3\textwidth]{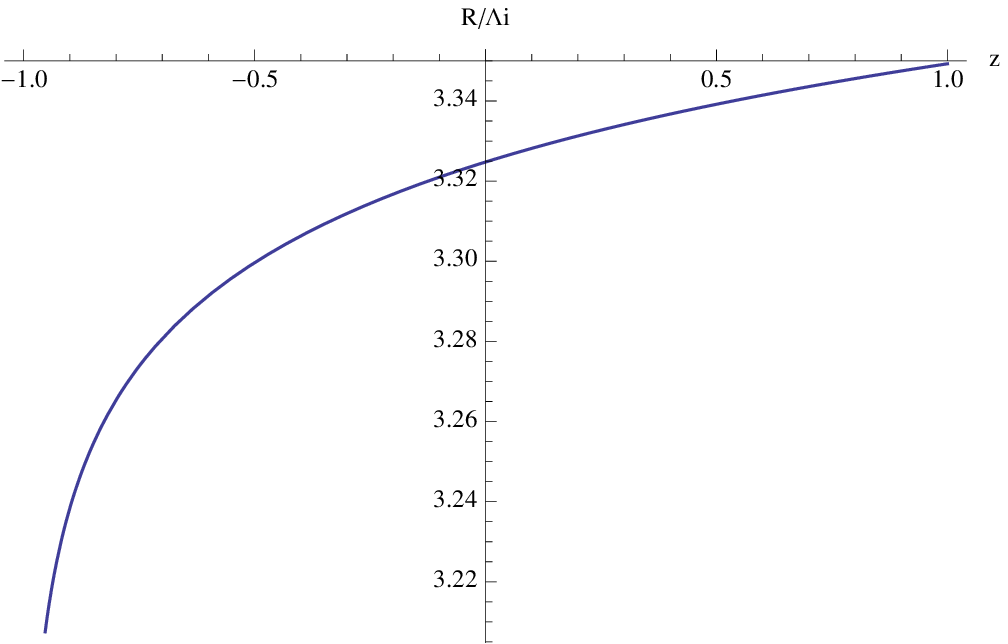}}
\centering
\subfigure[]{\includegraphics[width=0.3\textwidth]{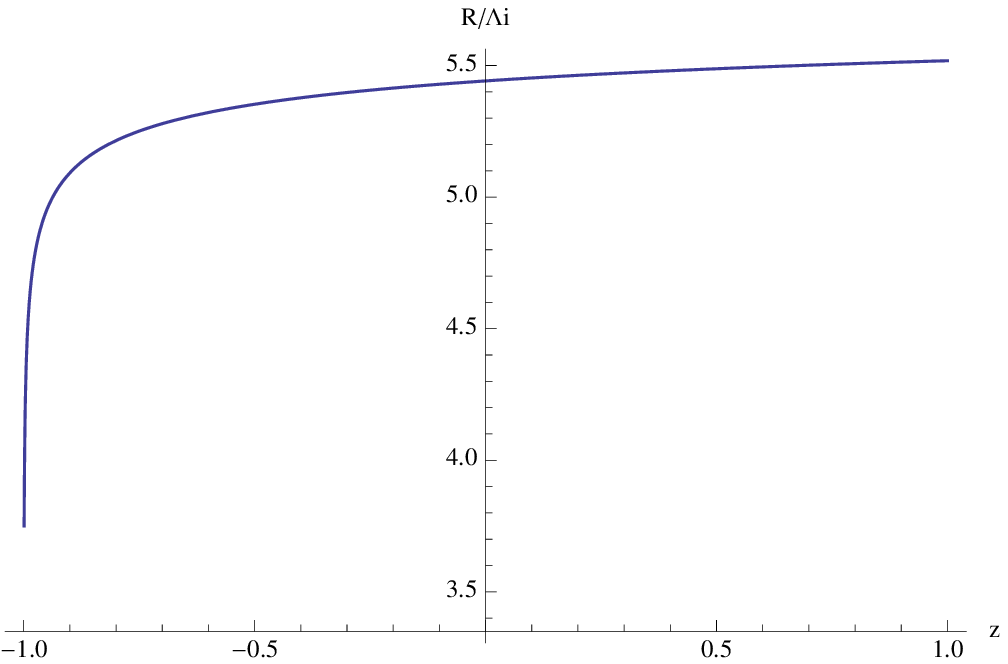}}
\centering
\subfigure[]{\includegraphics[width=0.3\textwidth]{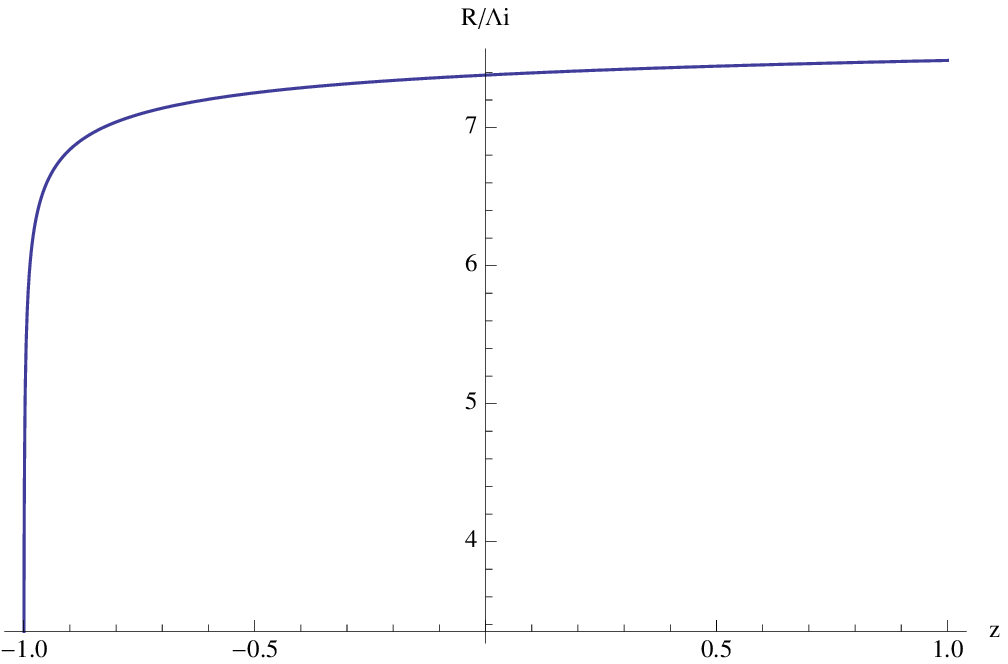}}
\caption{Cosmological evolution of $y_H$ [a-c] and $R/\Lambda_\mathrm{i}$ [d-f] as functions of the redshift 
$z$ in the region $-1<z<1$ for Model I with $\alpha=5/2$ [a-d], $\alpha=8/3$ [b-e] and $\alpha=11/4$ [c-f]. 
\label{0001}}
\end{figure}

In Fig.~\ref{001}, we illustrate the cosmological evolutions of $y_H$ and the corresponding modified gravity EoS parameter $\omega_{\mathrm{MG}}$ (defined as in Eq. (\ref{oo})) as functions of the redshift $z$ in the three cases. 
We can see, during inflation $\omega_{\mathrm{MG}}$ is indistinguishable from the value of -1 and $y_H$ tends to decrease very slowly with respect to 
$y_H=1, 3/2, 2$ for $\alpha=5/2, 8/3, 11/4$, so that the curvature can be the expected de Sitter one, $R_{\mathrm{dS}}(=4y_H)=4\Lambda_\mathrm{i}, 6\Lambda_\mathrm{i}, 8\Lambda_\mathrm{i}$. 
The expected values of $z_\mathrm{e}$ at the end of inflation may be derived from the number of $e$-folds in (\ref{stime}) during inflation 
and read $z_\mathrm{e} \simeq -0.47$ for $\alpha=5/2$; $z_\mathrm{e}\simeq -0.74$ for $\alpha=8/3$; $z_\mathrm{e}\simeq -0.39$ for $\alpha=11/4$. 
The numerical extrapolation yields 
\begin{eqnarray*}
y_H(z_\mathrm{e}) \Eqn{=} 0.83y_H(z_\mathrm{i})\,,\quad R(z_\mathrm{e})=0.825 R_{\mathrm{dS}}\,,\quad (\mathrm{for} \,\,\, \alpha=5/2)\nonumber\\
y_H(z_\mathrm{e}) \Eqn{=} 0.88y_H(z_\mathrm{i})\,,\quad R(z_\mathrm{e})=0.853 R_{\mathrm{dS}}\,,\quad (\mathrm{for} \,\,\, \alpha=8/3)\nonumber\\
y_H(z_\mathrm{e}) \Eqn{=} 0.92y_H(z_\mathrm{i})\,,\quad R(z_\mathrm{e})=0.911 R_{\mathrm{dS}}\,.\quad (\mathrm{for} \,\,\, \alpha=11/4)
\end{eqnarray*}
To confirm the exit from inflation,  in Fig.~\ref{0001}
we plot the cosmological evolutions of $y_H$ and $R/\Lambda_\mathrm{i}$ 
as functions of the redshift $z$ 
in the region $-1<z<1$, where $z_\mathrm{e}$ is included. 
The effective modified gravity energy density and the curvature decrease at the end of inflation and 
the physical processes described by the $\Lambda$CDM model can appear. 

\subsubsection*{Model II}

Next, we investigate Model II in Eq.~(\ref{sin}). 
Here, in order to satisfy the condition for no antigravity 
we choose $n=3$ and $R_\mathrm{i}=2\Lambda_\mathrm{i}$, so that $F'(R>0)>0$. 
We take $\bar{\gamma}=1$ again and we execute the same numerical evaluation for $\alpha=5/2, 13/5, 21/8$ in this model as that in the previous case for Model I. 
The corresponding de Sitter curvatures of inflation are 
$R_{\mathrm{dS}}=4\Lambda_\mathrm{i}, 5\Lambda_\mathrm{i}, 16\Lambda_\mathrm{i}/3$. Now, we obtain the factor in Eq.~(\ref{x}) for instability as 
$x=-0.086$, $-0.170$, and $-0.188$ for $\alpha=5/2$, $13/5$, and $21/8$, respectively. 
Hence, 
we set $\tilde{\chi}=10^{-3}\,y_0/(z_\mathrm{i}+1)^4$ for $\alpha=5/2$, 
$\tilde{\chi}=10^{-4}\,y_0/(z_\mathrm{i}+1)^4$ for $\alpha=13/5$, and 
$\tilde{\chi}=10^{-5}\,y_0/(z_\mathrm{i}+1)^4$ for $\alpha=21/8$. 
As a consequence, the numbers of $e$-folds during inflation result in 
$N=80$, $54$, and $61$. 
The initial conditions are the same as those in the previous case 
in (\ref{BC}). Furthermore, we put $z_\mathrm{i}=10^{34}$, $10^{22}$, and 
$10^{26}$ for $\alpha=5/2$, $13/5$, and $21/8$. 

\begin{figure}[!h]
\subfigure[]{\includegraphics[width=0.3\textwidth]{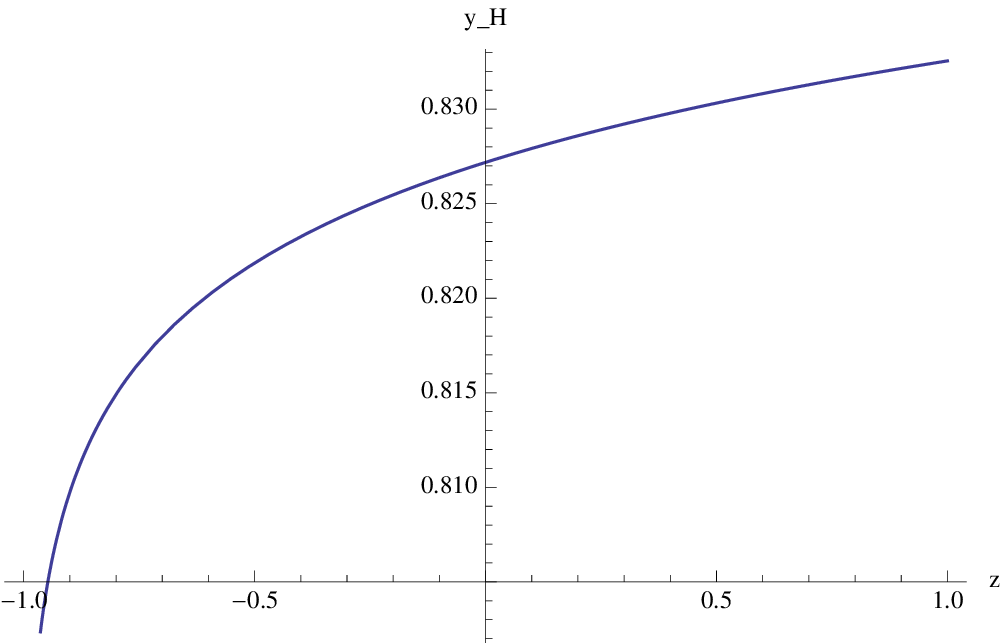}}
\centering
\subfigure[]{\includegraphics[width=0.3\textwidth]{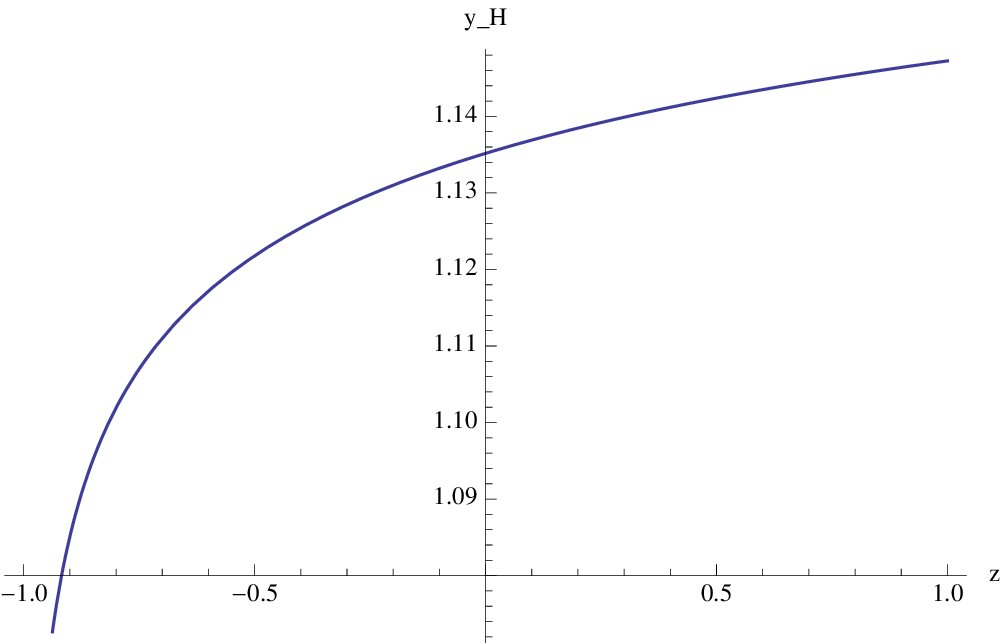}}
\centering
\subfigure[]{\includegraphics[width=0.3\textwidth]{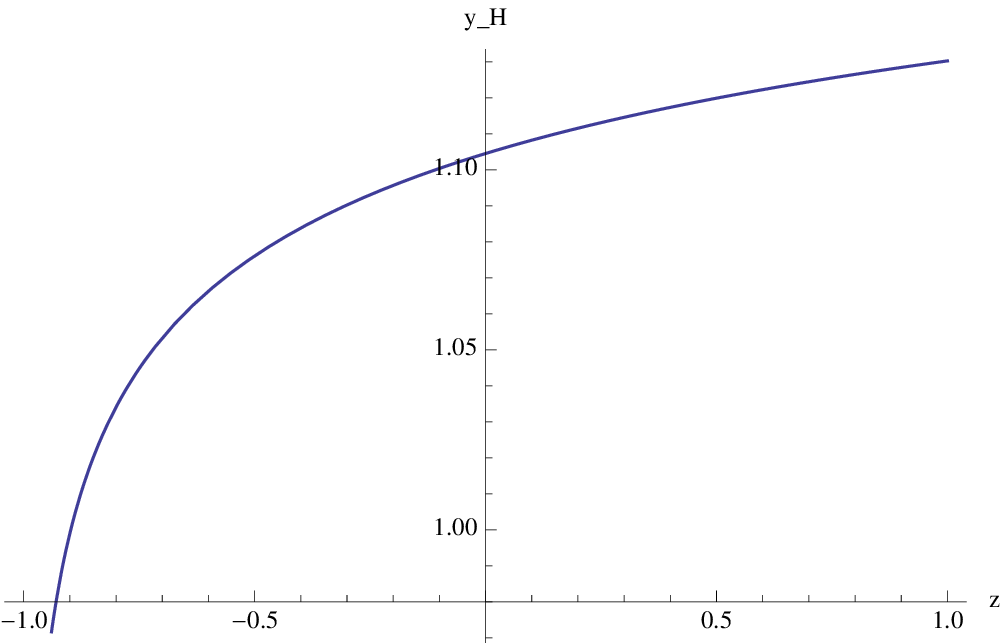}}
\quad
\subfigure[]{\includegraphics[width=0.3\textwidth]{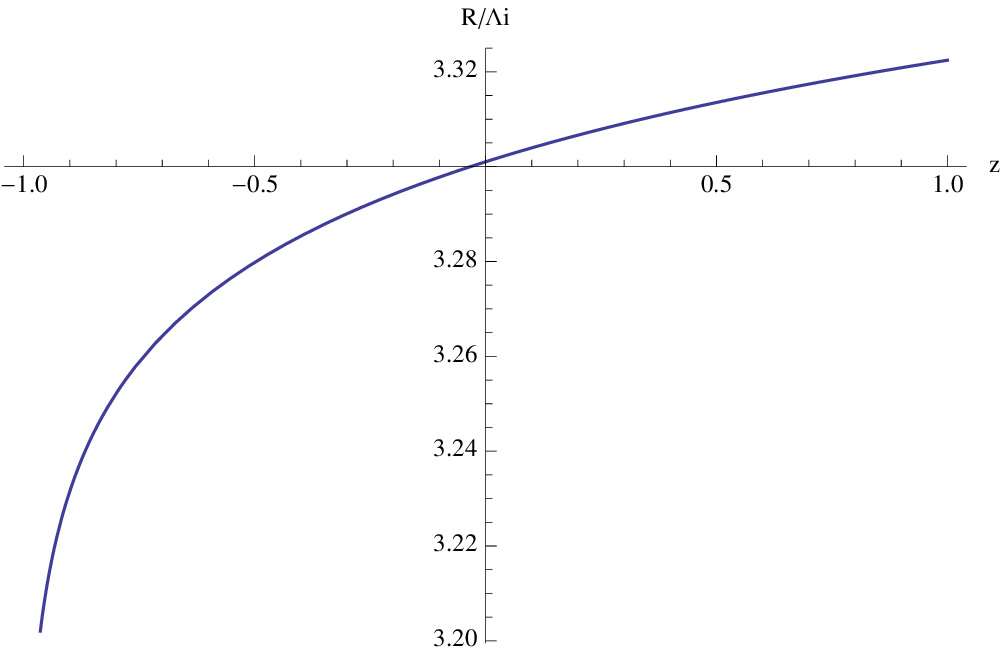}}
\centering
\subfigure[]{\includegraphics[width=0.3\textwidth]{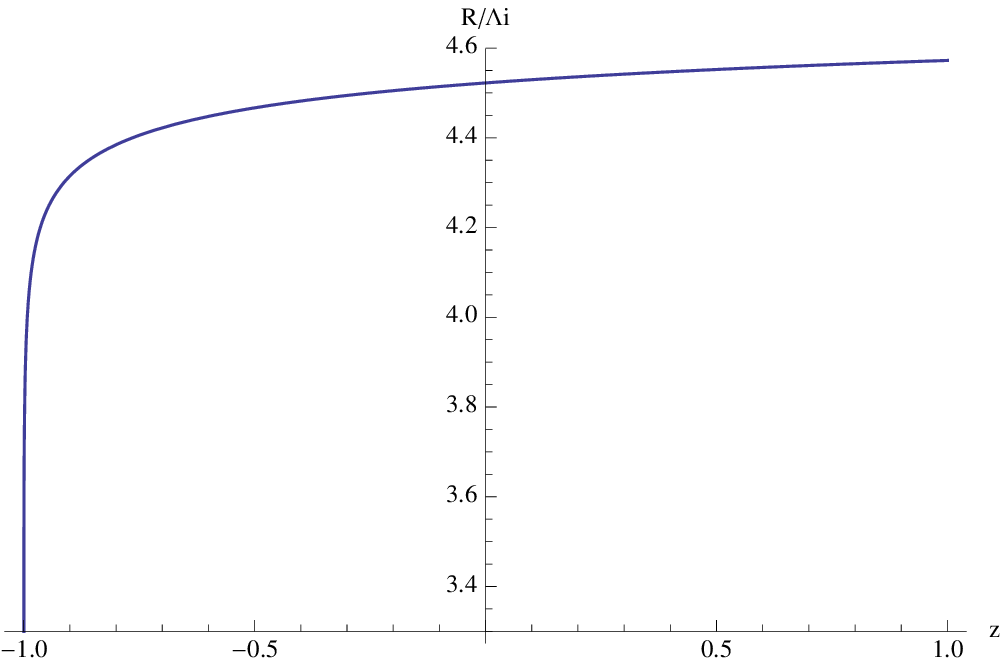}}
\centering
\subfigure[]{\includegraphics[width=0.3\textwidth]{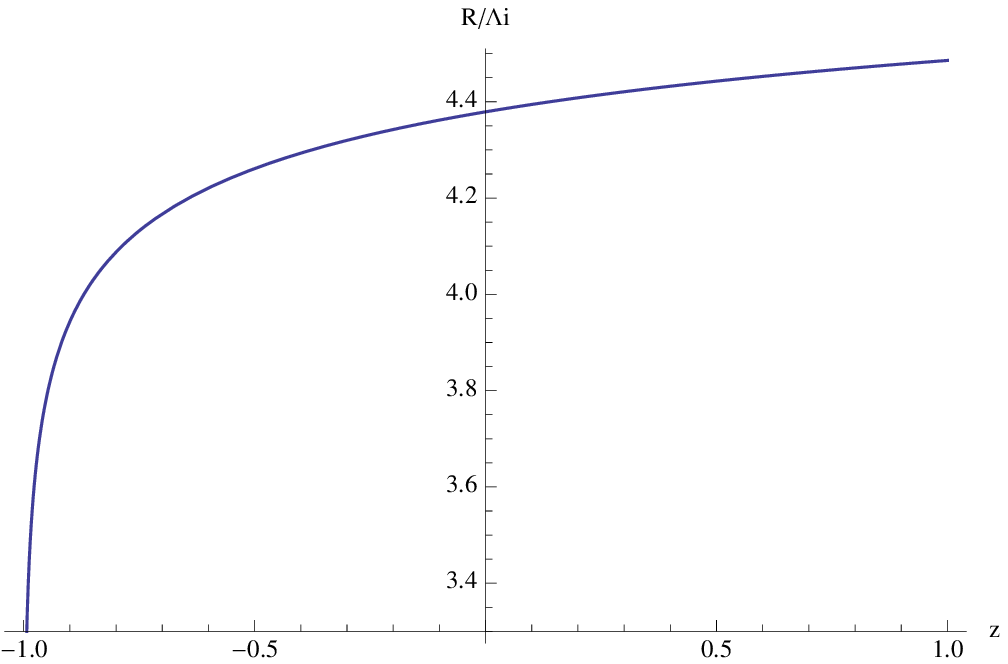}}
\caption{Cosmological evolution of $y_H$ [a-c] and $R/\Lambda_\mathrm{i}$ [d-f] as functions of the redshift 
$z$ in the region $-1<z<1$ for Model II with $\alpha=5/2$ [a-d], $\alpha=13/5$ [b-e] and $\alpha=21/8$ [c-f]. 
\label{0001bis}}
\end{figure}

Through the numerical extrapolation, we acquire 
the expected values of $z_\mathrm{e}$ at the end of inflation as 
$z_\mathrm{e}=-0.80$, $-0.97$, and $-0.71$, and 
the following values for the effective modified gravity energy density and the Ricci scalar: 
\begin{eqnarray*}
y_H(z_\mathrm{e}) \Eqn{=} 0.82y_H(z_\mathrm{i})\,,\quad R(z_\mathrm{e})=0.813 R_{\mathrm{dS}}\,,\quad(\mathrm{for} \,\,\, \alpha=5/2)\nonumber\\
y_H(z_\mathrm{e}) \Eqn{=} 0.84y_H(z_\mathrm{i})\,,\quad R(z_\mathrm{e})=0.884 R_{\mathrm{dS}}\,,\quad(\mathrm{for} \,\,\, \alpha=13/5)\nonumber\\
y_H(z_\mathrm{e}) \Eqn{=} 0.79y_H(z_\mathrm{i})\,,\quad R(z_\mathrm{e})=0.780 R_{\mathrm{dS}}\,.\quad(\mathrm{for} \,\,\, \alpha=21/8)
\end{eqnarray*}
For this model, in Fig.~\ref{0001bis}  we depict the cosmological evolutions of $y_H$ and $R/\Lambda_\mathrm{i}$ as functions of the redshift $z$ in the region $-1<z<1$ at the end of inflation. 
Again in this case, the effective modified gravity energy density and curvature decrease, 
and therefore inflation ends and then the physical processes described 
by the $\Lambda$CDM model can be realized. 

Here, we note that at the inflationary stage, radiation is negligible, 
as in the ordinary inflationary scenario. 
It causes the perturbations at the origin of instability. 
This point has been shown in a numerical way by using radiation, whose energy density is six order of magnitude smaller than that of dark energy.  

It should be emphasized that in this work, 
as a first step, we have concentrated on only the possibility 
of the realization of inflation, and 
hence that important issues in inflationary cosmology such as 
the graceful exit problem of 
inflation, the following reheating process, and the generation of the 
curvature perturbations, whose power spectrum has to be consistent with 
the anisotropies of the CMB radiation obtained from the 
Wilkinson Microwave Anisotropy Probe (WMAP) Observations~\cite{WMAP-Spergel, 
WMAP, Komatsu:2010fb}, are the crucial future works of our unified scenario 
between inflation and the late-time cosmic acceleration. 

In the future works, if we analyze the power spectrum of the curvature perturbations in our models, the next question becomes 
not what the total number of $e$-folds is, 
but how many $e$-folds one could obtain from the point 
when the power-law index of the primordial power spectrum $n_s$ 
is close to its observed value. 
It is presumed that 
since the equation of state $w$ at the inflationary stage 
is so close to the model, e.g., 
with $\alpha=11/4$, 
the number of $e$-folds from the point 
when $n_s \simeq 0.96$~\cite{WMAP, Komatsu:2010fb} until 
the end of inflation is much smaller. 
Accordingly, we should examine whether 
it is enough for the galaxy power spectrum to be reasonably close to 
the scale invariance of the power spectrum of the curvature perturbations. 
Moreover, as a more relevant question which remains is the mechanism for 
reheating. 
The problem is 
how the universe becomes the radiation dominated stage again 
after the inflationary period. 
In order to construct complete models of inflation, 
we need a discussion of the reheating mechanism and 
that of exactly how the power spectrum of anisotropies is transferred to 
the matter. 
These are significant future subjects in our studies. 

\section{Conclusions and general remarks}

In the present paper, 
we have examined a generic feature of viable $F(R)$ gravity models, 
in particular, exponential gravity as well as a power form model. 
We have shown that the behavior of higher derivatives of the Hubble parameter 
may be affected by large frequency oscillations of effective dark energy, 
and consequently solutions may become singular and unphysical at a high 
redshift. 
The analyzed models approach to a model with the cosmological constant in a 
manner different from each other, and hence it is reasonable to expect that 
the found results can be generalized to realistic $F(R)$ gravity models, 
in which the cosmological evolutions are similar to those in a model with the 
cosmological constant. 
To support our claim, in the first part of this paper 
we have explicitly demonstrated 
how the origin of the problem influences the stability conditions satisfied by 
these models in order to reproduce the realistic matter dominated era. 
Since the corrections to the Einstein equations at the small curvature 
regime may lead to undesired effects at the high curvature regime, 
we have reconstructed a correcting (compensating) term added to the models 
in order to stabilize the oscillations of the effective dark energy 
in the matter dominated era with retaining the viability properties. 
It is emphasized that all the results we have found in an analytical way via 
studying the perturbation theory are confirmed by 
the numerical analysis performed on the models under consideration. 
Moreover, a detailed investigation on the cosmological evolutions of the 
universe described by those models has been executed. 
In particular, we have demonstrated that our correction term does not cause 
any problem to the viability of the models, and that the obtained results are 
consistent with recent very accurate observational data of our current 
universe and easily pass the local tests of the solar system. 
Furthermore, we have shown that the effective crossing of the phantom divide, 
which characterizes the de Sitter epoch, occurs in the very far future. 
A way to avoid the crossing of phantom divide by using inhomogeneous 
fluids has also been explored. 

After the discovery of the accelerated expansion of our universe, a lot of 
theories are proposed in order to explain it. 
The issue of discriminating among all of these theories has become very 
important. 
The first step in order to distinguish between theories can be the study of 
their expansion history, but it has been revealed that sometimes different 
models exhibit the same (or very similar) expansion history. 
For this reason, the investigation of growth of the matter density 
perturbations by using the so-called growth index can provide a significant 
tool in order to distinguish among the different gravitational theories. 
In this context, the growth of the matter density perturbations has been 
examined for our models. 
Several ansatz for the growth index have been considered, and consequently 
it has been concluded that the 
choice of the growth index as $\gamma = \gamma_0 + \gamma_1 z$ 
is the most appropriate parameterization for these theories. 

In addition, in the second part of this paper 
we have discussed the inflationary cosmology in two exponential gravity 
models. 
It has explicitly been shown that different numbers of $e$-folds during 
inflation can be obtained by taking different model parameters 
in the presence of ultrarelativistic matter, the existence of which makes 
inflation to end and realize the exit from it. 
We have performed the numerical analysis of the inflationary stage 
in two viable exponential gravity models. 
It has been found that at the end of the inflation, the effective energy 
density and therefore the curvature of the universe become small. 
As a result, we have proved that it is possible to acquire 
a gravitational alternative scenario for a unified description of inflation in 
the early universe with the late-time cosmic acceleration due to the 
$\Lambda$CDM-like dark energy domination. 

It should be cautioned that 
in this work, we have constructed a unified description of 
inflation with the late-time cosmic acceleration in 
$F(R)$ gravity by examining the cosmological evolutions of 
inflation in Secs.~V and VI and the late-time cosmic acceleration 
in Sec.~III one by one, 
and therefore that the evolution equation expressing all the processes 
from inflation to the current accelerated cosmic acceleration 
has not been obtained yet. 
In order to obtain such a gravitational field equation, 
the detailed considerations 
on the reheating process after inflation is also necessary 
(for a very recent analysis, see, e.g,~\cite{Motohashi:2012tt}). 
Qualitatively, 
from our results it can presumably be considered that 
at the inflationary stage 
the EoS parameter $w_\mathrm{eff}$ is approximately equal to 
$-1$ and after that it becomes close to $1/3$ 
during the reheating stage because of the appearance of 
radiation, and after the radiation-dominated stage with 
$w_\mathrm{eff} \approx 1/3$ following the matter-dominated stage with 
$w_\mathrm{eff} \approx 0$, the dark energy dominated stage with 
$w_\mathrm{eff} \approx -1$ can be realized. 
If we successfully acquire the equation and solve it analytically or 
numerically, it would be possible to plot 
the evolution of the Hubble expansion rate $H$ or 
$w_\mathrm{eff}$ from the inflationary stage in 
the early universe to the present time. 
This is very interesting and significant task in our aim, 
hence it would be one of the important future works of our 
study. 

We also mention that 
as another important future work in terms of our present investigations, 
at the next step we plan to study cosmological 
perturbations~\cite{Kodama:1985bj, Mukhanov:1990me} 
in such resultant $F(R)$ gravity theories. 
We calculate the power spectrum of the cosmological 
perturbations as well as the tensor-to-scalar ratio 
in these models 
and compare those with the observational data 
from such as WMAP satellite~\cite{Komatsu:2010fb}, 
future PLANCK satellite~\cite{Planck-1, Planck-2}, 
QUIET~\cite{QUIET-1, Samtleben:2008rb}, 
B-Pol~\cite{B-Pol} and LiteBIRD~\cite{LiteBIRD} 
in terms of the polarization of the CMB radiation. 
Furthermore, it is meaningful to remark that 
the growth of the matter density perturbations in modified gravity 
affects the spectrum of weak lensing (for a concrete way of comparing the 
theoretical predictions with the observations, see~\cite{Amendola:2007rr}), 
and therefore more precise future observations of 
weak lensing effects have a potential to present the chance to find out 
the signal of the modification of gravity. 

It is considered that the consequences obtained in this work can be a clue of 
explore the features of dark energy as well as inflation. 
By developing this work further, 
it is strongly expected that we are able to construct a more sophisticated 
and realistic inflation model, in which the power spectrum of the 
curvature perturbations is consistent with the observations, 
the reheating mechanism is well understood, 
and the structure formation can be explained more naturally. 

\section*{Acknowledgments}

We would like to thank Mark Trodden for useful comments. 
K.B., S.D.O. and L.S. would like to appreciate the support and very kind hospitality 
at Eurasian National University, where the work was developed. 
K.B. also expresses his sincere gratitude to National Center for Theoretical 
Sciences and National Tsing Hua University very much for the very kind and 
warm hospitality, where the revision of this work was executed. 
A. L-R. is grateful to MICINN (Spain) for three months visiting grant at Trento university.
The work is supported in part 
by 
MICINN (Spain) project FIS2010-15640 and AGAUR (Catalonia) 2009SGR-994
(S.D.O. and A. L-R.).

\section*{Appendix A: Conformal transformation of exponential model for inflation}

In several cases, a suitable conformal frame to study inflation 
may be the so-called ``Einstein frame''. 
An $F(R)$ gravity theory can be rewritten in the scalar field theory 
form via the conformal transformation. 
We can rewrite the action in Eq.~(\ref{action}) by introducing a scalar
field which couples to the curvature. 
Of course, this is not exactly physically-equivalent formulation, 
but the formulation in the Einstein frame 
may be used to obtain some of intermediate results in simpler form 
(especially, the case that the matter is not taken into account). 

We introduce a scalar field $A$ into the action 
\begin{equation}
I_\mathrm{JF}=\frac{1}{2\kappa^{2}}\int_{\mathcal{M}}\sqrt{-g}\left[
F'(A) \, (R-A)+F(A)\right] d^{4}x\,.
\label{JordanFrame}
\end{equation}
Here, the subscript ``JF'' means ``the Jordan frame'' 
and we neglect the contribute of matter. By making the variation of the
action with respect to $A$, we have $A=R$. We define the scalar field $\sigma$ 
as
\begin{equation}
\sigma = -\frac{\sqrt{3}}{\sqrt{2\kappa^2}}\ln [F'(A)]\label{sigma}\,.
\end{equation}
We make the conformal transformation of the metric 
\begin{equation}
\tilde g_{\mu\nu}=\mathrm{e}^{-\sigma}g_{\mu\nu}\label{conforme}\,,
\end{equation}
for which we acquire the ``Einstein frame'' (EF) action of the scalar field 
$\sigma$~\cite{F-M-C}
%
\begin{eqnarray}
I_\mathrm{EF} \Eqn{=} \int_{\mathcal{M}} d^4 x \sqrt{-\tilde{g}} \left\{ \frac{\tilde{R}}{2\kappa^2} -
\frac{1}{2}\left(\frac{F''(A)}{F'(A)}\right)^2
\tilde{g}^{\mu\nu}\partial_\mu A \partial_\nu A - \frac{1}{2\kappa^2}\left(\frac{A}{F'(A)}
+ \frac{F(A)}{F'(A)^2}\right)\right\} \nonumber\\ 
\Eqn{=} 
\int_{\mathcal{M}} d^4 x \sqrt{-\tilde{g}} \left( \frac{\tilde{R}}{2\kappa^2} -
\frac{1}{2}\tilde{g}^{\mu\nu}
\partial_\mu \sigma \partial_\nu \sigma + V(\sigma)\right)\,,\label{EinsteinFrame}
\end{eqnarray}
%
where
\begin{equation}
V(\sigma)\equiv-\frac{1}{2\kappa^2}\left(\frac{A}{F'(A)} - \frac{F(A)}{F'(A)^2}\right)=-\frac{1}{2\kappa^2}\left\{\mathrm{e}^{\sigma}R(\mathrm{e}^{-\sigma})
-\mathrm{e}^{2\sigma}F[R(\mathrm{e}^{-\sigma})]\label{V(sigma)}\right\}\,.
\end{equation}
Here, $R(\mathrm{e}^{-\sigma})$ is the solution of
Eq.~(\ref{sigma}) with $A=R$, becoming $R$ a function of $\mathrm{e}^{-\sigma}$, and $\tilde{R}$ denotes the Ricci scalar evaluated with respect to the conformal metric $\tilde{g}_{\mu\nu}$. 
Furthermore, 
$\tilde{g}=\mathrm{e}^{-4\sigma}g$ is the determinant of conformal metric. 

As an example, 
we explore our unified model (\ref{total}) with $\gamma=1$. 
Since we are interested in the de Sitter solution, 
we take $\exp[-(R/R_\mathrm{i})^n]\rightarrow 0$ and neglect 
the cosmological constant $\Lambda$. 
In this case 
the potential $V(\sigma)$ reads 
\begin{equation}
V(\sigma)=-\frac{1}{2\kappa^2}\left[\tilde R\left(\frac{\e^{-\tilde \sigma}-1}{\alpha}\right)^{\frac{1}{\alpha-1}}\left(\e^{\tilde \sigma}-2\e^{2\tilde \sigma}\right)+\Lambda_\mathrm{i}\,\e^{2\tilde \sigma}\right]\,.
\end{equation}
According with Sec.~\ref{Inflation}, 
we put $\tilde R_\mathrm{i}=R_{\mathrm{dS}}$. 
It is clearly seen that for $R=R_{\mathrm{dS}}$, $\sigma_{\mathrm{dS}}=-\sqrt{3/(2\kappa^2)}\log(1+\alpha)$ and $V'(\sigma_{\mathrm{dS}})=0$, 
where the prime denotes the derivative with respect to the inflation field $\sigma$. 
Since $V''(\sigma_{\mathrm{dS}})>0$, 
the scalar potential has a minimum, that is a necessary condition for 
a slow-roll inflation. 
For slow-roll parameters, we have to require 
\begin{eqnarray}
\epsilon(\sigma) \Eqn{=} \frac{1}{2\kappa^2}\left(\frac{V'(\sigma)}{V(\sigma)}\right)^2\ll 1\,,\nonumber\\
|\eta(\sigma)| \Eqn{=} \frac{1}{\kappa^2}\left|\frac{V''(\sigma)}{V(\sigma)}\right|\ll 1\,.
\end{eqnarray}
By defining the energy density and pressure of $\sigma$ as 
$\rho_{\sigma}=\dot{\sigma}^2/2-V(\sigma)$ and 
$P_{\sigma}=\dot{\sigma}^2/2+V(\sigma)$, 
these conditions imply that the gravitational field equations in the flat 
FLRW space-time are given by $3H^2/\kappa^2=-V(\sigma)$, 
$3H\dot{\sigma}\simeq-V'(\sigma)$, and that $\ddot{a}(t)>0$, 
and hence guarantee a sufficiently long time inflation. 
In our case, since $V(\sigma_{\mathrm{dS}})\neq 0$, these two conditions are 
well satisfied around the de Sitter solution. 
Thus, since $\dot{\sigma}\simeq 0$, we find 
$H_{\mathrm{dS}}=R_{\mathrm{dS}}/\left[12(1+\alpha)\right]=\tilde R_{\mathrm{dS}}/12$.

\section*{Appendix B: Asymptotically phantom or quintessence modified gravity}

In general, 
realistic models of modified gravity are similar to GR with the cosmological constant, i.e., the dark energy fluid with the EoS parameter 
$\omega_{\mathrm{DE}}=-1$ and the de Sitter universe as the final scenario for 
the cosmological evolution. 
Since in principle quintessence/phantom-dark energy phases are not excluded by 
observations, it may be of some interest to try to reconstruct an $F(R)$ gravity theory where the quintessence or phantom dark energy 
(with a constant $\omega_{\mathrm{DE}}$) emerges. 
The big difficulty is due to the fact that in the dark energy density 
$\rho_{\mathrm{DE}}$ 
and pressure $P_{\mathrm{DE}}$ of modified gravity, 
the effective gravitational terms appear.
In this appendix, we reconstruct the form of $F(R)$ gravity which resembles to a fluid with $\omega_{\mathrm{DE}}$ being very close but not equal to $-1$.

If the energy density of 
a quintessence/phantom fluid is given by
\begin{equation}
\rho=\rho_0(z+1)^{3(1+\omega)}\,,
\label{A}
\end{equation}
where $\omega$ is the EoS parameter, 
the Hubble parameter reads 
\begin{equation}
H(z)=\sqrt{\frac{\kappa^2}{3}\rho}\simeq \sqrt{\frac{\kappa^2\rho_0}{3}}+\frac{1}{2}\sqrt{3\kappa^2\rho_0}(1+\omega)\log[z+1]\,. 
\end{equation}
Here, we have taken into account that $\omega$ is very close to $-1$. 
We can write $R$ as a function of the redshift as 
\begin{equation}
R(z)=\frac{1}{2}\kappa^2\rho_0\left[2+3(1+\omega)\log(z+1)\right]\left[1-3\omega+6(1+\omega)\log(z+1)\right]\,. 
\label{C}
\end{equation}
In addition, from Eq.~(\ref{rhoeffRG}), in vacuum we find
\begin{eqnarray}
\rho_{\mathrm{eff}} \Eqn{\equiv} \rho_{\mathrm{DE}}=
\frac{1}{2\kappa^{2}}
\Biggl\{
\left[ \left(\frac{d R(z)}{dz}\right)^{-1}\frac{d F(z)}{d z}R(z)-F(z)\right]
-6H^2(z)\left[\left(\frac{d R(z)}{dz}\right)^{-1}\frac{d F(z)}{d z}-1\right] 
\nonumber\\ 
&&
{}+6H^2(z)(z+1)\frac{d R(z)}{d z}\left[\left(\frac{d^2 R(z)}{dz^2}\right)^{-1}\frac{d F(z)}{d z}+\left(\frac{d R(z)}{dz}\right)^{-2}\frac{d^2 F(z)}{d z^2}
\right]\Biggl\}\,.
\label{B} 
\end{eqnarray}
Here, $F(R)$ model is expressed as a function of the redshift $F(z)$. 
By equating $\rho_{\mathrm{eff}}$ to $\rho$ of Eq.~(\ref{A}), 
we can find the $F(R)$ model realizing this cosmology. 
For $|\omega-1|\ll 0$, the solution of Eq.~(\ref{B}) is given by 
\begin{equation}
F(z)\simeq\frac{6\kappa^2\rho_0 \left[11+(34-9\omega)\omega\right]}{(5-3\omega)^2}-6\kappa^2\rho_0(1+\omega)\log(z+1)\,,
\end{equation}
{}From Eq.~(\ref{C}), we have 
\begin{equation}
z=-1+
\exp\left\{ 
\frac{\rho_0\kappa^2(5-3\omega)\pm (1+\omega)\sqrt{\rho_0\kappa^2\left[16 R+9\rho_0\kappa^2(1+\omega)^2\right]}}{12\rho_0\kappa^2(1+\omega)} \right\}\,,
\end{equation}
where the plus sign corresponds to the quintessence solution, whereas 
the minus sign does to the phantom one. 
We can now write the modified gravity model 
as a function of the Ricci scalar as 
\begin{equation}
F(R)=\frac{\rho_0\kappa^2 (257+183\omega+27\omega^2-27\omega^3)}{2(5-3\omega)^2}\pm\frac{\sqrt{\rho_0\kappa^2\left[16 R+9\rho_0\kappa^2(1+\omega)^2\right]}}{2}\,,
\end{equation}
where $\rho_0$ is a free parameter of the theory, and $\omega$ is the EoS parameter of dark energy coming from the modification of gravity and 
equivalent to $\omega_{\mathrm{DE}}$. 
In this way, 
we have reconstructed the form of $F(R)$ gravity that gives the quintessence or phantom fluid solution in the empty universe. 
Remind that this reconstruction is valid for $\omega_{\mathrm{DE}}$ close to $-1$.

\end{document}